\documentclass[10pt]{article}

\usepackage[utf8]{inputenc}
\usepackage{graphicx,tabularx,amsmath,amsthm,mathtools,amsfonts,bm,amssymb,setspace,color,float,caption,subcaption,verbatim,ulem}
\usepackage[pdfpagemode=UseNone,pdfstartview={XYZ null null null}]{hyperref}
\usepackage[longnamesfirst,numbers,square]{natbib}
\usepackage[margin=0.5in]{geometry} 
\usepackage{rotating, float}
\newcommand{\R}{\mathbb{R}}

\newcommand{\C}{\mathbb{C}}
\newcommand{\Z}{\mathbb{Z}}
\newcommand{\M}{\mathbb{M}}

\newcommand{\A}{\mathbb{A}}
\newcommand{\Laplacian}{\mathbb{L}}

\newcommand{\Ctot}{C_{\rm tot}}
\newcommand{\Gtot}{G_{\rm tot}}
\newcommand{\kcat}{k_{\rm cat}}
\newcommand{\kminus}{k^-}
\newcommand{\kon}{k^{\rm on}}
\newcommand{\koff}{k^{\rm off}}
\newcommand{\FF}{\mathcal{F}}
\newcommand{\GG}{\mathcal{G}}

\newcommand{\bU}{\bm{U}}
\newcommand{\bV}{\bm{V}}

\newcommand{\bF}{\bm{F}}

\newcommand{\x}{{\mathbf x}}
\newcommand{\n}{{\mathbf n}}
\newcommand{\e}{{\rm e}}
\newcommand{\tr}{{\rm tr}}

\newcommand{\FFc}{\frac{\partial\mathcal{F}}{\partial c}}
\newcommand{\FFg}{\frac{\partial\mathcal{F}}{\partial g}}
\newcommand{\FFC}{\frac{\partial\mathcal{F}}{\partial C}}
\newcommand{\GGc}{\frac{\partial\mathcal{G}}{\partial c}}
\newcommand{\GGg}{\frac{\partial\mathcal{G}}{\partial g}}
\newcommand{\GGG}{\frac{\partial\mathcal{G}}{\partial G}}

\newcommand{\Cavg}{C_{\text{avg}}}
\newcommand{\Gavg}{G_{\text{avg}}}

\begin{document}

\title{Pattern Formation in a Coupled Membrane-Bulk Reaction-Diffusion Model for Intracellular Polarization and Oscillations}

\author{Fr\'{e}d\'{e}ric Paquin-Lefebvre \textsuperscript{1,*}, Bin Xu \textsuperscript{2,*}, Kelsey L. DiPietro\textsuperscript{2,3}, Alan E. Lindsay\textsuperscript{2}, Alexandra Jilkine\textsuperscript{2}}

\date{}
\maketitle

\noindent \textbf{1} Department of Mathematics and Institute of Applied Mathematics, University of British Columbia, Vancouver, Canada.\\
\textbf{2} {Department of Applied and Computational Mathematics and Statistics, University of Notre Dame, Notre Dame, IN, USA, 46556.\\
\textbf{3}  Sandia National Laboratories, NM, USA, 46556.\\
\textbf{*} \textit{Denotes equal contribution.}}

\begin{abstract}
Reaction-diffusion systems have been widely used to study spatio-temporal phenomena in cell biology, such as cell polarization. Coupled bulk-surface models naturally include compartmentalization of cytosolic and membrane-bound polarity molecules. Here we study the distribution of the polarity protein Cdc42 in a mass-conserved membrane-bulk model, and explore the effects of diffusion and spatial dimensionality on spatio-temporal pattern formation. We first analyze a one-dimensional (1-D) model for Cdc42 oscillations in fission yeast, consisting of two diffusion equations in the bulk domain coupled to nonlinear ODEs for binding kinetics at each end of the cell. In 1-D, our analysis reveals the existence of symmetric and asymmetric steady states, as well as anti-phase relaxation oscillations typical of slow-fast systems. We then extend our analysis to a two-dimensional (2-D) model with circular bulk geometry, for which species can either diffuse inside the cell or become bound to the membrane and undergo a nonlinear reaction-diffusion process. We also consider a nonlocal system of PDEs approximating the dynamics of the 2-D membrane-bulk model in the limit of fast bulk diffusion. In all three model variants we find that mass conservation selects perturbations of spatial modes that simply redistribute mass. In 1-D, only anti-phase oscillations between the two ends of the cell can occur, and in-phase oscillations are excluded. In higher dimensions, no radially symmetric oscillations are observed. Instead, the only instabilities are symmetry-breaking, either corresponding to stationary Turing instabilities, leading to the formation of stationary patterns, or to oscillatory Turing instabilities, leading to traveling and standing waves. Codimension-two Bogdanov--Takens bifurcations occur when the two distinct instabilities coincide, causing traveling waves to slow down and to eventually become stationary patterns. Our work clarifies the effect of geometry and dimensionality on behaviors observed in mass-conserved cell polarity models.
\end{abstract}


\section{Introduction}
Cell growth and division require long-distance communication and coordination between two ends of the cell. To polarize, cells require the ability to form distinct cellular domains with different molecular components.  Establishment of cell polarity relies on local accumulation of signaling molecules on the membrane and has attracted considerable attention from mathematicians and physicists \cite{jilkine2011comparison,rappel2017mechanisms,halatek2018self}. Positive and negative feedback loops can result in spontaneous symmetry breaking of an initially homogeneous distribution of polarity regulators to form one or multiple clusters, as well as dispersal and reformation of these clusters in an oscillatory manner \cite{johnson2011}. Thus, polarization involves spatial diffusion processes coupled with biochemical reactions occurring within localized signaling compartments.

The master regulator of cell polarity in a variety of organisms, from yeast to humans, is the protein Cdc42 \cite{etienne2002,martin2014}. Previous mathematical modeling of Cdc42 in cell polarization has focused primarily on symmetry breaking and establishment of active Cdc42 cortical zones in budding yeast, often via a Turing mechanism (reviewed in \cite{goryachev2017}). Detailed and phenomenological models of cell polarization investigated the effects of single or multiple positive feedback loops and mass conservation on the formation of a unique polarity zone \cite{otsuji2007,goryachev2008, mori2011, jilkine2011, freisinger2013,chiou2018}. In some circumstances, two or more active Cdc42 domains can coexist for some time, and the basis for the switch from a single to multiple polarity zones is not yet fully understood \cite{wu2013,chiou2018}. Here we focus on a model of spatio-temporal oscillations of Cdc42 from pole-to-pole that have been observed in fission yeast \cite{das2012}, a model organism for understanding how cells integrate polarity and spatial coordination of growth.

The mathematical modeling of  cell polarization typically involves a system of reaction-diffusion equations for the concentrations of membrane-bound active Cdc42 molecules and cytosolic inactive Cdc42 molecules with no-flux or periodic boundary conditions \cite{goryachev2008, mori2008, mori2011, lo2014}. This approach does not take into effect the spatial segregation of polarity molecules in the cell. For Cdc42 and many other polarity molecules, most biochemical reactions of interest happen at the interface between the cytosol and the membrane. Because the active form of the protein is found on the membrane, while the inactive form is in the cytosol \cite{etienne2002}, a more appropriate formulation would be a membrane bulk-cytosol model with diffusion in the interior of the cell, and boundary conditions modeling the exchange between the membrane and the cytosol. Pattern formation in membrane-bulk models with applications to cell polarity have previously been analyzed in \cite{ratz2012turing, ratz2014, madzvamuse2015} through linear stability analysis and PDE numerical simulations. Stationary front solutions corresponding to a polarized cell were demonstrated to exist in both the reaction-diffusion \cite{mori2011} and membrane-bulk versions \cite{cusseddu2018} of the wave-pinning model of cell polarization. When considered on a 1-D bulk domain, coupled membrane-bulk reaction-diffusion systems become coupled PDE--ODE systems. Recent studies of coupled PDE--ODE models \cite{gomez2007, gou2015_SIAM, gou2017,xu2017} showed the possible collective synchronization of localized active units (or membranes) coupled through a linear bulk diffusion field, even if each unit is at rest when isolated from the group.

Our work focuses on the spatio-temporal dynamics of Cdc42 during cell growth of fission yeast. Fission yeast (\textit{S. pombe}) cells are rod-shaped and grow in length by tip extension, while maintaining a constant diameter of 3 $\mu$m. During their cell cycle, the cells switch from growing at one end to growing at both ends. The active Cdc42 molecules localize at the growth site. At the cell ends, Cdc42 exhibits temporal oscillations with a period around 5 minutes \cite{das2012}. Earlier mathematical models for cell polarity in fission yeast emphasize the role of positive feedbacks on symmetry breaking \cite{csikasz2008,cerone2012}. These models assume that some ``tip factors'' (delivered by microtubules) are present at the two ends of the fission yeast that leads to Cdc42 preferentially binding at the two ends of the cell \cite{kokkoris2014,martin2014}. Recent models for spatio-temporal oscillations of Cdc42 involve both positive and negative feedback loops, as well as time delays \cite{das2012,xu2016}. Several potential positive feedbacks are presented in a recent review paper \cite{goryachev2017} while potential negative feedbacks may involve a Cdc42 GEF (Guanine nucleotide exchange factor) \cite{das2015}. In \cite{xu2018}, we used a 1-D coupled PDE--ODE model to include possible positive and negative feedbacks between Cdc42 and its GEF, with our analysis of the resulting dynamics focusing on the fast bulk diffusion limit, for which the full model is reduced to an ODE model. We then explored in \cite{xu2018BMB} the effect of intrinsic noise on Cdc42 oscillations and compared the dynamics of the reduced ODE model with its stochastic counterpart.

In this work, we relax the fast bulk diffusion assumption and compare the dynamics of the coupled PDE--ODE system to its well-mixed ODE limit using linear stability analysis, time-dependent simulations and numerical bifurcation analysis. We study the effect of decreasing the bulk diffusion level on the stability of the symmetric steady state, at which both ends of the cell have the same amount of Cdc42. We find conditions for a Hopf bifurcation, and analyze how the Cdc42 and GEF bulk diffusion rates affect its criticality. We also explore the role of one key biochemical reaction rate, the GEF dissociation rate, on the emergence of oscillatory dynamics.  We present numerical simulations showing two different types of pole-to-pole oscillations, either weakly nonlinear or of relaxation type.

In the second part of the paper, we extend the 1-D bulk-cytosol to a two-dimensional circular domain, with pure diffusion in the interior, as well as surface diffusion and nonlinear reactions on the membrane. We also consider a reduced one-dimensional system of nonlocal reaction-diffusion equations that governs the dynamics of membrane-bound species when assuming a well-mixed bulk. Linear stability analysis is performed for both the full membrane-bulk model and its nonlocal reduction. Our aim is to investigate the symmetry-breaking effects of varying the bulk diffusion coefficients, the membrane diffusion coefficients and the GEF dissociation rate. Finally, we present numerical results that confirm our analysis and exhibit a variety of spatio-temporal patterns on a unit disk, such as stationary Turing patterns, traveling waves and standing waves.

\section{Mathematical models}\label{sec:models}

\subsection{One-dimensional PDE--ODE model}
\begin{figure}[t!]
\centering
\includegraphics[width=0.7\linewidth]{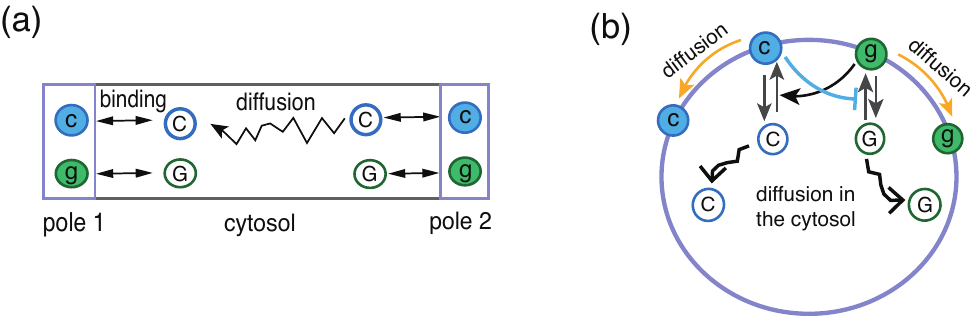}
\caption{(a) A one-dimensional PDE--ODE model with bulk diffusion in the interior given by equation \eqref{eq:pde} and binding kinetics at two well-mixed ends given by the ODE system \eqref{eq:ode}. (b) A two-dimensional model given by equation \eqref{eq:2Dmodel} and equation \eqref{eq:consv2dMAIN} with diffusion along the membrane. Filled circle: membrane-bound molecule. Open circle: cytosolic molecule. Blue: Cdc42. Green: GEF.}
\label{fig:2Dmodel}
\end{figure}

We first model the cell as a one-dimensional bulk domain separating two well-mixed compartments. This simplified geometry is based on the competition of two growth zones of active Cdc42, localized at the cell tips, for a common substrate, corresponding to inactive Cdc42, that diffuses in the cytosol and can become active at the two tips in the presence of a Cdc42 GEF. We also consider the spatial distribution of the Cdc42 GEF, found at each cell tip in its active form or in the cytosol in its inactive form. A schematic representation of this 1-D model is shown in Fig.~\ref{fig:2Dmodel}a.

Let $C(x,t)$ and $G(x,t)$ denote the concentrations of inactive Cdc42 and GEF in a cytosolic (or bulk) domain of length $L$, satisfying
\begin{equation}\label{eq:pde}
\frac{\partial C}{\partial t} = D_c \frac{\partial^2 C}{\partial x^2}\,, \quad \frac{\partial G}{\partial t} = D_g \frac{\partial^2 G}{\partial x^2}\,, \quad 0 < x < L.
\end{equation}
where $D_c$ and $D_g$ are two constant bulk diffusion coefficients. Near each of the two tips $(i=1,\,2)$, the binding and unbinding processes are modeled by flux boundary conditions given by, in $x=0$,
\begin{equation}\label{eq:BC_1}
\left. D_c\frac{\partial C}{\partial x}\right|_{x=0} = k^+(c_1(t),g_1(t)) C(0,t) - k^- c_1(t) \,, \quad \left. D_g\frac{\partial G}{\partial x}\right|_{x=0} = k^{\rm on}(c_1(t))G(0,t) - k^{\rm off} g_1(t)\,,
\end{equation}
while in $x=L$ the boundary conditions satisfy,
\begin{equation}\label{eq:BC_2}
\left. -D_c\frac{\partial C}{\partial x}\right|_{x=L} = k^+(c_2(t),g_2(t)) C(L,t) - k^- c_2(t) \,, \quad \left. -D_g\frac{\partial G}{\partial x}\right|_{x=L} = k^{\rm on}(c_2(t))G(L,t) - k^{\rm off} g_2(t)\,,
\end{equation}
where $c_i(t)$ and $g_i(t)$ denote the concentrations of active Cdc42 and GEF. We assume that active GEF promotes the binding of Cdc42 on each tip, and this results in the following association rate $k^+$:
\begin{equation}
k^+(c_i,g_i)=(k_0+\kcat c_i^2)g_i\,.
\end{equation}
Here we assume a form of positive feedback that was used in previous models of fission yeast \cite{csikasz2008} and \cite{cerone2012}. The nonlinearity $c_i^2$ is necessary for symmetry breaking, allowing an asymmetric steady state to emerge. Potential chemical reactions that can lead to such kinetics have been discussed in \cite{goryachev2017}. Some form of negative feedback effect is necessary to obtain the pole-to-pole oscillations observed in fission yeast \cite{das2012}. For negative feedback, we assume that active Cdc42 inhibits the binding of GEF and that the association rate $\kon$ is given by 
\begin{equation}
k^{\rm on}(c_i)=\frac{k^{\rm on}}{1+\kappa c_i^2}\,.
\end{equation}
We also note that this specific implementation has been discussed in our previous work~\cite{xu2018}, where it was found that negative feedback is more likely to act through inhibition of GEF association rather than upregulation of GEF dissociation. Finally, the dissociation rates $k^-$ and $k^{\rm off}$ are chosen to be linear.

For the rest of the paper, we will employ the following notation for the binding kinetics:
\begin{equation}\label{eq:kinetics}
\FF\left(C,c,g\right) = (k_0 + \kcat c^2) g C - \kminus c\,, \quad \GG\left(G,c,g\right) = \frac{\kon}{1+\kappa c^2} G - \koff g\,.
\end{equation}
Using this notation, we define the nonlinear ODEs regulating the active species dynamics as
\begin{equation}\label{eq:ode}
\begin{split}
\frac{d c_i(t)}{dt}&= \FF\left(C|_{x=(i-1)L},c_i(t),g_i(t)\right)\,, \quad i=1\,,2\,, \\
\frac{d g_i(t)}{dt}&= \GG\left(G|_{x=(i-1)L},c_i(t),g_i(t)\right)\,, \quad i=1\,,2\,.
\end{split}
\end{equation}
Here, we highlight that protein production and decay are neglected by our model formulation, and that only the exchange between the active and inactive forms is considered. This results in the total amount of Cdc42 and GEF being constant over the timescale on which cell polarization takes place. Hence, our model formulation implies the following conservation laws:
\begin{equation}\label{eq:mass_cons}
 c_1(t) + c_2(t) + \int_0^L C(x,t) dx = \Ctot\,, \quad g_1(t) + g_2(t) + \int_0^L G(x,t) dx = \Gtot\,,
\end{equation}
where $\Ctot$ and $\Gtot$ are the total amount of Cdc42 and GEF.

Finally, we mention the limit of fast bulk diffusion, corresponding to the well-mixed regime, for which the concentrations of inactive species $(C(x,t), G(x,t))$ become spatially homogeneous $(C(t),G(t))$. Using the conservation laws \eqref{eq:mass_cons}, we can reduce the full coupled PDE--ODE model to a system of four nonlinear ODEs given by
\begin{equation}\label{eq:well_mixed}
\begin{split}
\frac{d c_i(t)}{dt}&= \FF\left(C(t),c_i(t),g_i(t)\right)\,, \quad i=1,2,\\
\frac{d g_i(t)}{dt}&=\GG\left(G(t),c_i(t),g_i(t)\right)\,,\quad i=1,2.
\end{split}
\end{equation}
where $C(t)$ and $G(t)$ are  defined by
\begin{equation}\label{eq:leading_order}
 C(t) \equiv \frac{\Ctot - c_1(t) - c_2(t)}{L}\,, \quad G(t) \equiv \frac{\Gtot - g_1(t) - g_2(t)}{L}\,.
\end{equation}

The model presented here is dimensionless. A justification of the various parameters (and their values) can be found in Appendix~\ref{sec:appendix_params}.

\subsection{Two-dimensional membrane-bulk reaction-diffusion model}\label{subsec:2_D_membrane_bulk}
In this section we consider the model on a two-dimensional domain $\Omega\in\mathbb{R}^2$ with boundary $\partial\Omega$ corresponding to the cell membrane. The geometry of the model is shown in Fig.~\ref{fig:2Dmodel}b. We assume that inactive Cdc42 in the cytosol only undergoes diffusion, with all the nonlinear binding kinetics taking place near the boundary. We also consider the lateral diffusion of active Cdc42 on the boundary, with a diffusion coefficient $D_c^m$. Let $C(\x, t)$ denote the concentration of Cdc42 in the bulk, and $c(\x,t)$ denote the concentration of active Cdc42 on the membrane. These concentrations evolve according to a system of reaction-diffusion equations defined by
\begin{subequations}\label{eq:2Dmodel}
\begin{align}
\frac{\partial C(\x,t)}{\partial t}&=D_c\nabla^2C(\x,t),\quad \x\in\Omega,\\
\frac{\partial c(\x,t)}{\partial t}&=D_c^m\nabla_s^2 c(\x,t)+\FF[C(\x,t), c(\x, t), g(\x,t)],\quad \x\in\partial\Omega,\\
-D_c(\n\cdot\nabla C)&=\FF[C(\x,t), c(\x, t), g(\x,t)],\quad \x\in\partial\Omega\,,
\end{align}
\end{subequations}
where $\nabla_s^2$ is the Laplace-Beltrami operator and $\n$ represents the outward unit normal vector to $\partial \Omega$. Similar reaction-diffusion equations are defined for GEF, the second species; see below.
\begin{subequations}\label{eq:consv2dMAIN}
\begin{align}
\frac{\partial G(\x,t)}{\partial t}&=D_g\nabla^2G(\x,t),\quad \x\in\Omega,\\
\frac{\partial g(\x,t)}{\partial t}&=D_g^m\nabla_s^2 g(\x,t)+\GG[G(\x,t),c(\x,t),g(\x,t)],\quad \x\in\partial\Omega,\\
-D_g(\n\cdot\nabla G)&=\GG[G(\x,t),c(\x,t),g(\x,t)],\quad \x\in\partial\Omega.
\end{align}
\end{subequations}
The nonlinear functions $\FF$ and $\GG$ are as defined in \eqref{eq:kinetics}. Mass conservation arises from the formulation of the model, and we have the following conservation laws:
\begin{equation}\label{eq:consv2d}
\int_{\Omega} C(\x,t)d\x+\int_{\partial\Omega}c(\x,t)d\x = C_{\rm tot}\,, \quad \int_{\Omega} G(\x,t)d\x+\int_{\partial\Omega}g(\x,t)d\x=G_{\rm tot}\,.
\end{equation}
In Appendix \ref{subsec:appendix_mass}, we demonstrate that $\Ctot$ and $\Gtot$ must remain constant and show that the numerical method employed preserves mass over the timescale of cell polarization.

\subsection{Nonlocal PDE model on a circular domain}

Models of polarization often consider the cytosolic forms of Rho GTPases to be well-mixed \cite{goryachev2008, lo2014}.
We conclude this section with the introduction of a reduced nonlocal reaction-diffusion model governing the dynamics of membrane-bound species in the limit of fast bulk diffusion  $D \to\infty$. For simplicity, we consider a two-dimensional bulk domain $\Omega\in\mathbb{R}^2$  to be  a disk of radius $R$ and obtain a nonlocal reaction-diffusion model for membrane-bound species on a circle. In the well-mixed bulk, $C(t)$ and $G(t)$ are the spatially homogeneous concentrations of inactive Cdc42 and GEF, which are defined  using the conservation laws as
\begin{subequations}\label{eq:consv}
\begin{align}
C(t)&\equiv\frac{\Ctot}{|\Omega|}-\frac{1}{|\Omega|}\int_{\partial\Omega}c(\x,t)d\x=\Cavg-\frac{1}{\pi R}\int_0^{2\pi}c(\theta,t)d\theta\,,\\
G(t)&\equiv\frac{\Gtot}{|\Omega|}-\frac{1}{|\Omega|}\int_{\partial\Omega}g(\x,t)d\x=\Gavg-\frac{1}{\pi R}\int_0^{2\pi}g(\theta,t)d\theta\,,
\end{align}
\end{subequations}
where $\Cavg = \Ctot/(\pi R^2)$ and $\Gavg = \Gtot/(\pi R^2)$ are the average concentrations. Upon using polar coordinates to parametrize the Laplace-Beltrami operator, we write $\nabla_s^2 = R^{-2}\partial_{\theta\theta}$, with $\theta$ the angular variable, and find that the active species $c(\theta,t)$ and $g(\theta,t)$ must satisfy
\begin{subequations}\label{eq:nonlocal}
\begin{align}
\frac{\partial c(\theta,t)}{\partial t}&=\frac{D_c^m}{R^2}\frac{\partial^2 c(\theta,t)}{\partial \theta^2}+\FF(C(t),c(\theta,t),g(\theta,t))\,, \quad 0\leq \theta < 2\pi\,,\\
\frac{\partial g(\theta,t)}{\partial t}&=\frac{D_g^m}{R^2}\frac{\partial^2 g(\theta,t)}{\partial \theta^2}+\GG(G(t),c(\theta,t),g(\theta,t))\,, \quad 0\leq \theta < 2\pi\,.
\end{align}
\end{subequations}

\section{Results}

\subsection{One-dimensional PDE--ODE model}\label{sec:PDE_ODE}

\subsubsection{Steady states}\label{subsec:steady_states}
We first consider the steady states of the coupled PDE--ODE system \eqref{eq:pde}--\eqref{eq:mass_cons}. In the bulk domain, the Cdc42 and GEF profiles at a steady state must satisfy the following boundary value problems:
\begin{equation}
C''(x) = 0\,, \quad G''(x) = 0\,, \quad C'(0) = C'(L) = G'(0) = G'(L) = 0\,, \quad 0 < x < L\,.
\end{equation}
Hence, $C(x) \equiv C_a$ and $G(x) \equiv G_a$ are spatially homogeneous. From the conservation laws \eqref{eq:mass_cons} we obtain
\begin{equation}
C_a = \frac{\Ctot - c_{1a} - c_{2a}}{L}\,, \quad G_a = \frac{\Gtot - g_{1a} - g_{2a}}{L}\,,
\end{equation}
with the subscript $a$ referring to potential asymmetry $c_{1a} \neq c_{2a}$ and $g_{1a} \neq g_{2a}$ between both tips. Next, the active species steady state solutions satisfy the following system of nonlinear algebraic equations:
\begin{equation}\label{eq:steady_states}
\FF(C_a,c_{1a},g_{1a}) = 0\,, \quad \FF(C_a,c_{2a},g_{2a}) = 0\,, \quad \GG(G_a,c_{1a},g_{1a}) = 0\,, \quad \GG(G_a,c_{2a},g_{2a}) = 0\,,
\end{equation}
from which $g_{1a}$ and $g_{2a}$ are expressed as,
\begin{equation}
g_{1a} = \frac{L \kminus c_{1a}}{(k_0+\kcat c_{1a}^2)(\Ctot - c_{1a} - c_{2a})}\,, \quad g_{2a} = \frac{L \kminus c_{2a}}{(k_0+\kcat c_{2a}^2)(\Ctot - c_{1a} - c_{2a})}\,,
\end{equation}
with the corresponding Cdc42 values satisfying a system of two polynomial equations given by
\begin{equation}
H(c_{1a},c_{2a}) = 0\,, \quad H(c_{2a},c_{1a}) = 0\,,
\end{equation}
where $H(c_1,c_2)$ is defined by
\begin{equation}
\begin{split}
H(c_1,c_2) &\equiv \frac{\kon\Gtot}{L}(\Ctot - c_1 - c_2)(k_0 + \kcat c_1^2)(k_0 + \kcat c_2^2) - \kminus\kon c_1(k_0 + \kcat c_2^2) - \kminus\kon c_2(k_0 + \kcat c_1^2) \\ & - L \koff \kminus c_1(1+\kappa c_1^2) (k_0 + \kcat c_2^2)\,.
\end{split}
\end{equation}
Assuming the same binding kinetics for both cell tips causes reflection symmetry of solutions with respect to the midpoint $L/2$. As a result, interchanging $(c_{1a},g_{1a})$ with $(c_{2a},g_{2a})$ also satisfies \eqref{eq:steady_states}. 

The steady state is symmetric when both tips possess equal solution values $(c_s,g_s)$, with $g_s$ defined by
\begin{equation}
g_s = \frac{L \kminus c_s}{(k_0+\kcat c_s^2)(\Ctot - 2c_s)},
\end{equation}
while $c_s$ satisfies a third degree polynomial equation given by
\begin{equation}\label{eq:cubic}
\left( L^2\kminus\koff\kappa + 2\kon\kcat\Gtot \right)c_s^3 - \kon\kcat\Gtot\Ctot c_s^2 + \left( 2L\kminus\kon + L^2 \kminus\koff + 2 k_0\kon\Gtot \right)c_s - \kon k_0\Gtot\Ctot = 0.
\end{equation}
By Descartes' rule of signs, it can be shown that equation \eqref{eq:cubic} possesses either one or three positive real roots, and therefore there always exists a symmetric steady state. Finally, we point out that the existence of the steady states does not depend on the diffusivity level, and that the structure of solutions remains the same for the reduced ODE system \eqref{eq:well_mixed} in the well-mixed regime.

\subsubsection{Linear stability analysis}\label{subsec:linear_stability_analysis}

We now proceed to a linear stability analysis of the symmetric steady state, leaving the linear stability of a general asymmetric steady state to be addressed via numerical bifurcation analysis. Consider a perturbation of the symmetric steady state given by
\begin{equation}\label{eq:perturbation}
C(x,t) = C_s + \varphi(x)e^{\lambda t}\,, \quad G(x,t) = G_s + \psi(x)e^{\lambda t}\,, \quad 
\begin{pmatrix} c_1(t) \\ g_1(t) \end{pmatrix} = \begin{pmatrix} c_s \\ g_s \end{pmatrix} + \begin{pmatrix} u_1 \\ v_1 \end{pmatrix} e^{\lambda t}\,, \quad
\begin{pmatrix} c_2(t) \\ g_2(t) \end{pmatrix} = \begin{pmatrix} c_s \\ g_s \end{pmatrix} + \begin{pmatrix} u_2 \\ v_2 \end{pmatrix} e^{\lambda t}\,,
\end{equation}
where $\phi(x)$ and $\psi(x)$ are two eigenfunctions, while $(u_i,v_i)^T$ for $i=1,\,2$ are the eigenvectors of the linearized dynamics of active species at the left and right tips. Applying then the conservation laws \eqref{eq:mass_cons} yields
\begin{equation}\label{eq:mass_cons_eig}
u_1 + u_2 + \int_0^L \varphi(x) dx = 0, \quad v_1 + v_2 + \int_0^L \psi(x) dx = 0\,,
\end{equation}
and thus any perturbations can only result in a redistribution of total mass between active and inactive species. Next, upon inserting \eqref{eq:perturbation} within the PDE--ODE system \eqref{eq:pde}--\eqref{eq:ode} and linearizing about the symmetric steady state, we obtain an eigenvalue problem
\begin{equation}\label{eq:bvp}
0 = D_c \varphi_{xx} - \lambda \varphi\,, \quad 0 = D_g \psi_{xx} - \lambda \psi\,, \quad 0 < x < L\,, \\
\end{equation}
subject to boundary conditions at $x=0$ given by
\begin{equation}\label{eq:BC_1_eig}
D_c \varphi_x(0) = \FFC \varphi(0) + \FFc u_1 + \FFg v_1\,, \quad D_g \psi_x(0) = \GGG \psi(0) + \GGc u_1 + \GGg v_1\,,
\end{equation}
while on $x=L$ the boundary conditions are
\begin{equation}\label{eq:BC_2_eig}
D_c \varphi_x(L) = - \FFC \varphi(L) - \FFc u_2 - \FFg v_2\,, \quad D_g \psi_x(L) = - \GGG \psi(L) - \GGc u_2 - \GGg v_2\,.
\end{equation}
Finally, from the linearized ODE dynamics we find that the eigenvectors $(u_i,v_i)^T$ satisfy
\begin{equation}\label{eq:ode_eig}
\begin{pmatrix} \lambda - \FFc & - \FFg \\ - \GGc & \lambda - \GGg \end{pmatrix} \begin{pmatrix} u_1 \\ v_1 \end{pmatrix} = \begin{pmatrix} \FFC \varphi(0) \\ \GGG \psi(0) \end{pmatrix}\,, \qquad
\begin{pmatrix} \lambda - \FFc & - \FFg \\ - \GGc & \lambda - \GGg \end{pmatrix} \begin{pmatrix} u_2 \\ v_2 \end{pmatrix} = \begin{pmatrix} \FFC \varphi(L) \\ \GGG \psi(L) \end{pmatrix}\,.
\end{equation}
All the partial derivatives in equations \eqref{eq:BC_1_eig}--\eqref{eq:ode_eig} are evaluated at the symmetric steady state.

\subsubsection{Hopf bifurcation}\label{subsubsec:cond_hopf}
We first suppose that $\lambda \neq 0$. Hence, the only possible bifurcating eigenvalues are purely imaginary and lead to a Hopf bifurcation. Then, upon solving equation \eqref{eq:bvp}, we find that the eigenfunctions may be written as the sum of an even $(+)$ and an odd $(-)$ parts,
\begin{equation}\label{eq:phi_psi}
\varphi(x) = \varphi_+^0 \frac{\cosh\left(\omega_c\left(\frac{L}{2}-x\right)\right)}{\cosh\left(\omega_c \frac{L}{2}\right)} + \varphi_-^0 \frac{\sinh\left(\omega_c\left(\frac{L}{2}-x\right)\right)}{\sinh\left(\omega_c \frac{L}{2}\right)}\,, \quad 
\psi(x) = \psi_+^0 \frac{\cosh\left(\omega_g\left(\frac{L}{2}-x\right)\right)}{\cosh\left(\omega_g \frac{L}{2}\right)} + \psi_-^0 \frac{\sinh\left(\omega_g\left(\frac{L}{2}-x\right)\right)}{\sinh\left(\omega_g \frac{L}{2}\right)}\,,
\end{equation}
where $\omega_c = \sqrt{\lambda/D_c}$ and $\omega_g = \sqrt{\lambda/D_g}$. Next, the application of the boundary conditions \eqref{eq:BC_1_eig} and \eqref{eq:BC_2_eig} yields the following expressions for the even coefficients $\varphi_+^0$ and $\psi_+^0$:
\begin{equation}
\varphi_+^0 = -\frac{1}{2}\frac{\FFc(u_1+u_2) + \FFg(v_1+v_2)}{D_c\omega_c \tanh\left(\omega_c\frac{L}{2}\right) + \FFC}\,, \quad 
\psi_+^0 = -\frac{1}{2}\frac{\GGc(u_1+u_2) + \GGg(v_1+v_2)}{D_g\omega_g \tanh\left(\omega_g\frac{L}{2}\right) + \GGG}\,,
\end{equation}
while the odd coefficients $\varphi_-^0$ and $\psi_-^0$ are given by
\begin{equation}
\varphi_-^0 = -\frac{1}{2}\frac{\FFc(u_1-u_2) + \FFg(v_1-v_2)}{D_c\omega_c \coth\left(\omega_c\frac{L}{2}\right) + \FFC}\,, \quad
\psi_-^0 = -\frac{1}{2}\frac{\GGc(u_1-u_2) + \GGg(v_1-v_2)}{D_g\omega_g \coth\left(\omega_g\frac{L}{2}\right) + \GGG}\,.
\end{equation}
Since any solutions of the linearized system can be decomposed into an odd and an even parts, we treat the two different modes separately.

\paragraph{Odd mode $(-)$.} For the odd mode, we have that $u_2 = - u_1$, $v_2 = - v_1$ and thus
\begin{equation}\label{eq:phi_psi_odd}
\varphi(x) = -\dfrac{\left(\FFc u_1 + \FFg v_1\right)\sinh\left(\omega_c\left(\frac{L}{2}-x\right)\right)}{\left(D_c\omega_c\coth\left(\omega_c\frac{L}{2}\right) + \FFC\right)\sinh\left(\omega_c\frac{L}{2}\right)}\,, \quad
\psi(x) = -\dfrac{\left(\GGc u_1 + \GGg v_1\right)\sinh\left(\omega_g\left(\frac{L}{2}-x\right)\right)}{\left(D_g\omega_g\coth\left(\omega_c\frac{L}{2}\right) + \GGG\right)\sinh\left(\omega_g\frac{L}{2}\right)}\,,
\end{equation}
which automatically satisfies the constraints \eqref{eq:mass_cons_eig}. We then substitute \eqref{eq:phi_psi_odd} within \eqref{eq:ode_eig} and find that $(u_1,\,v_1)^T$ must be a non-trivial solution of
\begin{equation}
\begin{pmatrix}
\lambda - \dfrac{\FFc D_c \omega_c\coth\left(\omega_c\frac{L}{2}\right)}{D_c\omega_c\coth\left(\omega_c\frac{L}{2}\right) + \FFC} & 
- \dfrac{\FFg D_c\omega_c\coth\left(\omega_c\frac{L}{2}\right)}{D_c\omega_c\coth\left(\omega_c\frac{L}{2}\right) + \FFC} \\[5pt]
- \dfrac{\GGc D_g\omega_g\coth\left(\omega_g\frac{L}{2}\right)}{D_g\omega_g\coth\left(\omega_g\frac{L}{2}\right) + \GGG} & 
\lambda - \dfrac{\GGg D_g\omega_g\coth\left(\omega_g\frac{L}{2}\right)}{D_g\omega_g\coth\left(\omega_c\frac{L}{2}\right) + \GGG}
\end{pmatrix}
\begin{pmatrix} u_1 \\ v_1 \end{pmatrix} = \begin{pmatrix} 0 \\ 0 \end{pmatrix}\,,
\end{equation}
which, upon dividing through $\lambda \neq 0$, can be rearranged into
\begin{equation}
\begin{pmatrix}
\lambda + \frac{\omega_c}{\coth\left(\omega_c \frac{L}{2}\right)}\FFC - \FFc & - \FFg \\
- \GGc & \lambda + \frac{\omega_g}{\coth\left(\omega_g \frac{L}{2}\right)}\GGG - \GGg \\
\end{pmatrix} 
\begin{pmatrix} u_1 \\ v_1\end{pmatrix} = \begin{pmatrix} 0 \\ 0 \end{pmatrix}\,.
\end{equation}
The eigenvalue parameter $\lambda$ is therefore a root of the following characteristic equation:
\begin{equation}\label{eq:trans_odd}
 F_-(\lambda) \equiv \left[ \lambda + \frac{\omega_c}{\coth\left(\omega_c \frac{L}{2}\right)}\FFC - \FFc \right]\left[\lambda + \frac{\omega_g}{\coth\left(\omega_g \frac{L}{2}\right)}\GGG - \GGg\right] - \FFg\GGc = 0\,.
\end{equation}
In Section \ref{sec:results1-D}, equation \eqref{eq:trans_odd} will be solved numerically for a Hopf bifurcation, i.e. for a pair of roots $\lambda = \pm i\lambda_I$ with $\lambda_I > 0$. In the well-mixed regime, corresponding to the infinite bulk diffusion limit $D_{c,\,g} \to \infty$, equation \eqref{eq:trans_odd} reduces to
\begin{equation}\label{eq:trans_odd_WM}
\lim_{D_{c,\,g} \to \infty} F_-(\lambda) = \left| \lambda I - J_- \right| = 0\,,
\end{equation}
where $I$ is the $2\times 2$ identity matrix and $J_-$ is a Jacobian matrix defined by
\begin{equation}\label{eq:jac_odd}
 J_- = \begin{pmatrix} \FFc & \FFg \\[5pt] \GGc & \GGg \end{pmatrix}\,.
\end{equation}
In computing the limit, we have employed the following approximation of the hyperbolic cotangent function for a small argument:
\begin{equation*}
\coth\left(\omega_{c,\,g}\frac{L}{2}\right) \approx \frac{2}{L\omega_{c,\,g}}\,, \qquad \text{with} \qquad \omega_{c,g} = \sqrt{\frac{\lambda}{D_{c,g}}} \ll 1\,.
\end{equation*}
In the well-mixed regime, necessary conditions for the symmetric steady state to undergo a Hopf bifurcation are given by
\begin{equation}
\tr(J_-) = \FFc + \GGg = 0\,, \quad \det(J_-) = \FFc\GGg - \GGc\FFg > 0\,.
\end{equation}

\paragraph{Even mode $(+)$.} For the even mode, we have that $u_1=u_2$, $v_1=v_2$ and thus
\begin{equation}\label{eq:phi_psi_even}
\varphi(x) = -\dfrac{\left(\FFc u_1 + \FFg v_1\right)\cosh\left(\omega_c\left(\frac{L}{2}-x\right)\right)}{\left(D_c\omega_c\tanh\left(\omega_c\frac{L}{2}\right) + \FFC\right)\cosh\left(\omega_c\frac{L}{2}\right)}\,, \quad
\psi(x) = -\dfrac{\left(\GGc u_1 + \GGg v_1\right)\cosh\left(\omega_g\left(\frac{L}{2}-x\right)\right)}{\left(D_g\omega_g\tanh\left(\omega_c\frac{L}{2}\right) + \GGG\right)\cosh\left(\omega_g\frac{L}{2}\right)}\,.
\end{equation}
Upon substituting \eqref{eq:phi_psi_even} within \eqref{eq:ode_eig}, we find that $(u_1,\,v_1)^T$ is a non-trivial solution of 
\begin{equation}\label{eq:cond_even_lambda}
\begin{pmatrix}
\lambda - \dfrac{\FFc D_c\omega_c\tanh\left(\omega_c \frac{L}{2}\right)}{D_c\omega_c\tanh\left(\omega_c \frac{L}{2}\right) + \FFC} & 
- \dfrac{\FFg D_c\omega_c\tanh\left(\omega_c \frac{L}{2}\right)}{D_c\omega_c\tanh\left(\omega_c \frac{L}{2}\right) + \FFC} \\[10pt]
- \dfrac{\GGc D_g\omega_g\tanh\left(\omega_g \frac{L}{2}\right)}{D_g\omega_g\tanh\left(\omega_g \frac{L}{2}\right) + \GGG} & 
\lambda - \dfrac{\GGg D_g\omega_g\tanh\left(\omega_g \frac{L}{2}\right)}{D_g\omega_g\tanh\left(\omega_g \frac{L}{2}\right) + \GGG} \\
\end{pmatrix}
\begin{pmatrix} u_1 \\ v_1 \end{pmatrix} = \begin{pmatrix} 0 \\ 0 \end{pmatrix}\,.
\end{equation}
Similarly as for the odd mode, we can divide through $\lambda \neq 0$ in \eqref{eq:cond_even_lambda} to obtain
\begin{equation}
\begin{pmatrix}
\lambda + \frac{\omega_c}{\tanh\left(\omega_c \frac{L}{2}\right)}\FFC - \FFc & - \FFg \\[5pt]
- \GGc & \lambda + \frac{\omega_g}{\tanh\left(\omega_g \frac{L}{2}\right)}\GGG - \GGg \\
\end{pmatrix} \begin{pmatrix} u_1 \\ v_1\end{pmatrix} = \begin{pmatrix} 0 \\ 0 \end{pmatrix}\,,
\end{equation}
which corresponds to the linear system obtained from applying the constraints \eqref{eq:mass_cons_eig} onto an even eigenfunction. The characteristic equation is therefore given by
\begin{equation}\label{eq:trans_even}
F_+(\lambda) = \left[ \lambda + \frac{\omega_c}{\tanh\left(\omega_c \frac{L}{2}\right)}\FFC - \FFc \right] \left[ \lambda + \frac{\omega_g}{\tanh\left(\omega_g \frac{L}{2}\right)}\GGG - \GGg \right] - \FFg \GGc = 0\,.
\end{equation}
Once again, in the limit of fast bulk diffusion, the conditions for a Hopf bifurcation associated with the even mode is easily derived. Upon taking the limit $D_{c,\,g} \to \infty$ in equation \eqref{eq:trans_even} and using the following approximation of the hyperbolic tangent function:
\begin{equation*}
\tanh\left(\omega_{c,\,g}\frac{L}{2}\right) \approx \omega_{c,\,g}\frac{L}{2}\,, \qquad \text{with} \quad \omega_{c,g}=\sqrt{\frac{\lambda}{D_{c,g}}} \ll 1\,,
\end{equation*}
we obtain a quadratic equation given by
\begin{equation}\label{eq:trans_even_WM}
\lim_{D_{c,\,g} \to \infty} F_+(\lambda) = \left| \lambda I - J_+ \right| = 0\,, 
\end{equation}
where the Jacobian matrix $J_+$ is defined by
\begin{equation}\label{eq:jac_even}
J_+ = \begin{pmatrix} 
-\frac{2}{L}\FFC + \FFc & \FFg \\ 
\GGc & - \frac{2}{L}\GGG + \GGg 
\end{pmatrix} \,.
\end{equation}
In the well-mixed regime, necessary conditions for a Hopf bifurcation associated with the even mode are given by $\tr(J_+) = 0$ and $\det(J_+) > 0$. We will however see that no Hopf bifurcation of the even mode occurs for the nonlinear functions and the parameter values used in our study, and thus only anti-phase oscillations are observed in both the well-mixed and finite diffusion regimes. The absence of in-phase oscillations was  also noted in a PDE--DDE model with mass conservation \cite{xu2016}. Whether a mass-conserved PDE-ODE(DDE) model can have in-phase oscillations remains an open question.

\subsubsection{Zero-crossing eigenvalues}\label{subsubsec:zero_lambda}
We now investigate the occurrence of zero-crossing eigenvalues leading to either pitchfork or saddle-node bifurcations. We remark that a rigorous treatment of the eigenvalue problem defined in equations \eqref{eq:mass_cons_eig}--\eqref{eq:ode_eig} would require the introduction of a branch cut in the complex $\lambda\text{-plane}$ along $\infty < \Re(\lambda) \leq 0$ and $\Im(\lambda) = 0$. Fortunately, the limit of $\lambda \to0$, with $\lambda$ real, has the same effect as the infinite bulk diffusion limit: all the square roots from the characteristic equations \eqref{eq:trans_odd} and \eqref{eq:trans_even} cancel, and no continuous spectrum is introduced. Therefore, the solution for $\lambda = 0$ can be readily obtained by taking the well-defined limit $\lambda \to 0$.

\paragraph{Odd mode $(-)$.} After taking the limit $\lambda \to 0$ in equation \eqref{eq:trans_odd}, we obtain the following condition for the linearized system to have a zero-crossing eigenvalue associated with the odd mode:
\begin{equation} 
\lim_{\lambda \to 0} F_-(\lambda) = \det(J_-) = \FFc\GGg - \GGc\FFg = 0.
\end{equation}
When this condition is satisfied, the system undergoes a pitchfork bifurcation with two branches of asymmetric steady states emerging from the symmetric steady state.

\paragraph{Even mode $(+)$.} We proceed similarly for the even mode. The symmetric steady state undergoes a saddle-node bifurcation when the following condition is satisfied:
\begin{equation}
\lim_{\lambda \to 0} F_+(\lambda) = \det(J_+) = \frac{4}{L^2}\FFC\GGG - \frac{2}{L}\left(\FFc\GGG + \GGg\FFC\right) + \left(\FFc\GGg - \GGc\FFg\right) = 0\,.
\end{equation}

We highlight that bulk diffusion has no effect on the existence of steady states and on bifurcations resulting from zero-crossing eigenvalues. However, as will be seen in the next section, diffusion does influence the stability of symmetric steady states and possible oscillatory dynamics.

\subsubsection{Numerical bifurcation analysis}\label{sec:results1-D}

Our previous work on the well-mixed version of the model pointed out the importance of membrane residence times of GEF proteins (set by $1/\koff$) in determining whether Cdc42 proteins were at steady state or undergoing pole-to-pole oscillations \cite{xu2018}. Hence, we select $\koff$ and the bulk diffusion coefficients as bifurcation parameters, and numerically investigate their effects on the stability of the symmetric steady state ($c_1=c_2, \, g_1=g_2$). Details of the numerical bifurcation analysis are given in Appendix~\ref{sec:numerical_bifurcation_analysis}.

\begin{figure}[H]
\centering
\begin{subfigure}{0.36\linewidth}
\includegraphics[width=\linewidth]{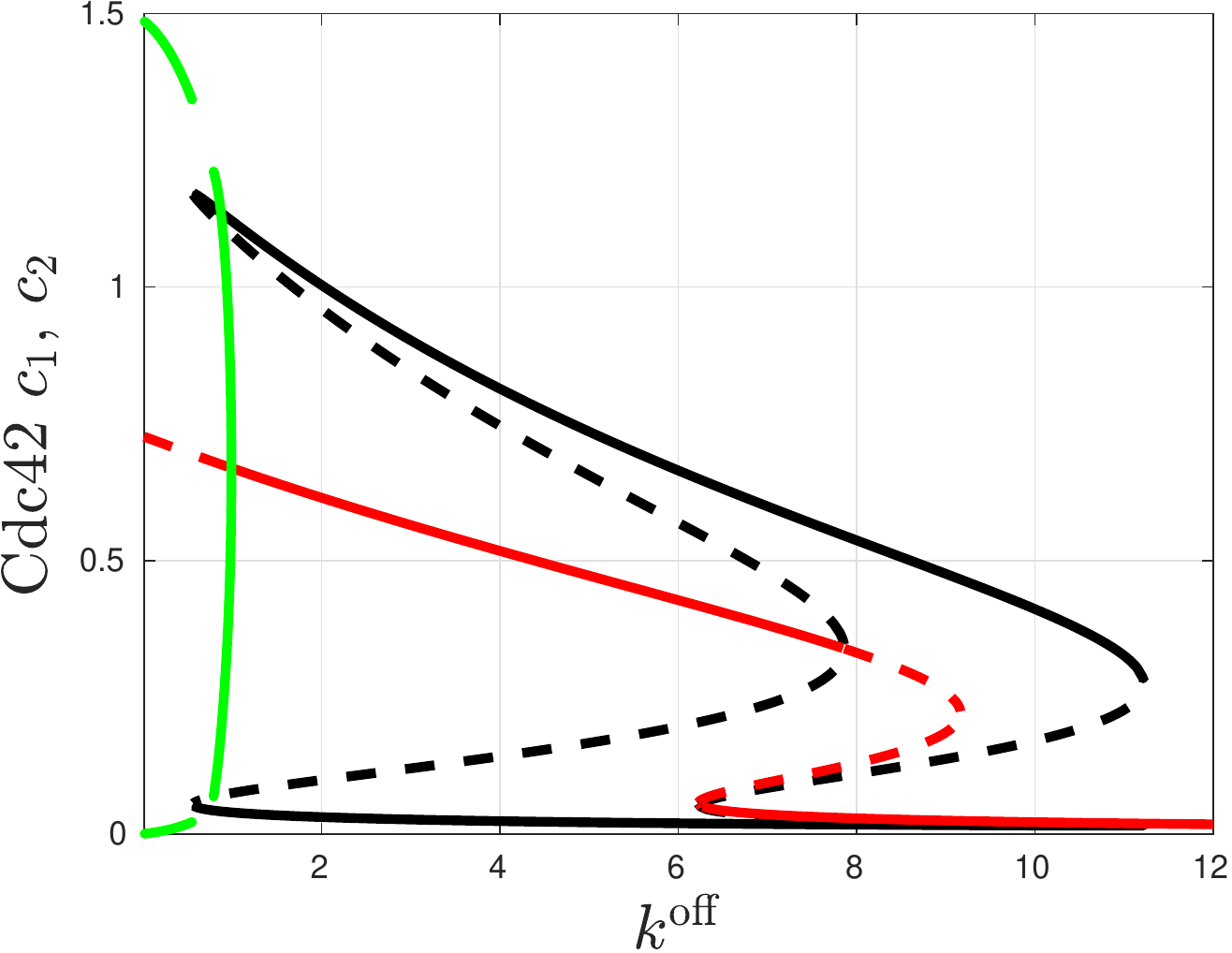}
\caption{Active Cdc42 levels.}
\end{subfigure}
\begin{subfigure}{0.36\linewidth}
\includegraphics[width=\linewidth]{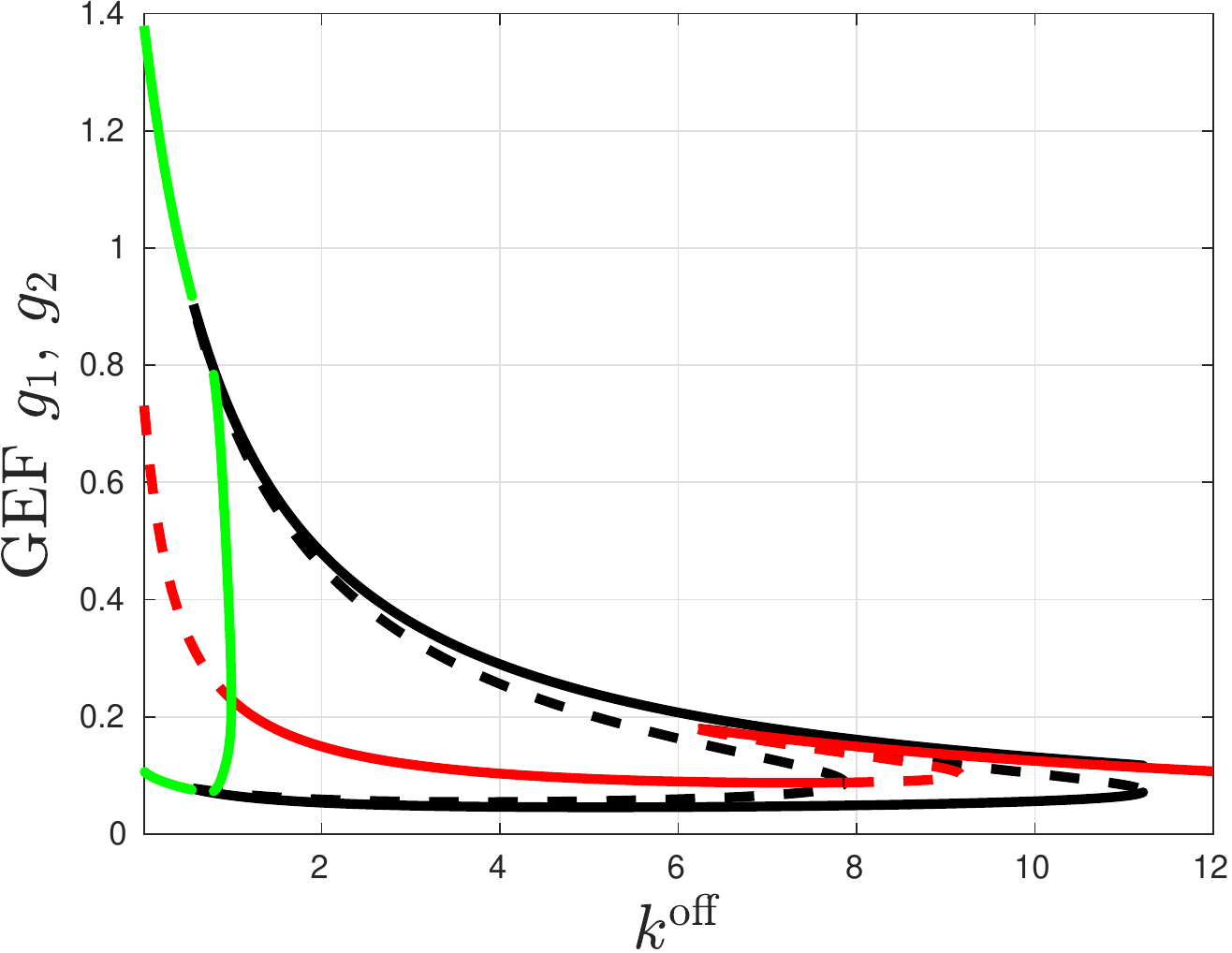}
\caption{Active GEF levels.}
\end{subfigure}
\caption{\label{fig:bif_diag_koff_WM} Symmetric steady state (\textcolor{red}{red}), asymmetric steady state (\textcolor{black}{black}) and periodic solution branches (\textcolor{green}{green}) for the reduced ODE model in the well-mixed regime, as a function of $\koff$. Other parameter values are given in Table~\ref{table:parm}. A full line indicates linear stability, while unstable branches are dashed.}
\end{figure}

We start with the dynamics of the ODE model (\ref{eq:ode}). The intricate succession of bifurcations that results when increasing $\koff$ is presented in Fig.~\ref{fig:bif_diag_koff_WM}. The symmetric steady state is unstable for low $\koff$ values, with linear stability gained through a supercritical Hopf bifurcation when $\koff \approx 0.99$. Further beyond this threshold, two fold bifurcations in $\koff \approx 6.23$ and $\koff \approx 9.17$ cause hysteresis and coexistence of up to three symmetric steady states. We observe that bistability of symmetric steady states occurs in the range $(6.24, 7.86)$, with such thresholds corresponding to subcritical pitchfork bifurcations connecting branches of asymmetric and symmetric steady states. Finally, stable and unstable branches of asymmetric steady states are connected via fold bifurcations at $\koff \approx 0.56$ and $\koff \approx 11.22$.

\begin{figure}[H]
\centering
\begin{subfigure}{0.31\linewidth}
\includegraphics[width=\linewidth]{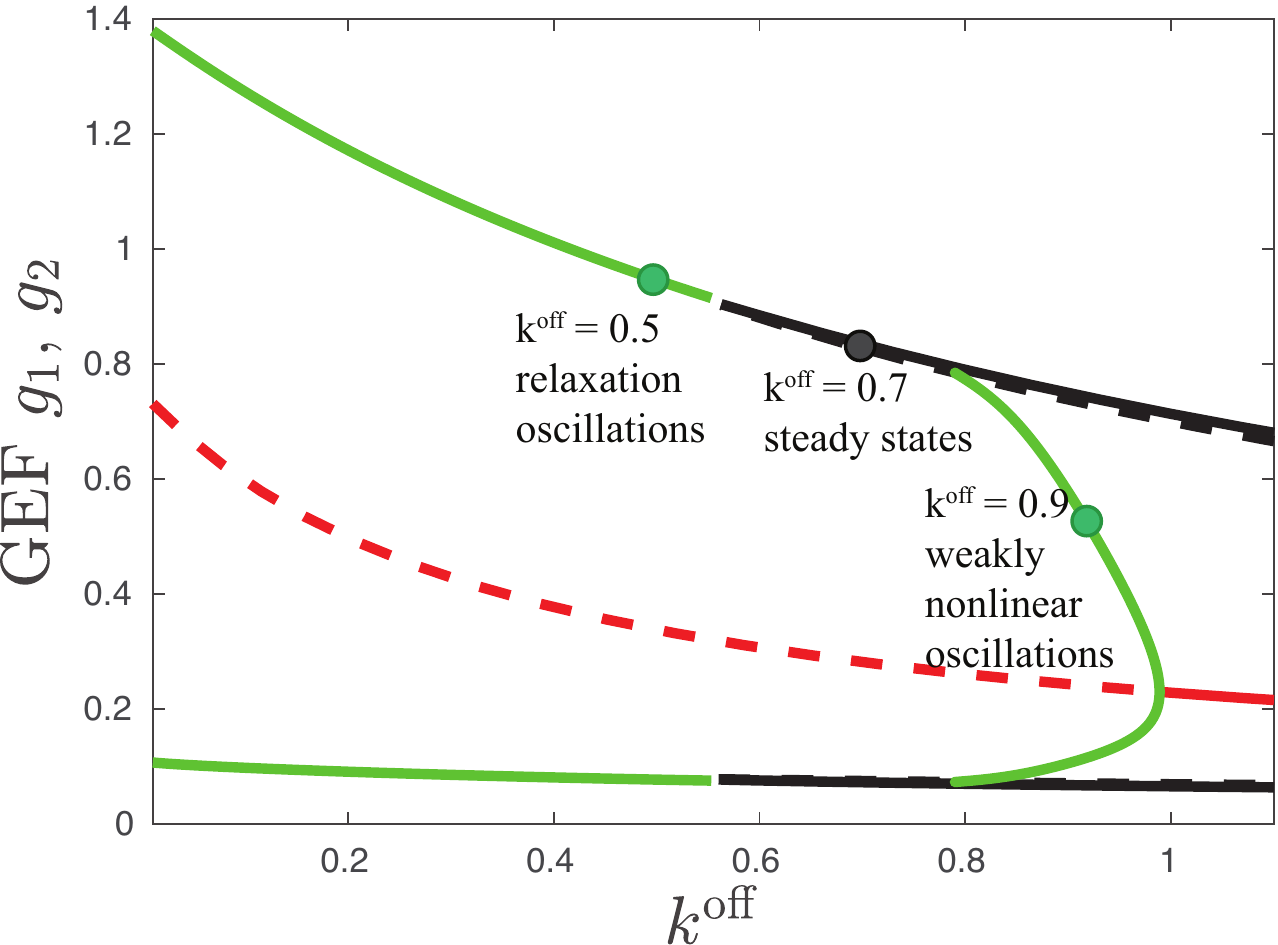} 
\caption{Active GEF levels.}
\end{subfigure}
\begin{subfigure}{0.31\linewidth}
\includegraphics[width=\linewidth]{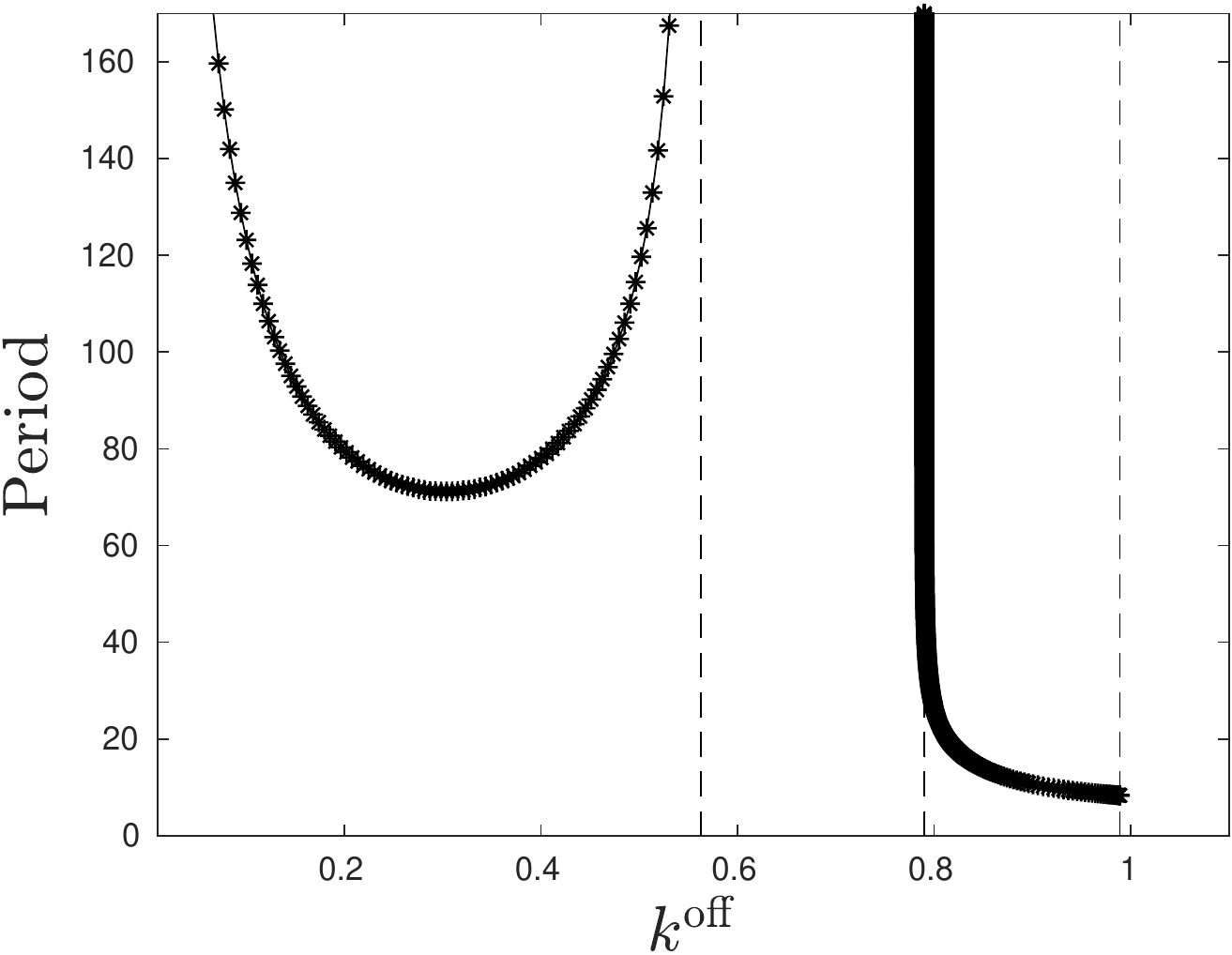} 
\caption{Oscillatory period.}
\end{subfigure}
\begin{subfigure}{0.31\linewidth}
\includegraphics[width=\linewidth]{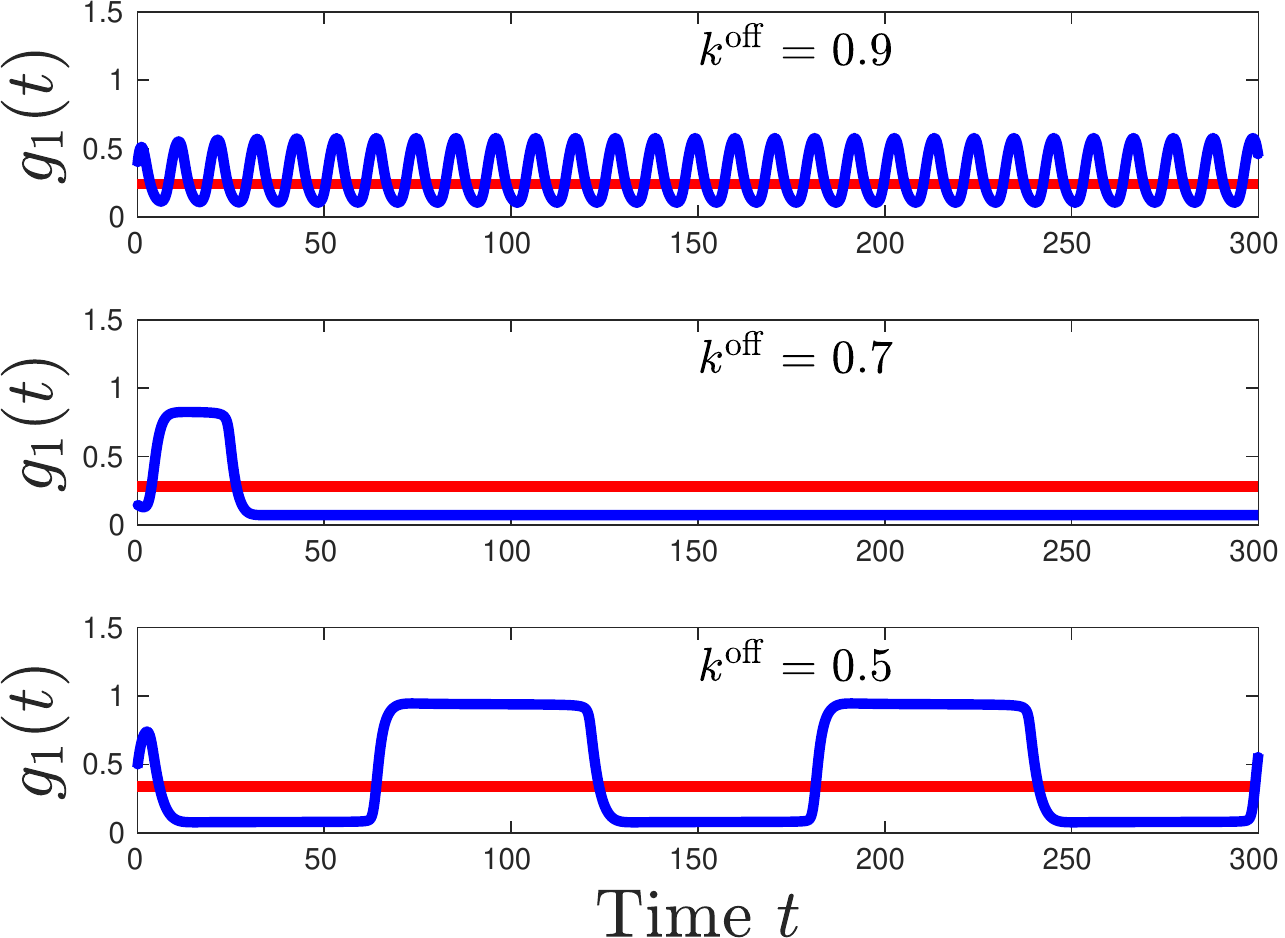} 
\caption{$g_1(t)$ for $\koff = 0.9,\, 0.7, \, 0.5$.}
\end{subfigure}
\caption{\label{fig:hopf_WM} Oscillatory dynamics for the reduced ODE model (\ref{eq:well_mixed}) in the well-mixed regime. (a) A zoomed-in view of the bifurcation diagram in Fig.~\ref{fig:bif_diag_koff_WM}(b). (b) Period of oscillations in the $\koff$ regimes $(0, 0.56)$ and $(0.79, 0.99)$. For the first branch, the period has a minimum near $\koff \approx 0.3$ and rapidly increases near both $\koff=0$ and $\koff \approx 0.56$, suggesting the occurrence of homoclinic bifurcations as limit cycles collide with asymmetric equilibria. For the branch of periodic solutions emerging from a supercritical Hopf bifurcation at $\koff \approx 0.99$, the period increases rapidly at a suspected homoclinic bifurcation near $\koff \approx 0.79$. (c) Numerical solutions for $\koff=0.9, 0.7, 0.5$. For a high rate $\koff=0.9$, weakly nonlinear oscillations are observed. For an intermediate rate $\koff=0.7$, numerical solutions evolve to an asymmetric steady state. As $\koff$ further decreases to $0.5$, highly nonlinear relaxation oscillations are observed. For each of the three simulations, the initial conditions correspond to the symmetric steady state (red horizontal line) plus a small perturbation stimulating the odd mode.}
\end{figure}

For the parameter regime considered, stable anti-phase oscillations are observed in the limit of fast bulk diffusion when $\koff$ is small ($\koff < 1$). A closer view of this regime is shown in panel (a) of Fig.~\ref{fig:hopf_WM}, where we observe that oscillations emerge either via supercritical Hopf or homoclinic bifurcations. Numerical simulations exhibiting the role of the asymmetric equilibrium on the shape of the oscillations are shown in panel (c) of the same figure. From $\koff = 0.9$ to $\koff = 0.5$, we observe a transition from weakly to highly nonlinear oscillations. The intermediate range $\koff \, \in \, (0.56, 0.79)$ is characterized by the absence of stable oscillations, with all the trajectories attracted to the asymmetric steady states.

\begin{figure}[H]
\centering
\begin{subfigure}{0.36\linewidth}
\includegraphics[width=\linewidth]{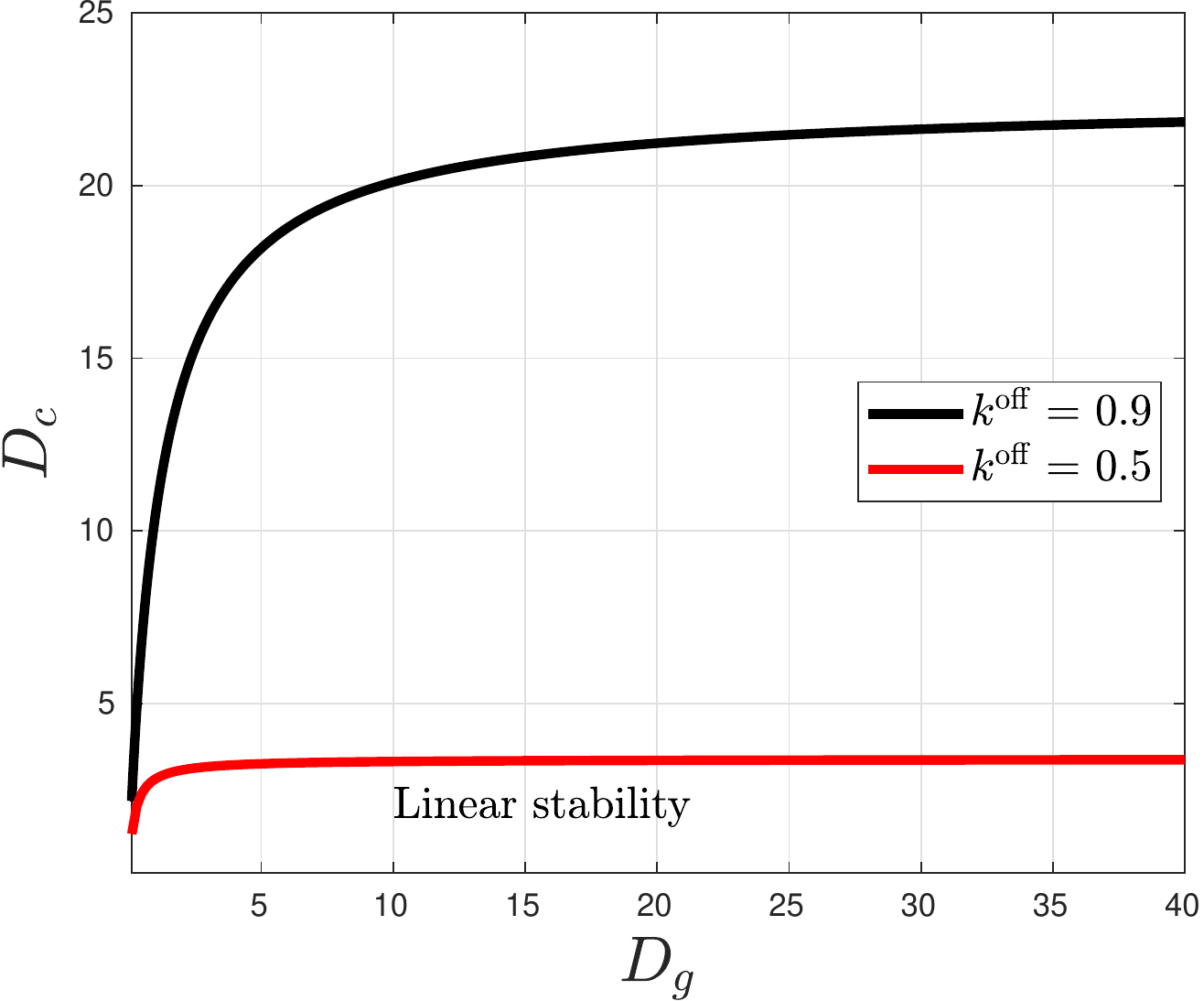}
\caption{$(D_g,\, D_c)$ with $\koff = 0.5,\, 0.9$.}
\end{subfigure}
\begin{subfigure}{0.36\linewidth}
\includegraphics[width=\linewidth]{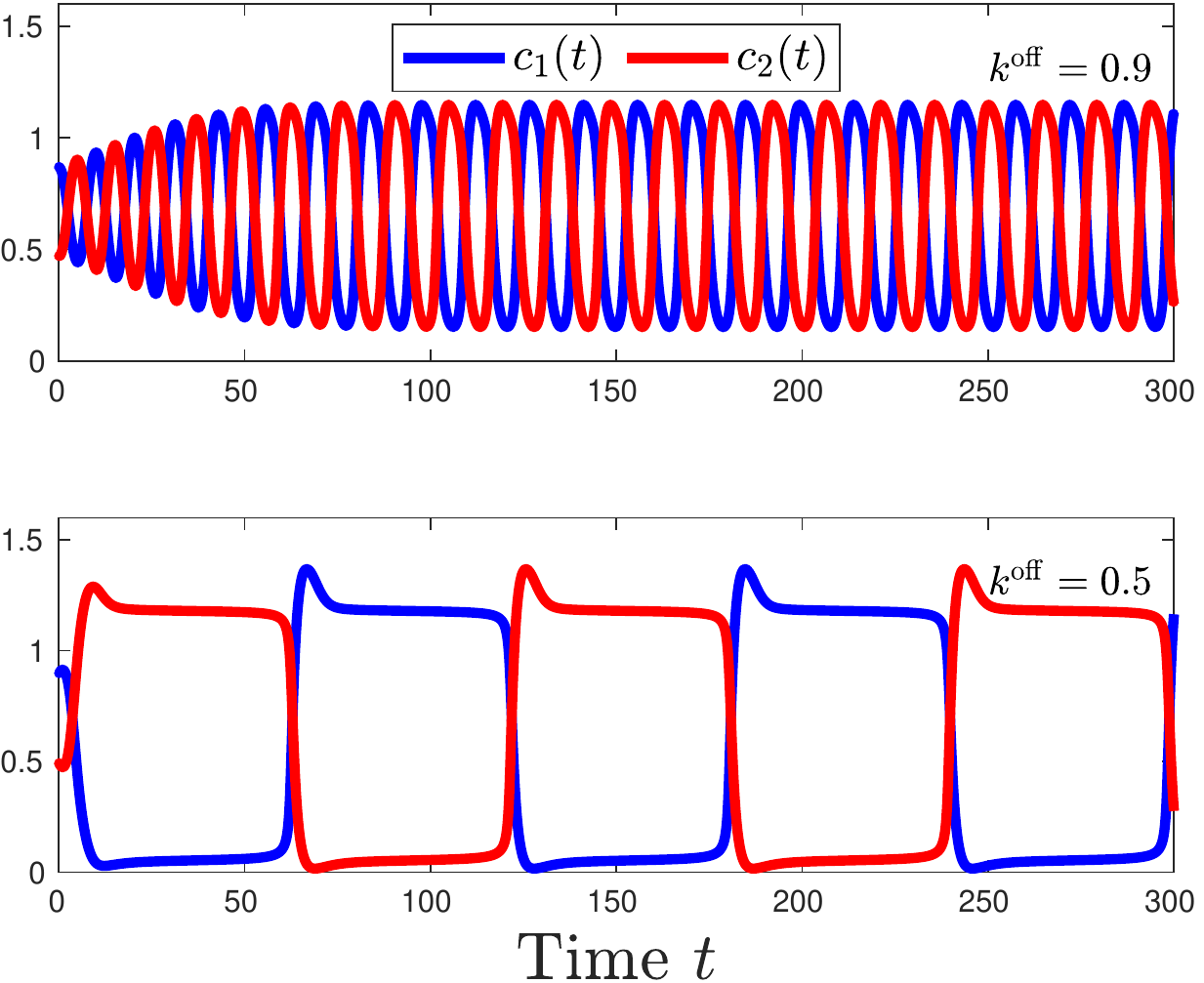}
\caption{$D_g = 1,\, D_c = 20$.}
\end{subfigure}
\caption{\label{fig:Dg_Dc} Panel (a): Hopf stability boundaries in the $D_c$ versus $D_g$ plane in the full PDE--ODE model for $\koff = 0.9$ (\textcolor{black}{black}) and $\koff = 0.5$ (\textcolor{red}{red}). Other parameter values are given in Table~\ref{table:parm}. Linear stability regions are located below each curve. Panel (b): for $D_c = 20$ and $D_g = 1$, increasing the membrane residence time results in a transition from weakly to highly nonlinear oscillations, a finding consistent with the dynamics in the well-mixed regime. The phase shift of half a period between $c_1(t)$ (\textcolor{blue}{blue}) and $c_2(t)$ (\textcolor{red}{red}) is a clear indication of anti-phase oscillations.}
\end{figure}

We consider next the effect of finite bulk diffusion on the emergence of oscillatory dynamics, focusing on the parameter regime where there is a unique symmetric steady state. We numerically solve the eigenvalue relation \eqref{eq:trans_odd} for a Hopf bifurcation associated with the anti-phase mode, i.e. for a pair of roots $\lambda = \pm i\lambda_I$ with $\lambda_I > 0$. The corresponding Hopf stability boundaries in the $D_c$ versus $D_g$ parameter plane are given in panel (a) of Fig.~\ref{fig:Dg_Dc}, where the region of linear stability of the symmetric steady state is located below the curves. We point out that fast cytosolic diffusion is sufficient to achieve oscillations and that decreasing $\koff$ shrinks the linear stability region. Since $1/\koff$ sets the membrane dwelling time for the GEF, we conclude that increasing this quantity pushes the system into oscillatory dynamics even for relatively modest diffusion coefficients. Numerical simulations exhibiting anti-phase oscillations with qualitatively different shapes are shown in panel (b) of the same figure.

\begin{figure}[H]
\centering
\begin{subfigure}{0.31\linewidth}
\includegraphics[width=\linewidth]{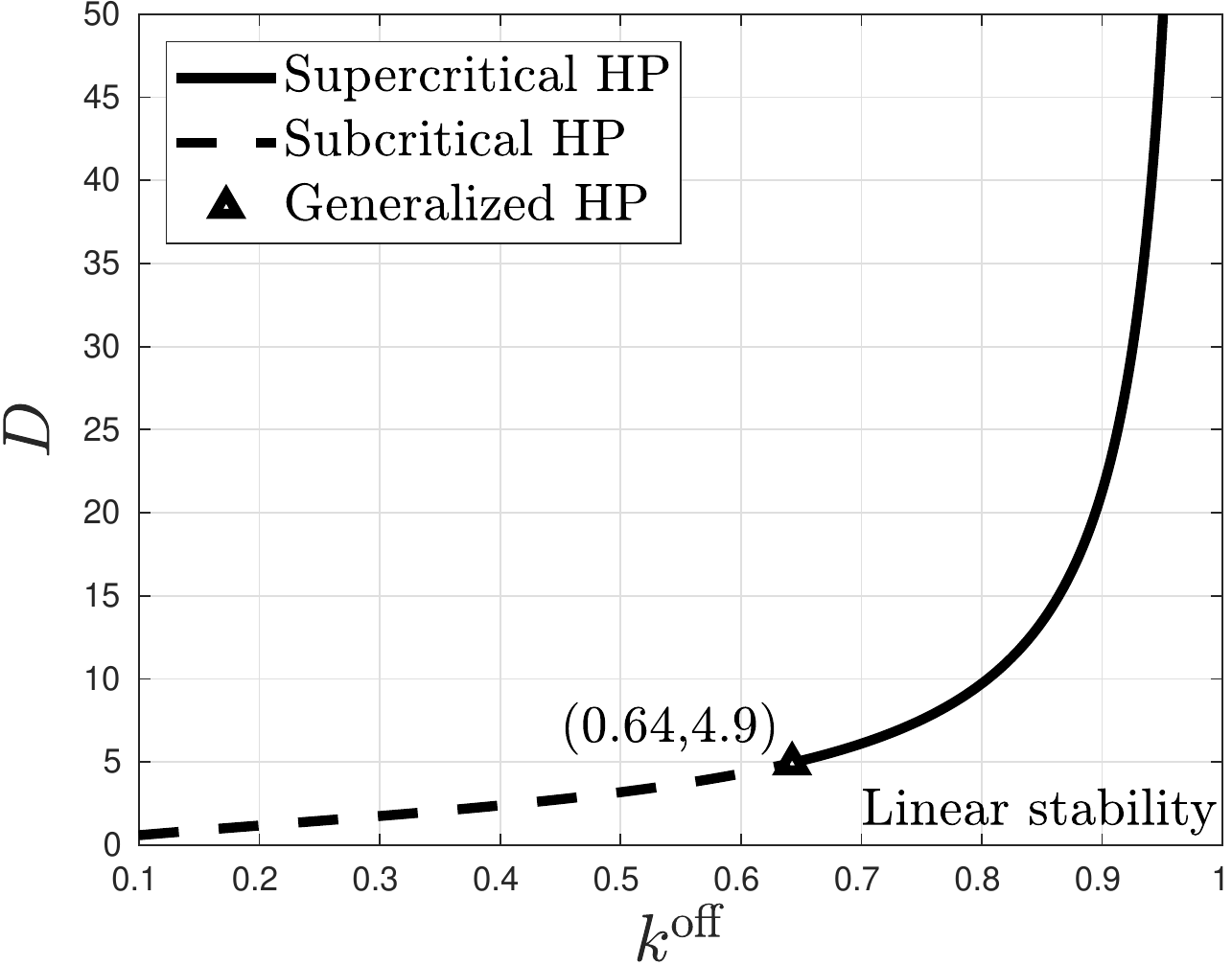}
\caption{$(\koff, D)$ with $D \equiv D_c = D_g$.}
\end{subfigure}
\begin{subfigure}{0.31\linewidth}
\includegraphics[width=\linewidth]{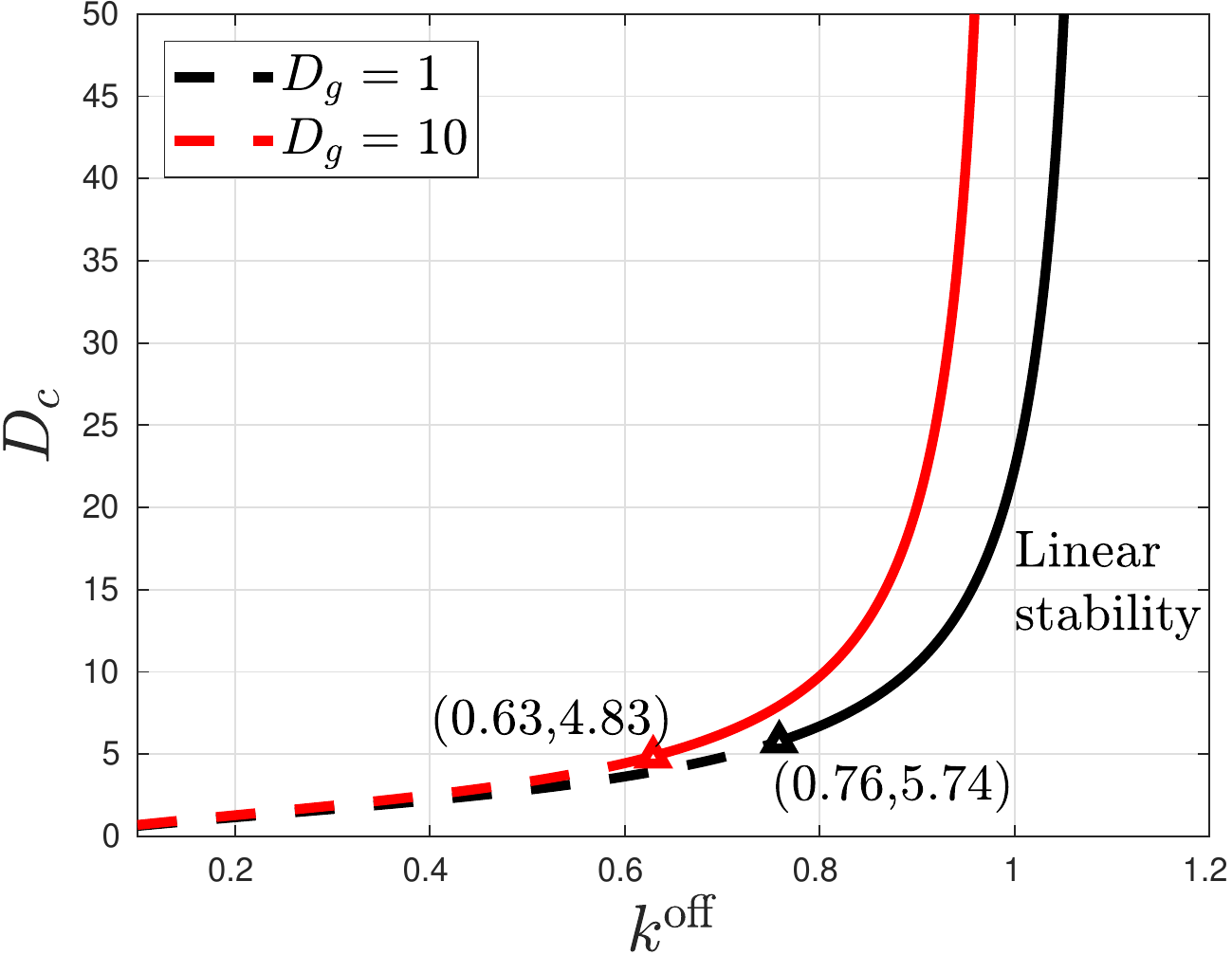} 
\caption{$(\koff, D_c)$ with $D_g=1,\, 10$.}
\end{subfigure}
\begin{subfigure}{0.31\linewidth}
\includegraphics[width=\linewidth]{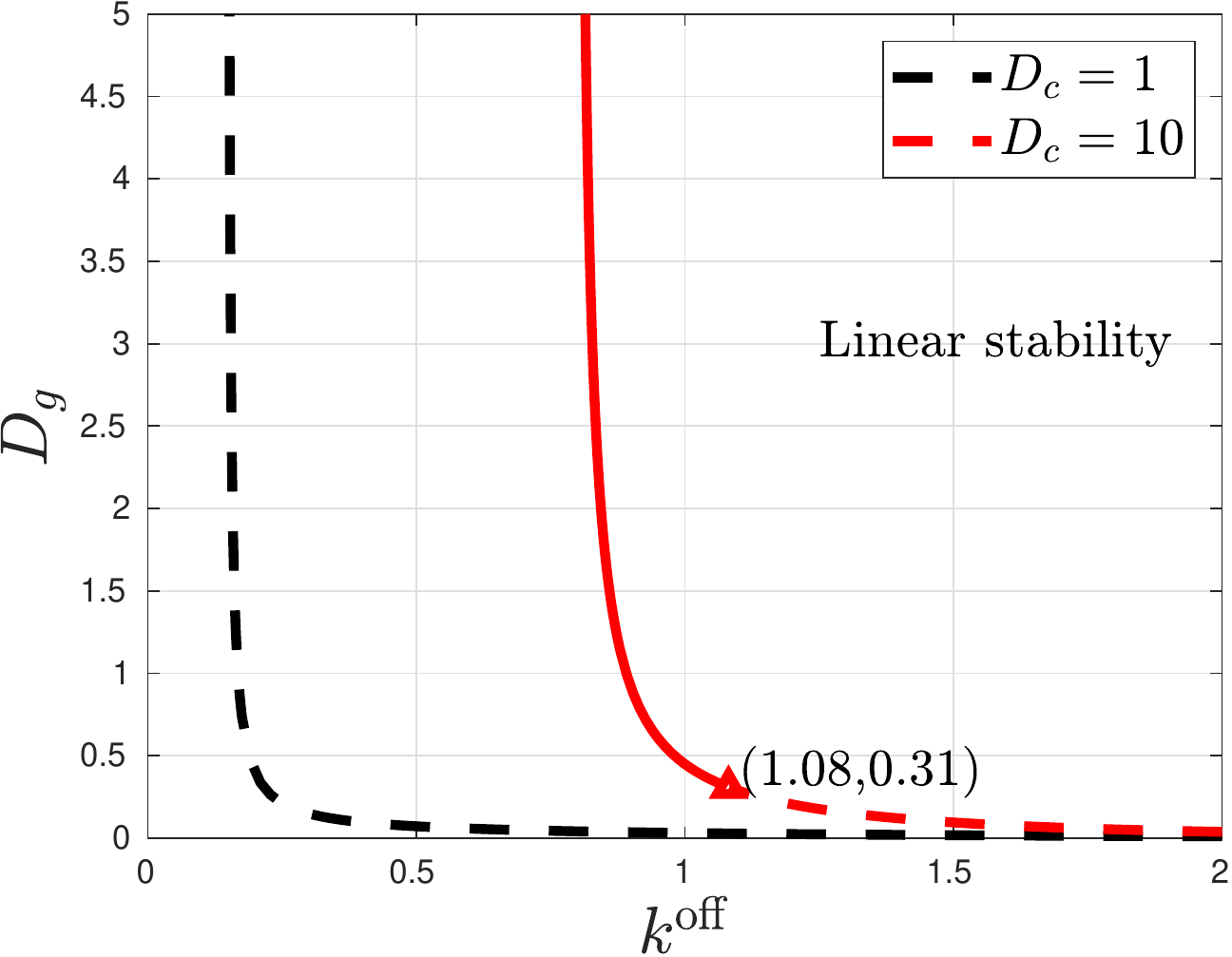}
\caption{$(\koff, D_g)$ with $D_c=1,\, 10$.}
\end{subfigure}
\caption{\label{fig:stability_boundaries} Hopf stability boundaries for the symmetric steady state in the full PDE--ODE model in the $D$ versus $\koff$ (panel (a)), $D_c$ versus $\koff$ (panel (b)) and $D_g$ versus $\koff$ (panel (c)) parameter planes. Full and dashed curves indicate respectively super- and subcritical Hopf bifurcations, while triangular points indicate a criticality change (generalized Hopf point). Other parameter values are given in Table~\ref{table:parm}.}
\end{figure}

Shown next in Fig.~\ref{fig:stability_boundaries} are stability diagrams for the symmetric steady state that use $\koff$ as a bifurcation parameter. We note that in panel (a), the bulk diffusion coefficients are set to be equal, while in panels (b) and (c) only $D_c$ or $D_g$ is allowed to vary. The branching behavior of the Hopf bifurcation, i.e. whether it is super- or subcritical, was resolved numerically with the software AUTO (cf.~\cite{doedel2007}) and a nonlocal formulation of the PDE--ODE model (see Appendix~\ref{sec:numerical_bifurcation_analysis}). Letting $D \equiv D_c = D_g$ or letting $D_c$ free with $D_g$ fixed yields qualitatively similar stability boundaries, with the second case seemingly converging to the first one as $D_g$ increases. In both cases, the bifurcation is supercritical when the diffusivity level is above some threshold and it is subcritical in the low diffusion regime. However, stability boundaries in the $D_g$ versus $\koff$ parameter plane have a qualitatively different shape, with the critical diffusivity threshold being inversely proportional to $\koff$. Here, increasing $D_c$ tenfold pushes the system further into instability and near the onset of oscillations. Hence, we conclude that $D_c$ has a stronger effect on the oscillatory dynamics than $D_g$.

\begin{figure}[H]
\centering
\begin{subfigure}{0.31\linewidth}
\includegraphics[width=\linewidth]{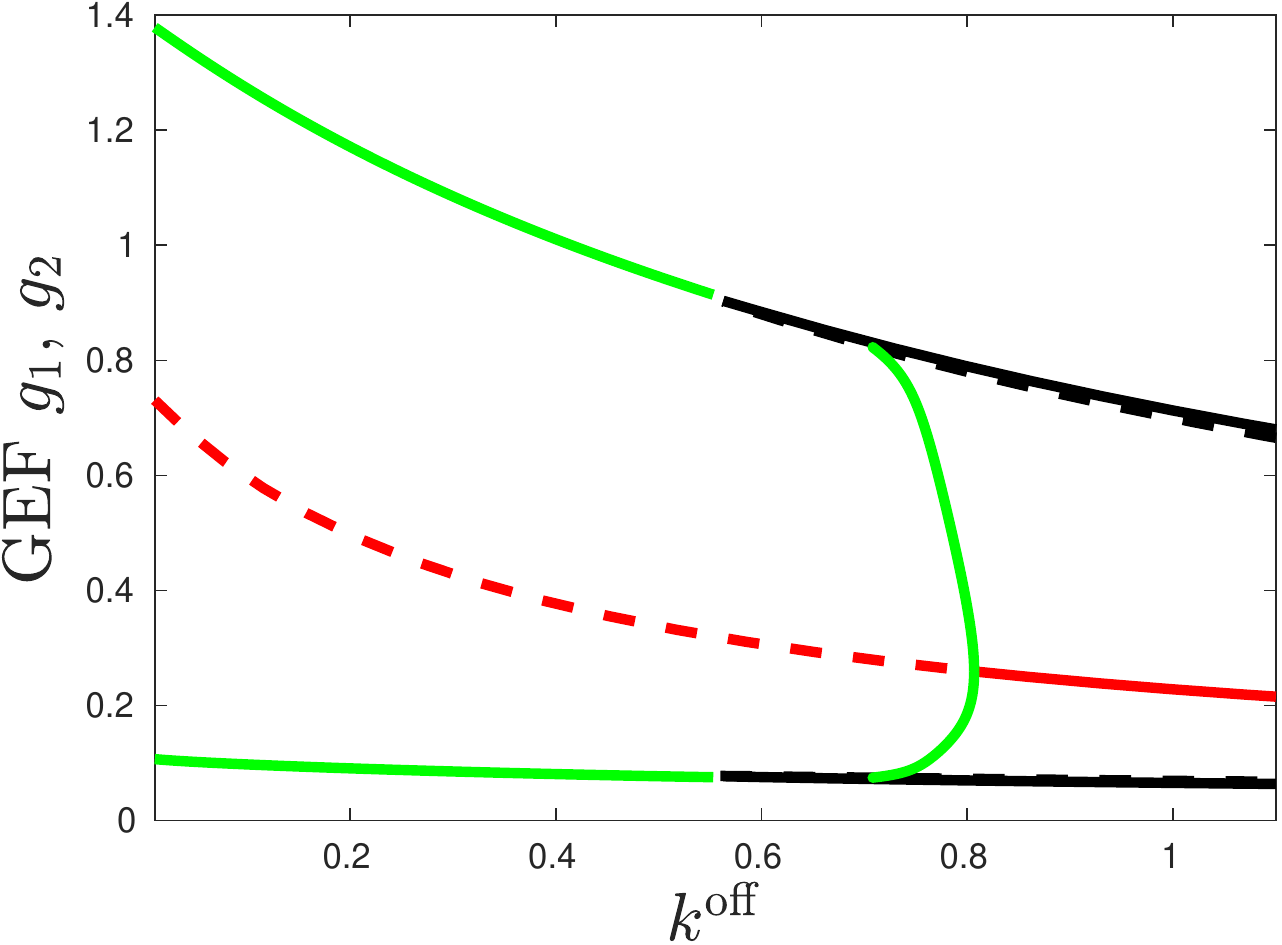} 
\caption{Active GEF levels, $D=10$}
\end{subfigure}
\begin{subfigure}{0.31\linewidth}
\includegraphics[width=\linewidth]{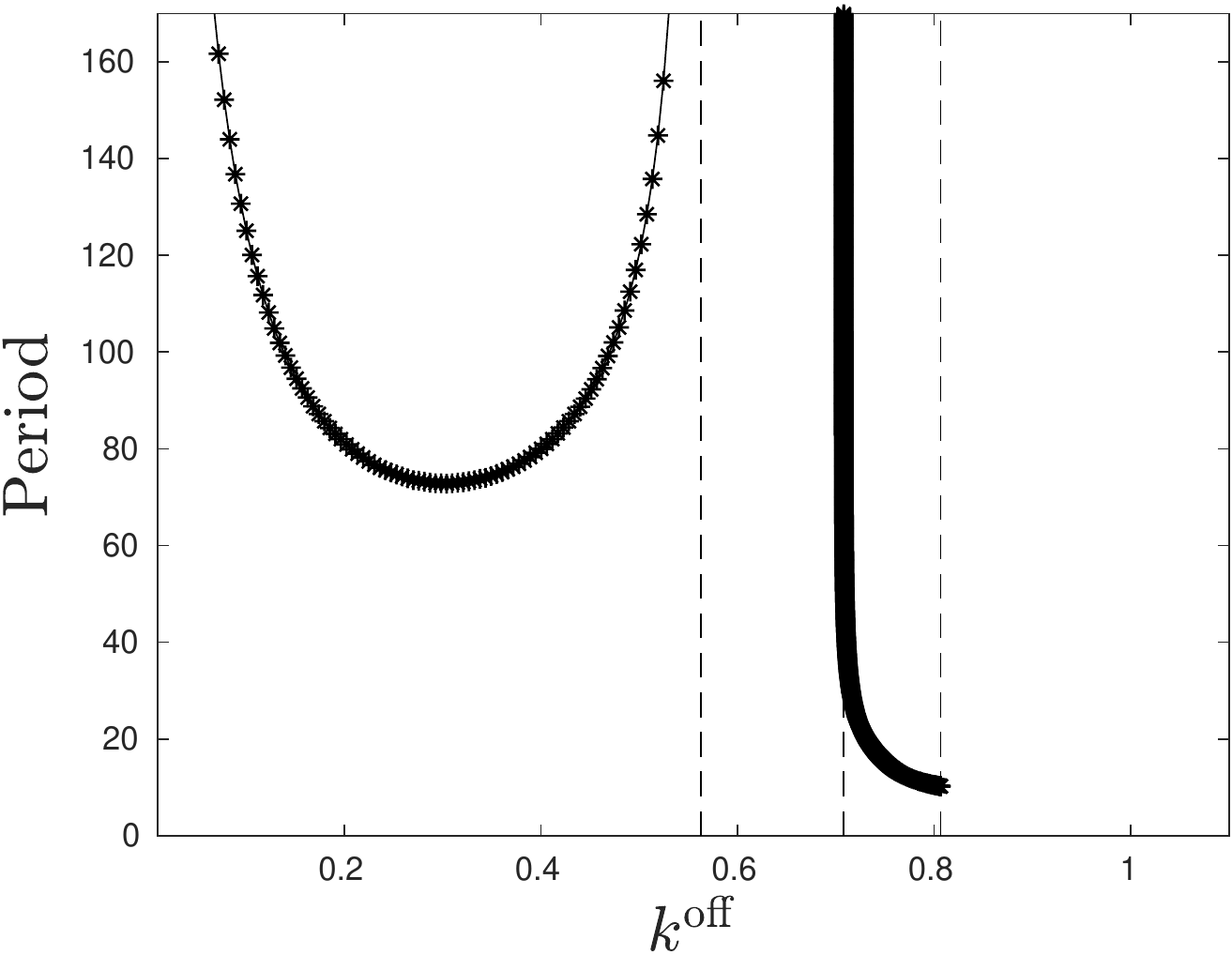}
\caption{Oscillatory period, $D=10$.}
\end{subfigure}
\begin{subfigure}{0.31\linewidth}
\includegraphics[width=\linewidth]{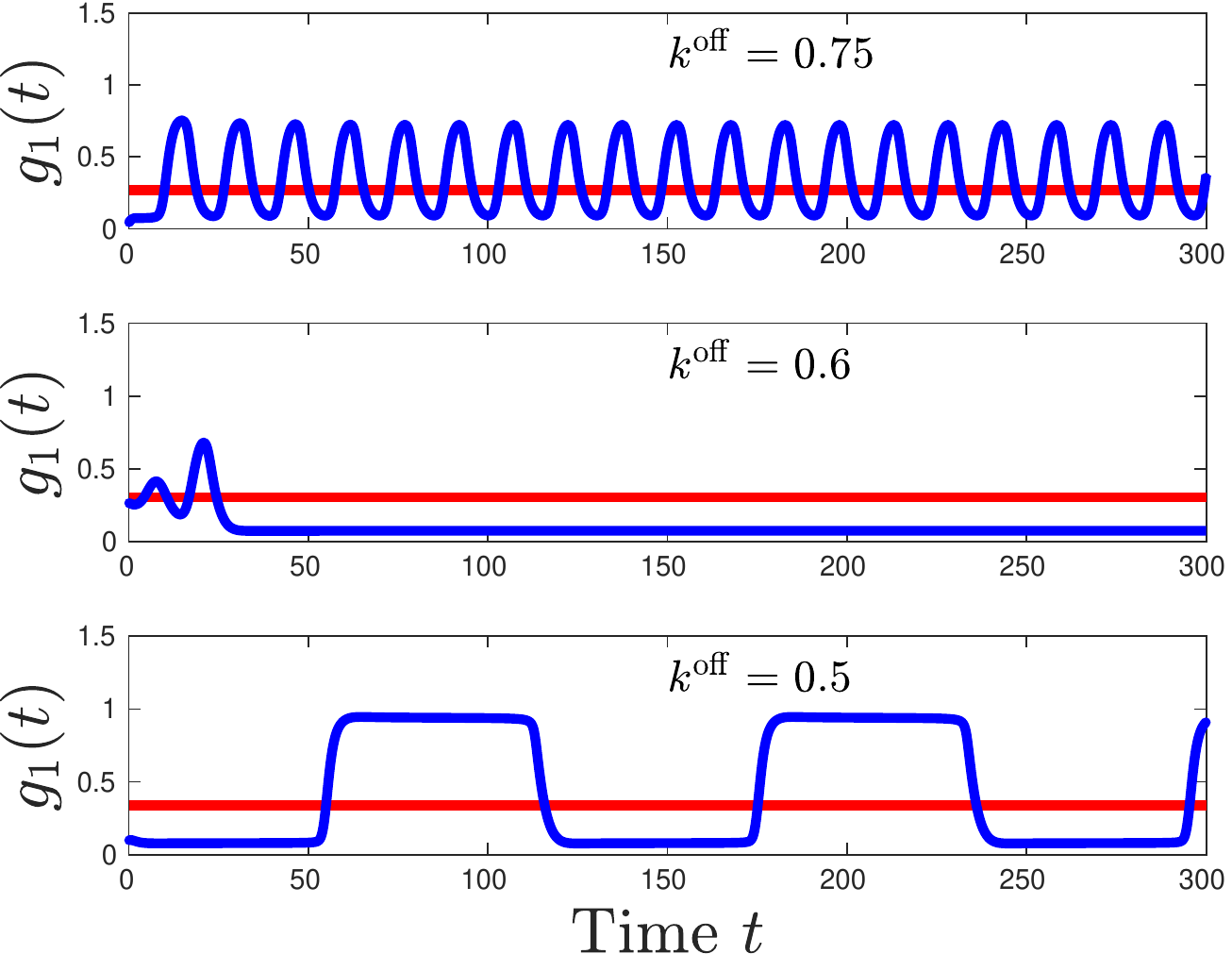}
\caption{$D=10$ and $\koff = 0.75,\, 0.6,\, 0.5$.}
\end{subfigure}
\begin{subfigure}{0.31\linewidth}
\includegraphics[width=\linewidth]{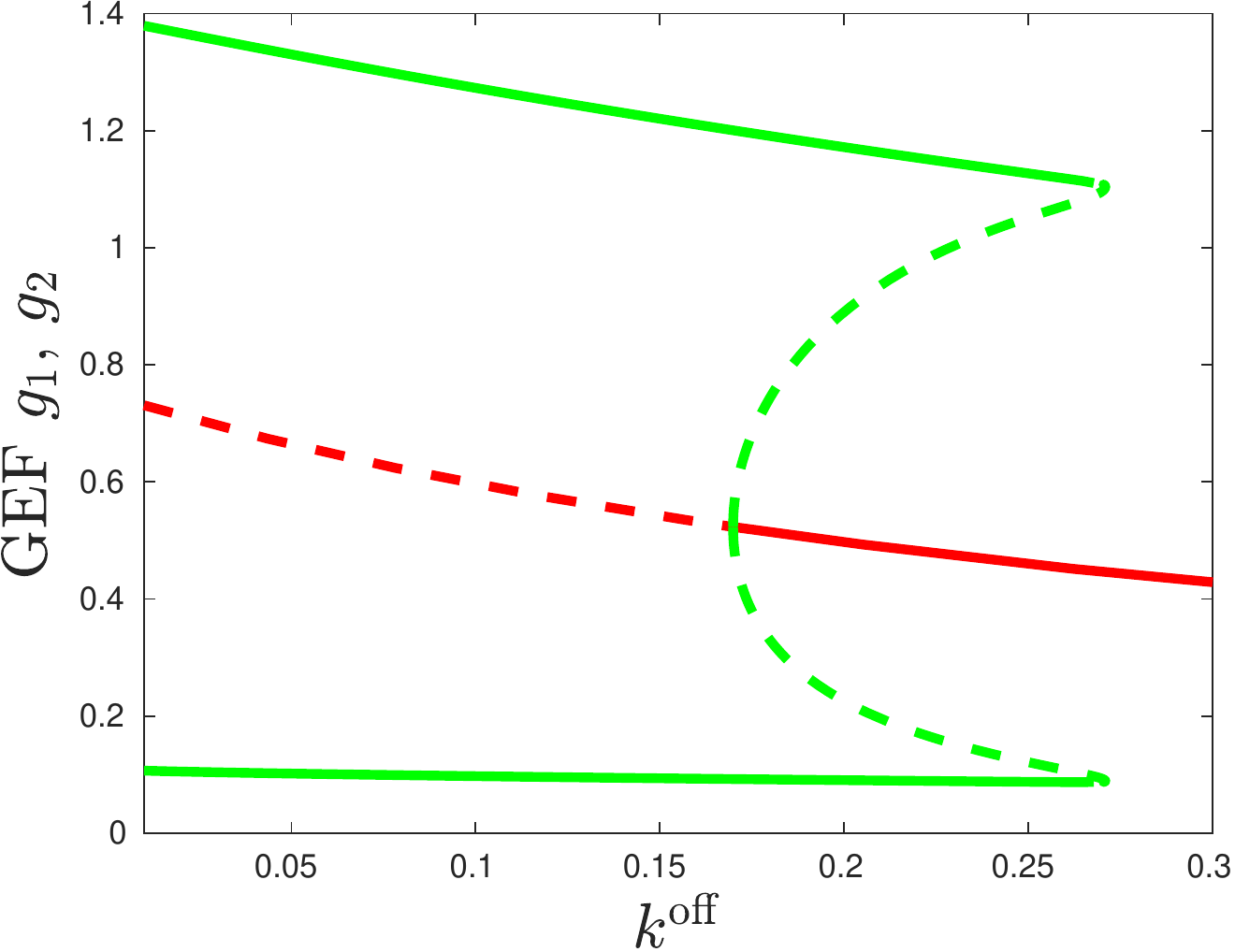}
\caption{Active GEF levels, $D=1$.}
\end{subfigure}
\begin{subfigure}{0.31\linewidth}
\includegraphics[width=\linewidth]{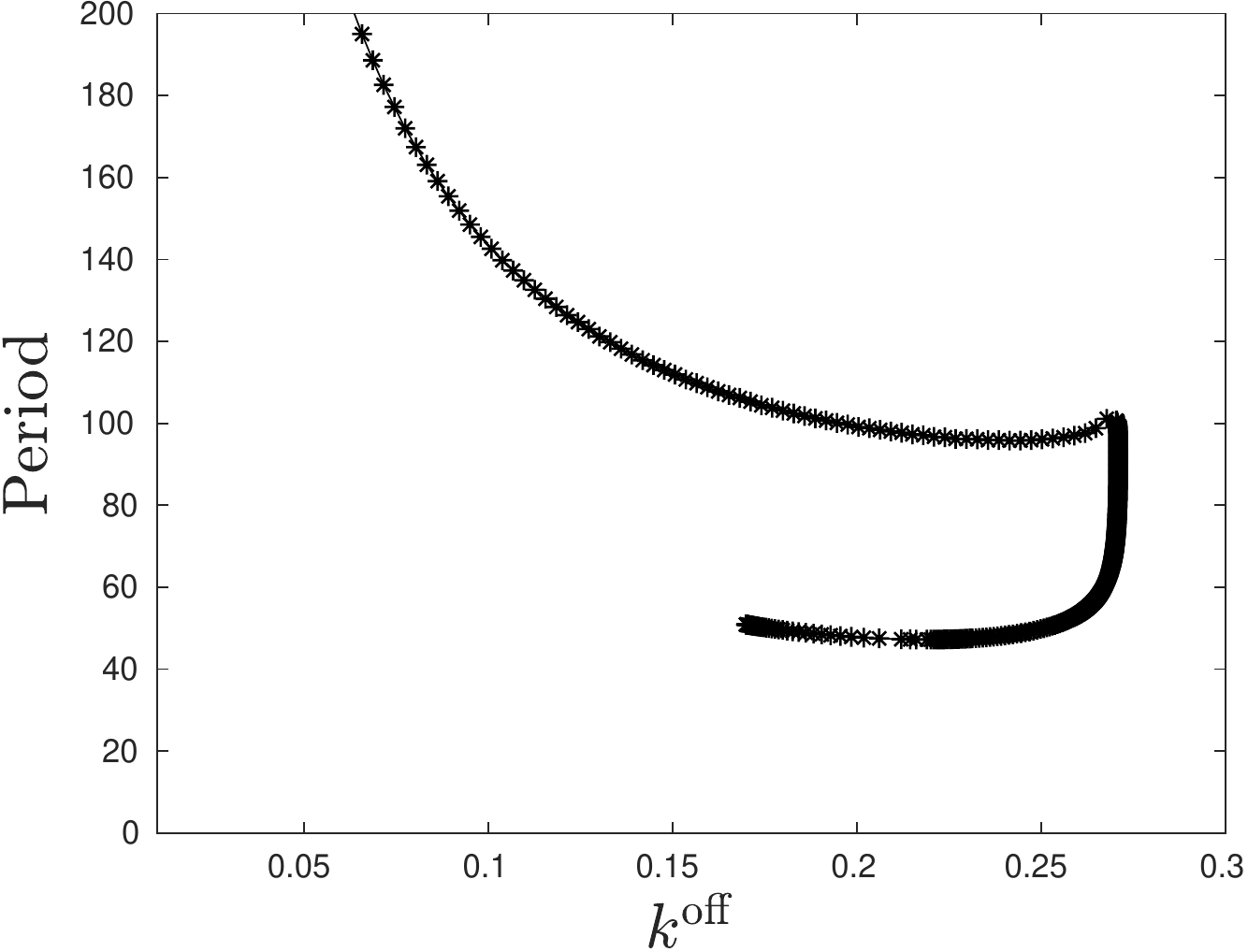}
\caption{Oscillatory period, $D=1$.}
\end{subfigure}
\begin{subfigure}{0.31\linewidth}
\includegraphics[width=\linewidth]{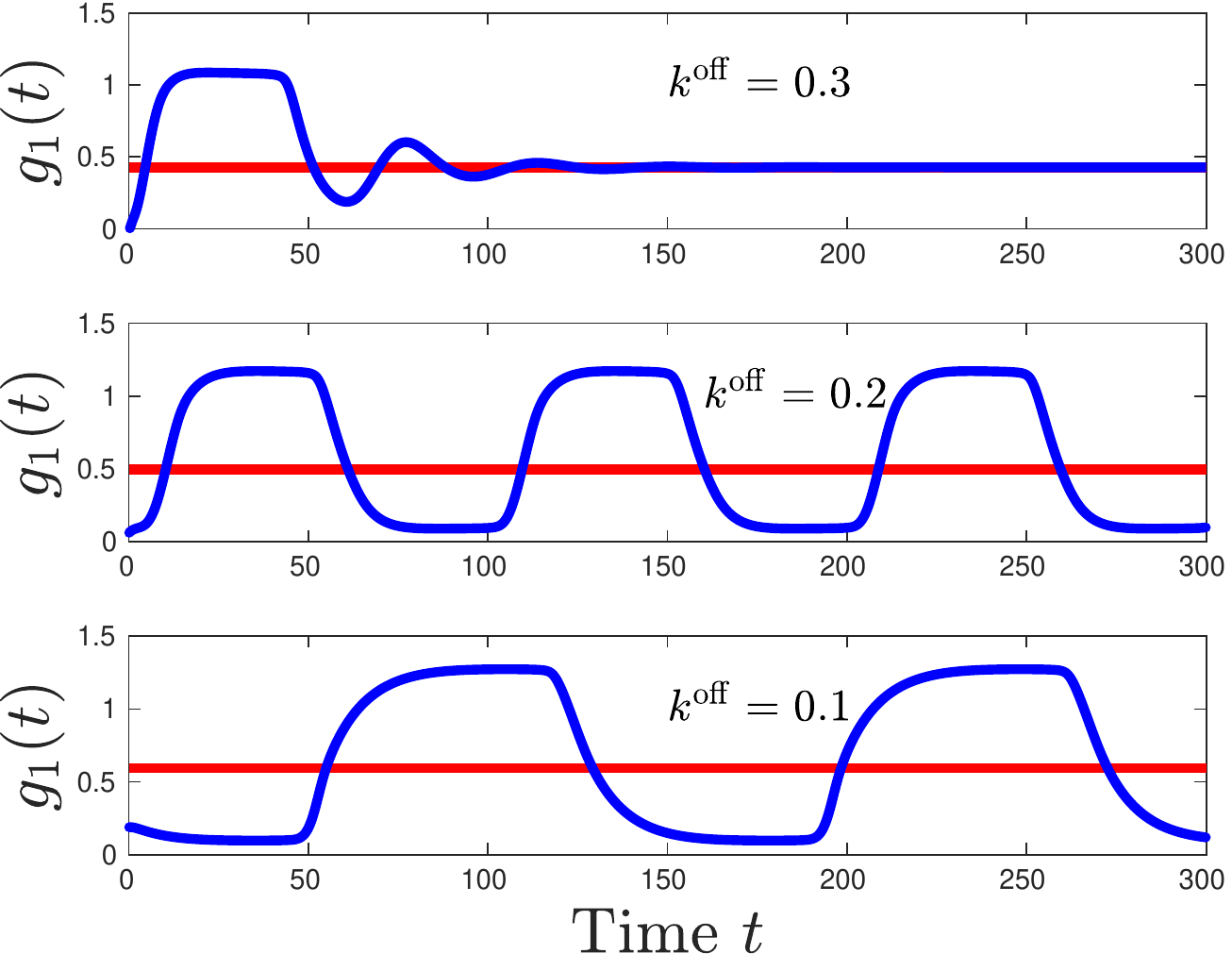}
\caption{$D=1$ and $\koff = 0.3,\, 0.2,\, 0.1$.}
\end{subfigure}
\caption{\label{fig:hopf_finite_diffusion} Oscillatory dynamics in the full PDE--ODE model for $D_c = D_g = 10$ (panels (a)-(c)) and for $D_c = D_g = 1$ (panels (d)-(f)). Other parameter values are given in Table~\ref{table:parm}. The case $D=10$ yields qualitatively similar dynamics to that of the well-mixed regime, with the Hopf bifurcation being supercritical. However, when $D=1$ the Hopf bifurcation is subcritical and a hard loss of stability of the symmetric steady state is observed; see panel (f).}
\end{figure}

Next, we compute global bifurcation diagrams along two horizontal slices from panel (a) of Fig.~\ref{fig:stability_boundaries}. The results are shown in Fig.~\ref{fig:hopf_finite_diffusion} below for $D=10$ (upper panels (a)-(c)) and $D=1$ (lower panels (d)-(f)). When $D=10$, the symmetric steady state loses stability through a supercritical Hopf bifurcation at $\koff \approx 0.81$. Further below this threshold, the stable weakly nonlinear anti-phase oscillations are annihilated by the asymmetric steady state at $\koff \approx 0.71$, with the oscillatory period undergoing a sharp increase to infinity. As in the well-mixed regime, relaxation oscillations exist in the range $\koff \, \in \, (0,0.56)$.

As predicted by panel (a) of Fig.~\ref{fig:stability_boundaries}, we find that the symmetric steady state undergoes a subcritical Hopf bifurcation on the horizontal slice $D=1$, near $\koff \approx 0.17$. Furthermore, the branch of limit cycles emerging from the Hopf bifurcation gains stability at a fold point near $\koff \approx 0.27$. Finally, panel (f) of Fig.~\ref{fig:hopf_finite_diffusion} illustrates a typical hard loss of stability near a subcritical Hopf bifurcation. Notice the absence of weakly nonlinear oscillations and the direct transition to relaxation oscillations.

\begin{figure}[H]
\centering
\begin{subfigure}{0.31\linewidth}
\includegraphics[width=\linewidth]{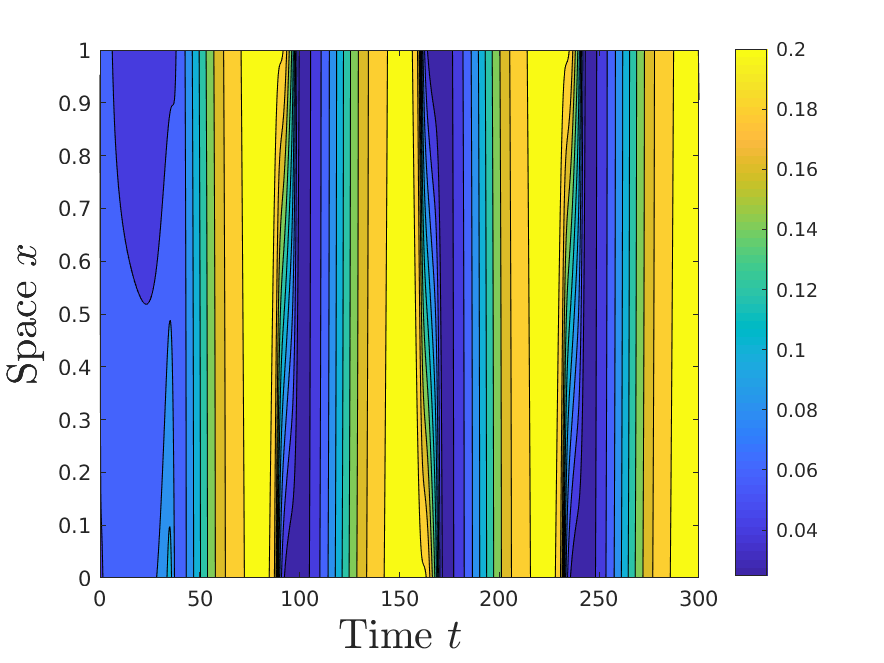}
\caption{Inactive Cdc42 $C(x,t)$.}
\end{subfigure}
\begin{subfigure}{0.31\linewidth}
\includegraphics[width=\linewidth]{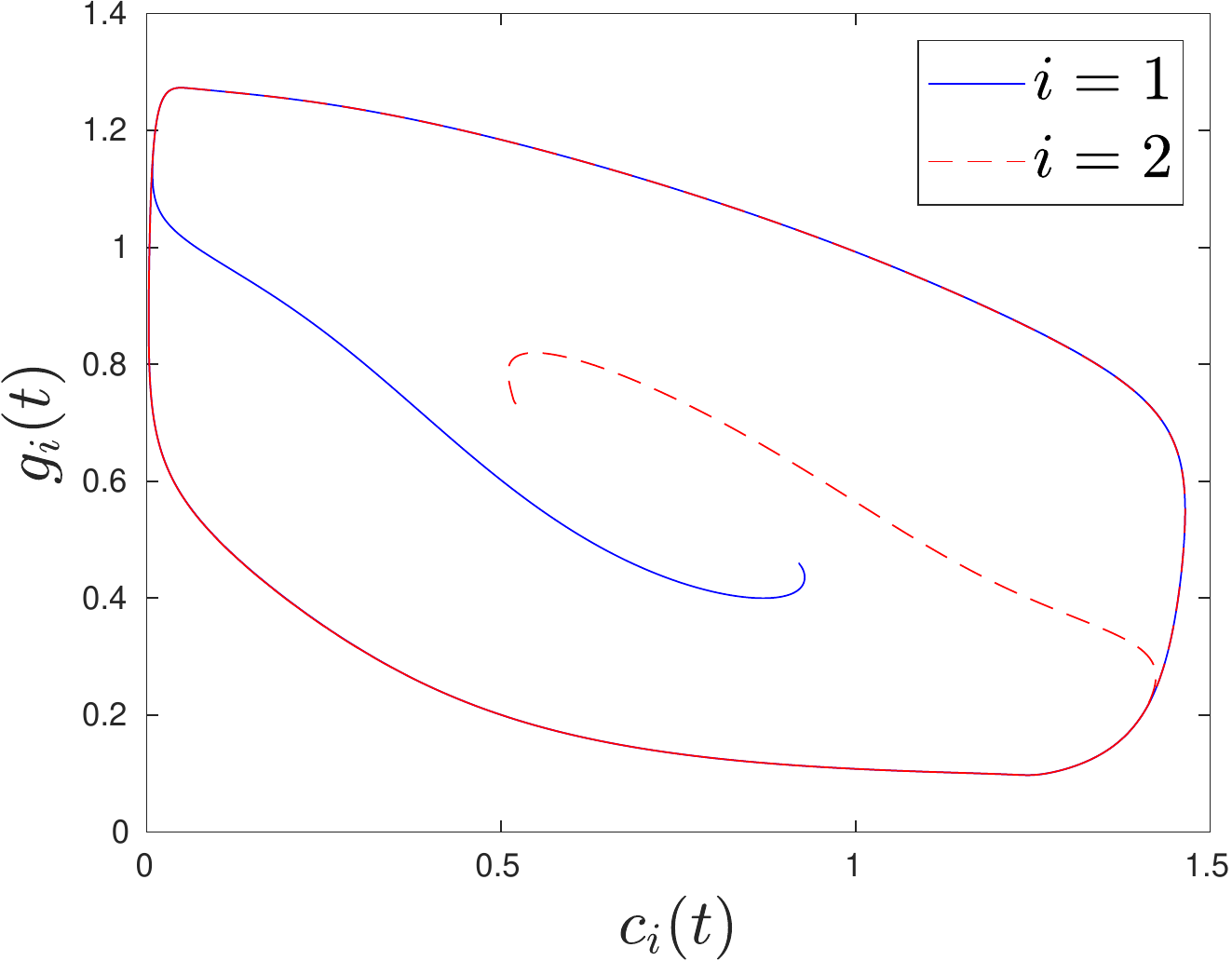}
\caption{Phase plane trajectories.}
\end{subfigure}
\begin{subfigure}{0.31\linewidth}
\includegraphics[width=\linewidth]{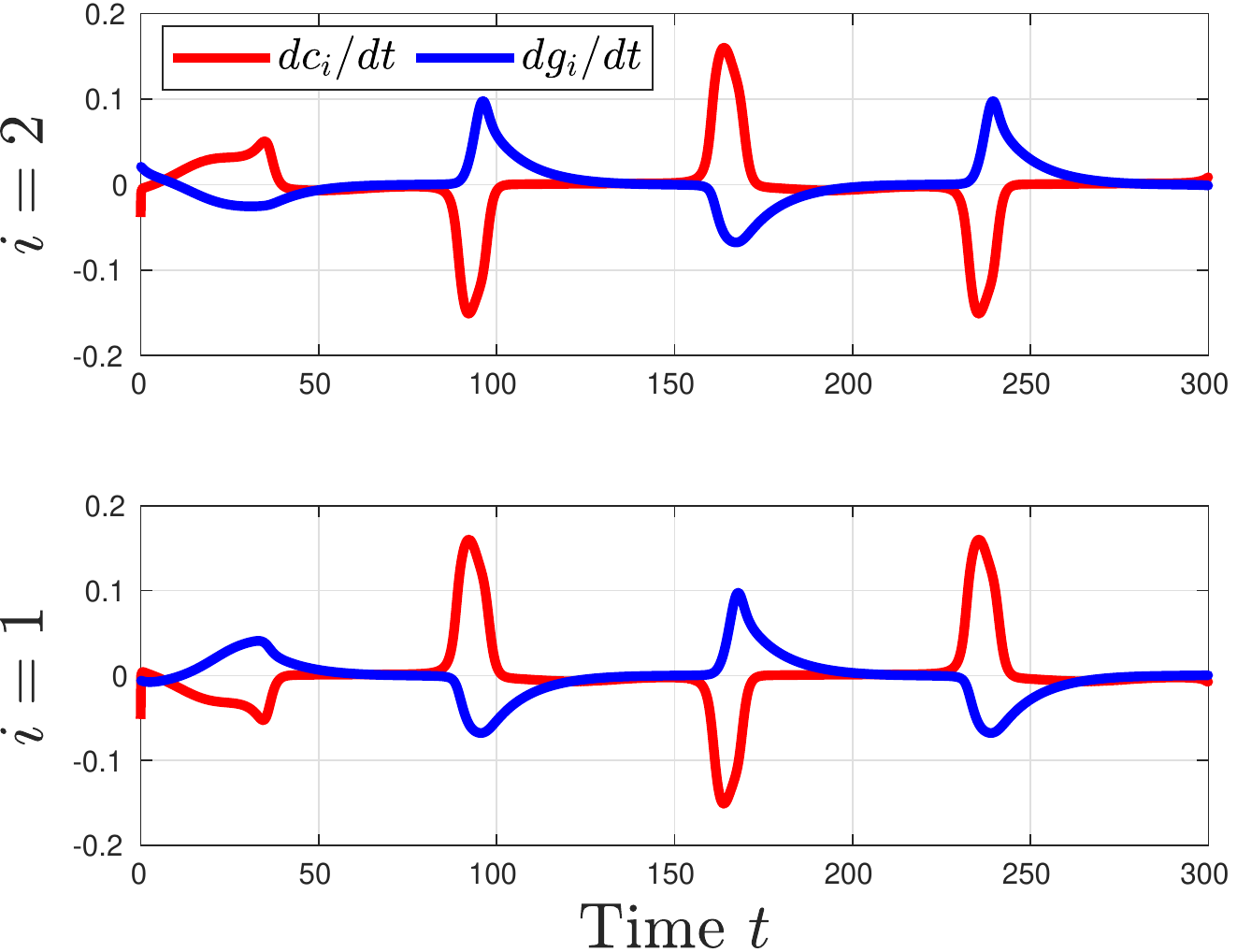}
\caption{Flux on/off each pole.}
\end{subfigure}
\caption{\label{fig:pole_to_pole} Relaxation oscillations in the full PDE--ODE model for $D_c = D_g = 1$, $\koff = 0.1$ and other parameter values as in Table~\ref{table:parm}. Panel (a): kymograph of inactive Cdc42 level, with time on the horizontal axis and space on the vertical axis (see Supplemental Movie S1). Panel (b): trajectories in the phase plane $(c_i(t),g_i(t))$. Panel (c): Cdc42 (blue) and GEF (red) fluxes on/off each pole as a function of time.}
\end{figure}

We conclude this section with Fig.~\ref{fig:pole_to_pole}, which gives some explanations behind the emergence of anti-phase relaxation oscillations in the limit of long GEF membrane residence time. Relaxation oscillations are characterized by long periods of rest followed by sudden sharp variations and are typical of slow-fast systems. Since $\koff < \kminus$, GEF dissociation is slow compared to Cdc42 dissociation, and active GEF can thus be seen as a slow variable while active Cdc42 as a fast variable. This results in a more rapid flux on (or off) each pole for Cdc42 (see panel (c)).

\subsection{Nonlocal PDE model on a circular domain}\label{sec:nonlocal}
In this section, we consider a two-dimensional membrane-bulk model with circular geometry and well-mixed bulk dynamics. This results in the reduced nonlocal PDE model given by system~\eqref{eq:nonlocal}. Our approach combines linear stability analysis and numerical simulations to predict and explore the formation of spatio-temporal patterns near stability thresholds and away from them, in the highly nonlinear regime.

We first define the spatially uniform (or trivial) steady state as $C(r,\theta) \equiv C^*,\, G(r,\theta) \equiv G^*,\, c(\theta) \equiv c^*$ and $g(\theta) \equiv g^*$. Similarly to the one-dimensional PDE--ODE model, $C^*,\, G^*$ and $g^*$ satisfy
\begin{equation*}
C^*=\Cavg-\gamma c^*\,, \quad G^*=\Gavg-\gamma g^*\,, \quad g^*=\frac{k^- c^*}{[k_0 + \kcat (c^*)^2](\Cavg-\gamma c^*)}\,,
\end{equation*}
where $\Gavg=\Gtot/|\Omega|$, $\Cavg=\Ctot/|\Omega|$ and $\gamma = |\partial\Omega|/|\Omega|$ are respectively the average masses and the ratio of the perimeter over the area. The steady state concentration of active Cdc42 is then found to satisfy the following cubic equation:
\begin{equation}\label{eq:ss2D}
[\koff\kappa k^- +\gamma\kon\kcat\Gavg](c^*)^3-\kon\kcat\Gavg\Cavg(c^*)^2 + [\koff k^-+\kon\gamma(k^-+k_0\Gavg)]c^*-\kon k_0\Gavg\Cavg=0,
\end{equation}
where $\gamma=2/R$ for a circle of radius $R$. In the one-dimensional case, the average masses are defined as $\Cavg=\Ctot/L$ and $\Gavg=\Gtot/L$, and equation \eqref{eq:ss2D} reduces to equation \eqref{eq:cubic} derived for the symmetric steady state of the PDE--ODE model.

\subsubsection{Linear stability analysis}
We consider the following nonuniform, spatially periodic, perturbations of the trivial steady state:
\begin{equation}\label{eq:pertb}
 c(\theta, t)=c^* + u_n\e^{\lambda t + in\theta}\,, \quad g(\theta, t)=g^* + v_n\e^{\lambda t + in\theta}\,, \quad n\neq 0\,,
\end{equation}
where once again $\lambda \in \C$ is the eigenvalue parameter. After substitution of \eqref{eq:pertb} within the system of nonlocal PDEs \eqref{eq:nonlocal} and upon linearizing around the trivial steady state, we find that the eigenvector $(u_n,v_n)^T$ must satisfy
\begin{equation}\label{eq:eigen_nonlocal}
\left[ \lambda I - J_n\right] \begin{pmatrix} u_n \\ v_n \end{pmatrix} = \begin{pmatrix} 0 \\ 0 \end{pmatrix},
\end{equation}
where the Jacobian matrix $J_n$ is defined by
\begin{equation}\label{eq:J_n}
J_n = \begin{pmatrix} - D_c^m\frac{n^2}{R^2} + \FFc & \FFg \\ \GGc & - D_g^m\frac{n^2}{R^2} + \GGg \end{pmatrix} = 
\begin{pmatrix} -D_c^mn^2/R^2+2\kcat c^*g^*C^*-k^- &C^*[k_0+\kcat (c^*)^2]\\[5pt] -{2\kappa\kon c^*G^*}/{[1+\kappa(c^*)^2]^2} &-D_g^mn^2/R^2-\koff \end{pmatrix}. 
\end{equation}
For each mode $n$, the corresponding pair of eigenvalues is given by 
\begin{equation*}
\lambda_{\pm}(n)=\frac{\tr J_n\pm\sqrt{\tr^2J_n-4\det J_n}}{2},
\end{equation*}
where the trace and the determinant are
\begin{equation*}
\tr J_n=-n^2D_c^m/R^2-n^2D_g^m/R^2+2\kcat c^*g^*C^*-(k^{-}+\koff)\,, \quad \det J_n=[n^2D_c^m/R^2-2\kcat c^*g^*C^*+k^-][n^2D_g^m/R^2+\koff]+\eta\,,
\end{equation*}
with $\eta$ defined as
\begin{equation}\label{eq:eta}
\eta = \frac{2\kappa\kon c^*[k_0+\kcat (c^*)^2]C^*G^*}{ {[1+\kappa(c^*)^2]^2}}.
\end{equation}
As usual, linear stability requires a negative trace and a positive determinant of the Jacobian matrix for each $n$, therefore yielding the following inequality to be satisfied by $D_c^m$:
\begin{align}\label{cond:nonlocal}
&D_c^m > \max\left\{-D_g^m+\frac{R^2}{n^2}(2\kcat c^*g^*C^*-k^{-}-\koff)\,, -\frac{\eta R^2/n^2}{D_g^mn^2/R^2+\koff}+\frac{R^2}{n^2}(2\kcat c^*g^*C^*-k^-)\right\}.
\end{align}
This inequality places a bound on the minimal diffusion coefficient of Cdc42 for the linear stability of the trivial steady state with respect to non-spatially homogeneous perturbations. This boundary is plotted in Fig.~\ref{fig:stb_nonlocal} and verified numerically.

Notice that because of the integral terms within the nonlocal PDE system \eqref{eq:nonlocal}, we cannot simply set $n=0$ in \eqref{eq:J_n} to recover the Jacobian matrix associated with spatially uniform perturbations. A similar observation was made by R{\"a}tz and R{\"o}ger in their stability analysis of a coupled membrane-bulk reaction-diffusion model for polarity of small GTPases \cite{ratz2012turing}. Furthermore, stability with respect to spatially uniform perturbations, or equivalently, in the absence of membrane diffusion, is required for Turing diffusion-driven instability (cf.~\cite{turing1952}). For this special case, the Jacobian matrix of the linearized system satisfies
\begin{equation}\label{eq:J_0}
J_0 = \begin{pmatrix} -\FFC\gamma + \FFc & \FFg \\ \GGc & -\GGG\gamma + \GGg \end{pmatrix}
\end{equation}
and necessary conditions for diffusion-driven instability are given by
\begin{equation}\label{eq:stability_condition}
\tr J_0 = \FFc+\GGg - \gamma\left( \FFC + \GGG \right)<0,\quad \det J_0 = \gamma^2 \FFC \GGG - \gamma\left( \FFc \GGG + \GGg \FFC \right) + \FFc\GGg-\FFg\GGc>0,
\end{equation}
which have been verified numerically for the parameter values employed in our study (see Appendix \ref{sec:appendix_nonlocal}). Restoring membrane diffusion, two different symmetry-breaking mechanisms may cause loss of stability of the trivial steady state: the stationary or the oscillatory Turing instabilities, which we describe below.
\begin{itemize}
\item The \textbf{stationary Turing instability}, which happens when the Jacobian matrix possesses an eigenvalue at the origin: $\det J_n = 0$, with $n \neq 0$. We note that because of circular symmetry (or $O(2)$ symmetry), an unstable mode always comes in a pair $\pm n$, and thus the stationary Turing instability corresponds to a pitchfork bifurcation.
\item The \textbf{oscillatory Turing instability}, which happens when the Jacobian matrix possesses a pair of critical eigenvalues on the imaginary axis: $\tr J_n = 0$ and $\det J_n > 0$, with $n \neq 0$. This instability corresponds to a Hopf bifurcation with $O(2)$ symmetry (cf.\cite{vangils1986}), and traveling waves as opposed to stationary patterns are expected to form.
\end{itemize}
More intricate bifurcations resulting from the interaction of two spatial modes may also be detected when performing linear stability analysis. One example is the Bogdanov--Takens bifurcation with $O(2)$ symmetry (cf.~\cite{dangelmayr1987}), which occurs when the stationary and oscillatory Turing instabilities coincide for the same (nonzero) spatial modes $\pm n$. At such a bifurcation point, we have $\tr J_n = 0$ and $\det J_n = 0$, and thus the linearized system possesses two zero-eigenvalues of multiplicity two.

\subsubsection{Spatio-temporal pattern formation in the nonlocal PDE model}\label{sec:nonlocal_stb}

Here, we compare our linear stability analysis against numerical simulations of the nonlocal system of PDEs \eqref{eq:nonlocal}. Simple finite differences are employed to spatially discretize diffusion terms while nonlocal integral terms are handled using the trapezoidal quadrature rule. As initial conditions, we take in this section
\begin{equation}
c(\theta,0) = c^* + u_n\cos\left(n\theta+\varphi\right)\,, \qquad g(\theta,0) = g^* + v_n\cos\left(n\theta+\varphi\right)\,,
\end{equation}
where the eigenvector $(u_n,\,v_n)^T$ is a non-trivial solution of \eqref{eq:eigen_nonlocal} and $\varphi$ is an arbitrary angular phase shift.

\begin{figure}[H]
\centering
\includegraphics[width=0.9\linewidth]{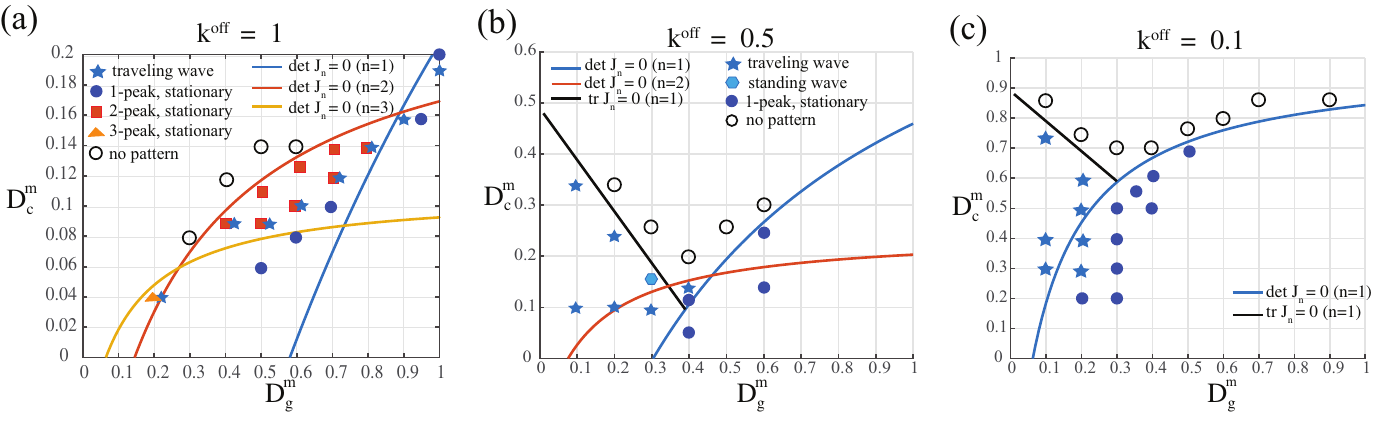}
\caption{\label{fig:stb_nonlocal} Stability boundaries for the nonlocal PDE model in the $(D_g^m, D_c^m)$ parameter plane with $\koff = 1$ (panel (a)), $\koff=0.5$ (panel (b)) and $\koff=0.1$ (panel (c)) and other parameters the same as in Table \ref{table:parm}. For illustrative purposes, parameter values where we observe a stationary pattern with a single peak, a stationary pattern with two peaks, a traveling and a standing waves are labeled by a filled circle, a square, a star and a polygon. A stationary pattern that evolves into a traveling wave is indicated by a star superposed on a square. Some secondary stability boundaries associated with higher spatial modes have been omitted.}
\end{figure}

In Fig.~\ref{fig:stb_nonlocal}, we show three stability diagrams in the $(D_g^m, D_c^m)$ parameter plane for different $\koff$ values. As expected, the trivial steady state is linearly stable when Cdc42 diffuses fast. Fig.~\ref{fig:stb_nonlocal} also confirms the destabilizing effect of long GEF membrane residence times, since the linear stability region shrinks as $\koff$ decreases. In each diagram, the different spatio-temporal patterns observed via numerical simulations are labeled with a variety of symbols. For instance, the star and the filled circle respectively indicate a traveling wave and a stationary pattern, while the empty circle shows the absence of patterns. This non-exhaustive parameter exploration reveals that the spatio-temporal dynamics thresholds clearly correspond to the stability boundaries shown in Fig.~\ref{fig:stb_nonlocal}, suggesting that in the limit of fast bulk diffusion, all the identified bifurcations are supercritical.

\begin{figure}[H]
\centering
\begin{subfigure}{0.31\linewidth}
\includegraphics[width=\linewidth]{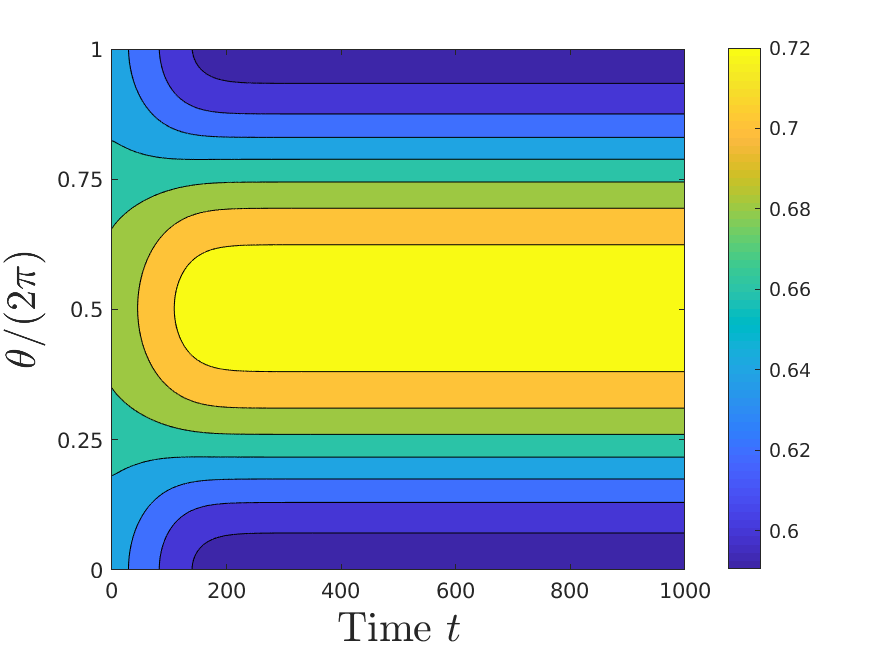}
\caption{$D_g^m = 1,\, D_c^m = 0.2$}
\end{subfigure}
\begin{subfigure}{0.31\linewidth}
\includegraphics[width=\linewidth]{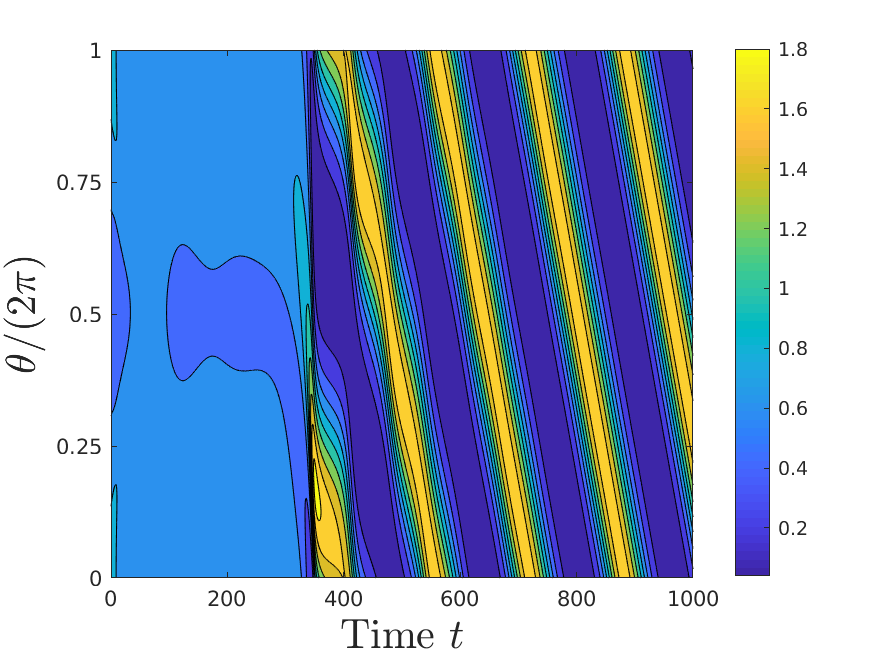}
\caption{$D_g^m = 1,\, D_c^m = 0.18$}
\end{subfigure}
\begin{subfigure}{0.31\linewidth}
\includegraphics[width=\linewidth]{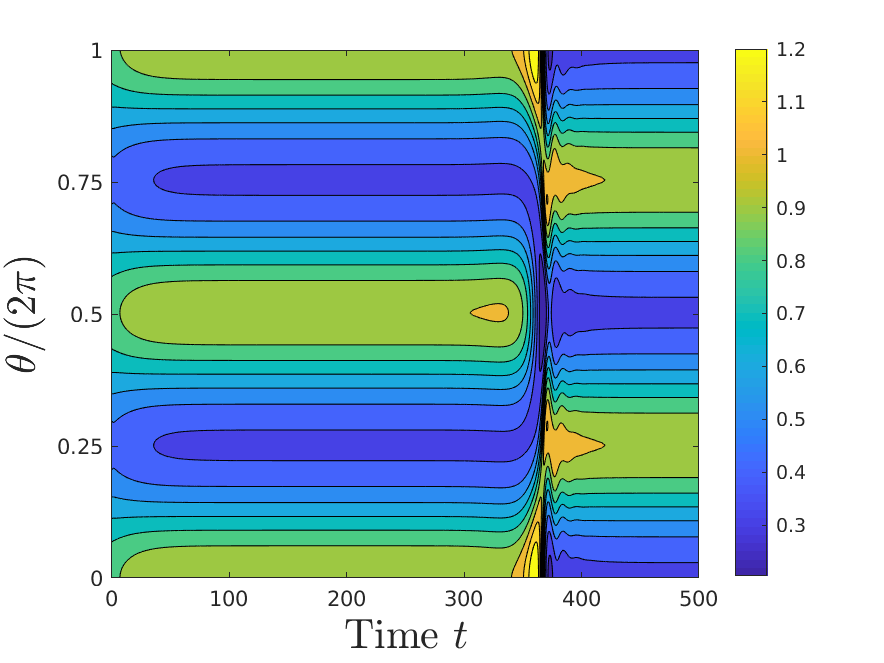}
\caption{$D_g^m = 0.5,\, D_c^m = 0.11$}
\end{subfigure}
\begin{subfigure}{0.31\linewidth}
\includegraphics[width=\linewidth]{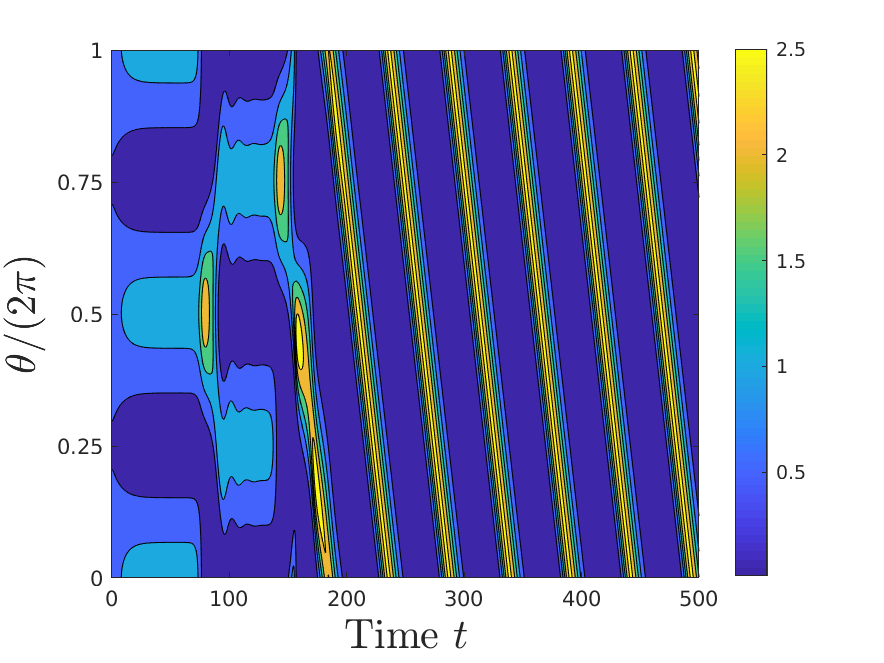}
\caption{$D_g^m = 0.5,\, D_c^m = 0.1$}
\end{subfigure}
\begin{subfigure}{0.31\linewidth}
\includegraphics[width=\linewidth]{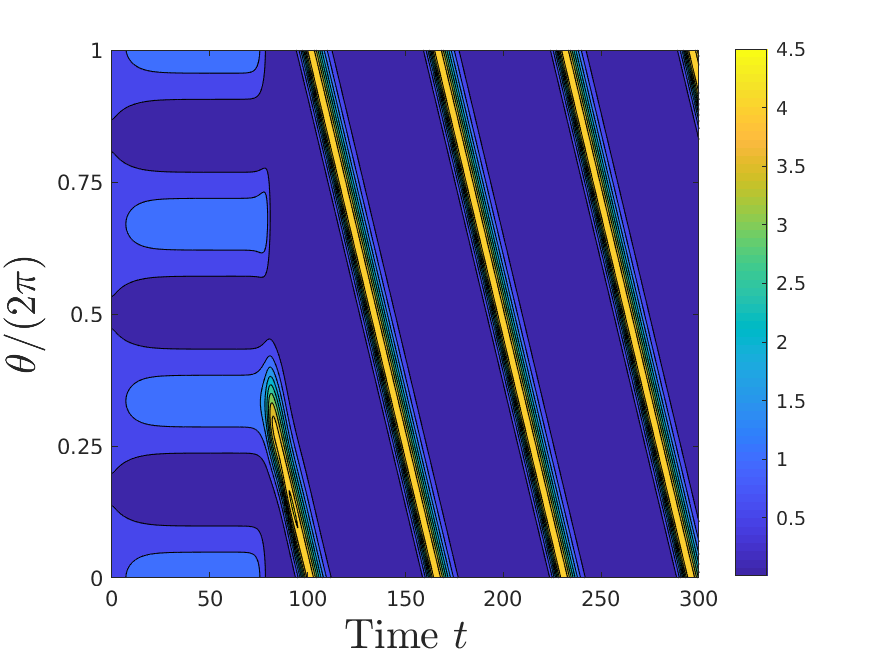}
\caption{$D_g^m = 0.2,\, D_c^m = 0.04$}
\end{subfigure}
\begin{subfigure}{0.31\linewidth}
\includegraphics[width=\linewidth]{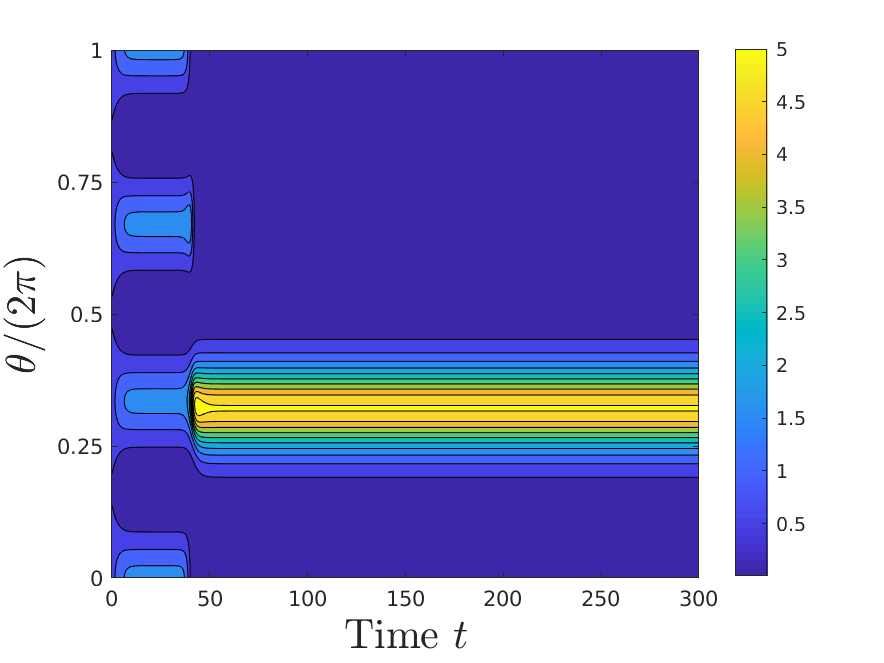}
\caption{$D_g^m = 0.4,\, D_c^m = 0.04$}
\end{subfigure}
\caption{\label{fig:WM_pattern} A gallery of numerical simulations of the nonlocal model with $\koff = 1$ and membrane diffusion coefficients taken from Fig.~\ref{fig:stb_nonlocal}(a). Shown above are kymographs of membrane-bound active Cdc42, with time on the horizontal axis and space on the vertical axis. Panel (a): weakly nonlinear stationary pattern near a mode $n=1$ pitchfork stability boundary (see Supplemental Movie S2). Panel (b): for a slightly lower $D_c^m$ value, the same pattern evolves into a traveling wave. Panel (c): quasi-stationary pattern with two peaks near a mode $n=2$ pitchfork stability boundary. Panel (d): transition to a traveling wave from an initial pattern with two peaks. Panel (e): transition to a traveling wave from an initial pattern with three peaks. Panel (f): transition to a localized stationary pattern with a single peak from an initial pattern with three peaks.}
\end{figure}

Some numerically computed spatio-temporal patterns obtained when $\koff = 1$ are shown in Fig.~\ref{fig:WM_pattern}. For this case, the trivial steady state is expected to lose stability via pitchfork bifurcations associated with the modes $n=1,2$ and $3$ (indicated by the blue, red and golden curves). Nevertheless, our simulations revealed that some patterns that are initially stationary can evolve into traveling waves, and this phenomenon occurs in the absence of any oscillatory Turing instabilities (see panels (b), (d) and (e)). One potential explanation comes from the vicinity of multiple stability curves, which causes two or more spatial modes to interact and ultimately leads to a stationary pattern undergoing a secondary Hopf bifurcation. We also observed ``quasi-stationary'' patterns with multiple peaks. For instance in Fig.~\ref{fig:WM_pattern}(c), two peaks are seen to collapse and split again immediately after, resulting in a different pattern configuration. Finally, when the ratio $D_c^m/D_g^m$ is very small, a large number of modes become unstable and linear stability analysis cannot be employed to predict the final number of peaks of a stationary pattern. This phenomenon precisely corresponds to what happens in panel (f), where for $D_c^m/D_g^m = 0.1$, an initial pattern with three peaks rapidly evolves into a highly localized pattern with a single peak.

\begin{figure}[H]
\centering
\begin{subfigure}{0.31\linewidth}
\includegraphics[width=\linewidth]{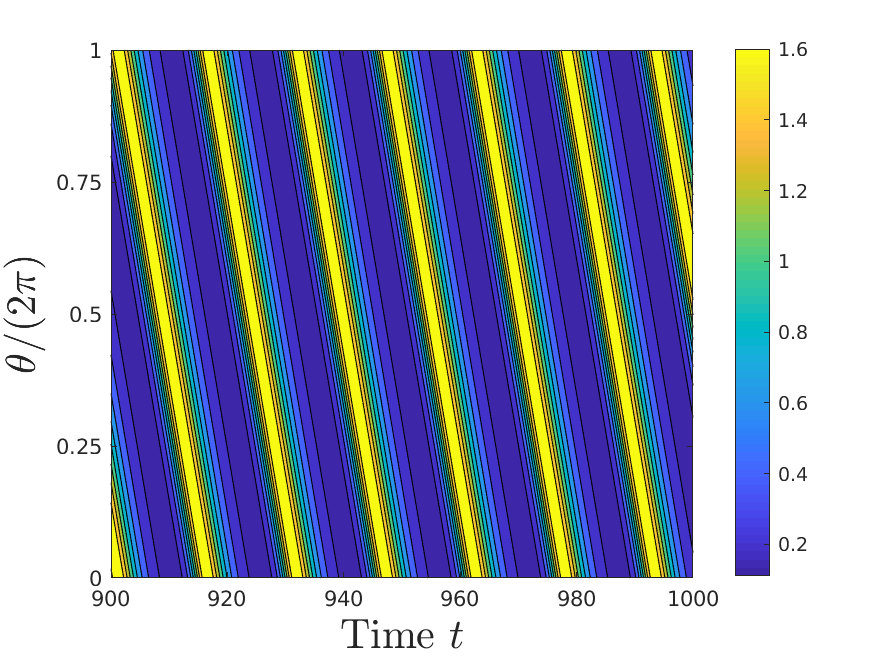}
\caption{$D_g^m = 0.2,\, D_c^m=0.28$}
\end{subfigure}
\begin{subfigure}{0.31\linewidth}
\includegraphics[width=\linewidth]{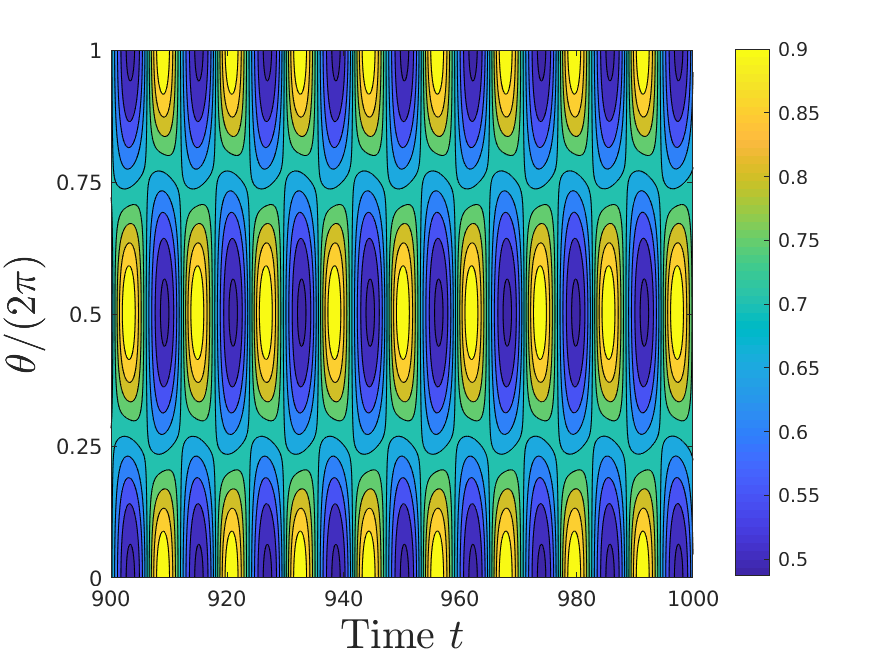}
\caption{$D_g^m = 0.2,\, D_c^m=0.28$}
\end{subfigure}
\begin{subfigure}{0.31\linewidth}
\includegraphics[width=\linewidth]{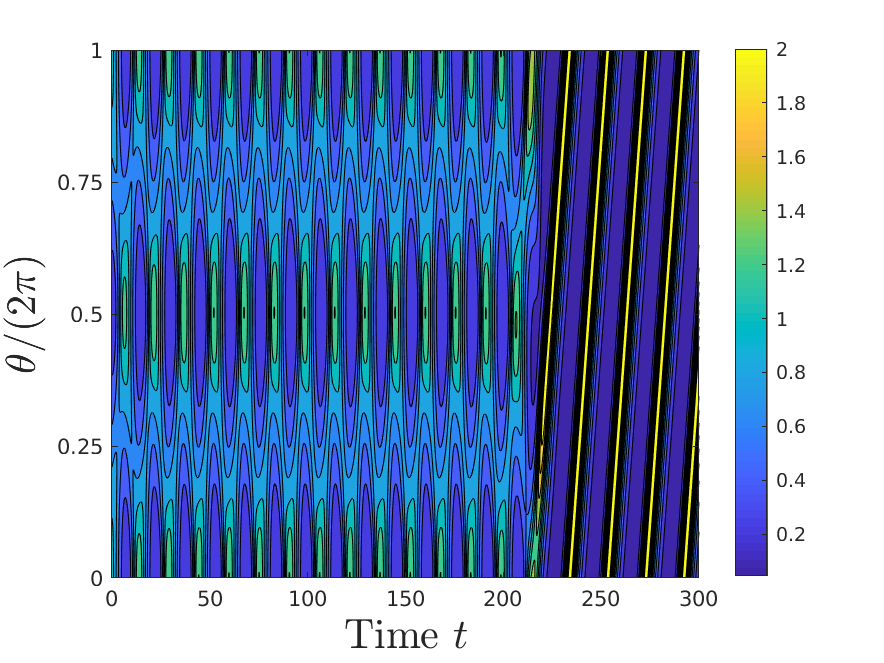}
\caption{$D_g^m = 0.2,\, D_c^m=0.2$}
\end{subfigure}
\caption{\label{fig:WM_wave} Numerically computed traveling waves with $\koff = 0.5$ and membrane diffusion coefficients taken from Fig.~\ref{fig:stb_nonlocal}(b). In panels (a-b), there is coexistence of rotating and standing waves near the mode $n=1$ Hopf stability boundary. Further below this threshold, in $D_c^m = 0.2$, standing waves are only transient and evolve into rotating waves. Each solution is shown in Supplemental Movies S3, S4 and S5.}
\end{figure}

At smaller $\koff$ values, as indicated by the black curve in panels (b-c) of Fig.\ref{fig:stb_nonlocal}, the loss of stability of the trivial steady state can arise via a Hopf bifurcation (see Appendix \ref{sec:appendix_nonlocal}). This is similar to what is seen in the one-dimensional PDE--ODE model from Section \ref{sec:PDE_ODE}, where oscillatory dynamics also require a small GEF dissociation rate. Near those oscillatory Turing instabilities, the formation of traveling waves is expected. We distinguish between rotating and standing waves, with the latter type often referred to as pole-to-pole oscillations. As shown in Fig.~\ref{fig:WM_wave}, our numerical experiments revealed that the two types coexist near the Hopf stability boundaries, while away from them in the highly nonlinear regime the standing waves are only transient.

Finally, for both $\koff = 0.5$ and $\koff = 0.1$, we notice in Fig.~\ref{fig:stb_nonlocal} the presence of Bogdanov--Takens bifurcation points associated with the $n=1$ spatial mode and indicated by the branching of the black from the blue curves. Traveling waves are expected to interact with stationary patterns in the neighborhood of such a bifurcation point, and further details about its effect on the dynamics will be given in Section \ref{sec:2D_membrane_bulk} in the context of finite bulk diffusion.

\subsection{Two-dimensional membrane-bulk reaction-diffusion model}\label{sec:2D_membrane_bulk}

In this section, we consider the full two-dimensional membrane-bulk reaction-diffusion model with circular geometry. Our aim is to determine the effects of a finite bulk diffusion field on the spatio-temporal dynamics from Section \ref{sec:nonlocal}. Linear stability analysis is performed first, followed then by numerical simulations.

\subsubsection{Linear stability analysis}

We consider a perturbation of the spatially uniform steady state given by
\begin{equation}\label{eq:pert_disk}
C(r,\theta,t)=C^*+\e^{\lambda t}\Phi(r,\theta)\,, \quad G(r,\theta,t)=G^*+\e^{\lambda t}\Psi(r,\theta)\,, \quad c(\theta,t)=c^*+\e^{\lambda t}\phi(\theta)\,, \quad g(\theta,t)=g^*+\e^{\lambda t}\psi(\theta),
\end{equation}
where $\Phi(r,\theta),\,\Psi(r, \theta)$ and $\phi(\theta),\, \psi(\theta)$ are respectively the bulk and membrane eigenfunctions, while $(C^*,\, G^*,\, c^*,\, g^*)$ is such as introduced in Section \ref{sec:nonlocal}. Next, upon inserting \eqref{eq:pert_disk} within equations \eqref{eq:2Dmodel}--\eqref{eq:consv2dMAIN} and after linearization, we obtain the following eigenvalue problem in the bulk:
\begin{subequations}\label{eq:linearsys2d}
\begin{align}
& D_c\left[\Phi_{rr}(r,\theta) +\frac{1}{r}\Phi_r(r,\theta)+\frac{1}{r^2}\Phi_{\theta\theta}(r,\theta) \right] - \lambda\Phi(r,\theta) = 0,\quad 0<r<R,\quad 0<\theta<2\pi,\\
& D_g\left[\Psi_{rr}(r,\theta) +\frac{1}{r}\Psi_r(r,\theta)+\frac{1}{r^2}\Psi_{\theta\theta}(r,\theta) \right] - \lambda\Psi(r,\theta) = 0,\quad 0<r<R,\quad 0<\theta<2\pi,
\end{align}
and on the membrane:
\begin{align}\label{eq:linear_memb}
& \frac{D_c^m}{R^2}\phi_{\theta\theta}(\theta) + \FFC\Phi(R,\theta) + \left(\FFc - \lambda\right)\phi(\theta) + \FFg\psi(\theta) = 0\,, \quad 0<\theta<2\pi\,, \\
& \frac{D_g^m}{R^2}\psi_{\theta\theta}(\theta) + \GGG\Psi(R,\theta) + \GGc\phi(\theta) + \left(\GGg - \lambda\right)\psi(\theta) = 0\,, \quad 0<\theta<2\pi\,,
\end{align}
with boundary conditions given by
\begin{align}\label{eq:linear_BC2D}
&-D_c\Phi_r(R,\theta)=\FFC\Phi(R,\theta)+\FFc\phi(\theta)+\FFg\psi(\theta),\quad -D_g\Psi_r(R,\theta)=\GGG\Psi(R,\theta)+\GGc\phi(\theta)+\GGg\psi(\theta).
\end{align}
Here each of the partial derivatives is evaluated at the spatially uniform steady state. Also, the conservation laws \eqref{eq:consv2d} yield
\begin{equation}\label{eq:zero_mass}
\int_0^{2\pi}\int_0^R \Phi(r,\theta)rdrd\theta + \int_0^{2\pi} \phi(\theta)R d\theta = 0\,, \quad \int_0^{2\pi}\int_0^R \Psi(r,\theta)rdrd\theta + \int_0^{2\pi} \psi(\theta)R d\theta = 0.
\end{equation}
\end{subequations}

Our aim is to derive, for each spatial mode, the characteristic equations satisfied by the eigenvalue parameter. We consider first the general case where $\lambda$ is nonzero, followed then by the zero-crossing eigenvalue case.

\paragraph{Nonzero eigenvalue.} When $\lambda \neq 0$, separation of angular and radial variables yields the following expressions for the eigenfunctions:
\begin{equation}\label{eq:ansatz}
\Phi_n(r,\theta)=A_n \frac{I_n(\omega_c r)}{I_n(\omega_c R)} e^{in\theta}\,, \quad \Psi_n(r,\theta)=B_n \frac{I_n(\omega_g r)}{I_n(\omega_g R)} e^{in\theta}\,, \quad \phi_n(\theta) = u_n e^{in\theta}\,, \quad \psi_n(\theta) = v_n e^{in\theta}\,,
\end{equation}
where $\omega_c=\sqrt{\lambda/D_c}$, $\omega_g=\sqrt{\lambda/D_g}$ and $I_n(z)$ for $n \in \Z$ are the usual modified Bessel functions. Next, from the linearized boundary conditions \eqref{eq:linear_BC2D}, $A_n$ and $B_n$ are readily found to satisfy
\begin{equation*}
A_n = - \frac{\FFc u_n + \FFg v_n}{D_c \omega_c \frac{I_n'(\omega_c R)}{I_n(\omega_c R)} + \FFC}\,, \quad B_n = - \frac{\GGc u_n + \GGg v_n}{D_g \omega_g \frac{I_n'(\omega_g R)}{I_n(\omega_g R)} + \GGG}.
\end{equation*}
Then upon substituting \eqref{eq:ansatz} within the linearized membrane reaction-diffusion system, we derive the following linear system to be satisfied by the eigenvector $(u_n, v_n)^T$:
\begin{equation}\label{eq:nonzero}
\begin{pmatrix}
\lambda + \frac{n^2 D_c^m}{R^2} - p_n(\lambda) \FFc & - p_n(\lambda) \FFg \\
- q_n(\lambda) \GGc & \lambda + \frac{n^2 D_g^m}{R^2} - q_n(\lambda) \GGg
\end{pmatrix}
\begin{pmatrix} u_n \\ v_n \end{pmatrix} = \begin{pmatrix} 0 \\ 0 \end{pmatrix}\,,
\end{equation}
where $p_n(\lambda)$ and $q_n(\lambda)$ are defined by
\begin{equation*}
p_n(\lambda) = \frac{D_c\omega_cI_n'(\omega_cR)}{D_c\omega_cI_n'(\omega_cR) + \FFC I_n(\omega_cR)}\,, \quad q_n(\lambda) = \frac{D_g\omega_gI_n'(\omega_gR)}{D_g\omega_gI_n'(\omega_gR)+\GGG I_n(\omega_gR)}.
\end{equation*}
Finally, setting the determinant of the matrix in \eqref{eq:nonzero} to zero yields the following characteristic equation:
\begin{equation}\label{eq:lambdaFull}
F_n(\lambda) \equiv \Bigg[\lambda+\frac{n^2 D_c^m}{R^2}-p_n(\lambda)\FFc\Bigg] \Bigg[\lambda+\frac{n^2 D_g^m}{R^2}-q_n(\lambda)\GGg\Bigg] - p_n(\lambda)q_n(\lambda)\FFg\GGc = 0.
\end{equation}
For an oscillatory Turing instability, purely imaginary roots of $F_n(\lambda)$ for $n \neq 0$ are necessary. Fortunately when treating the spatially uniform mode, the integral constraints \eqref{eq:zero_mass} and the linear system \eqref{eq:nonzero} with $n=0$ lead to an equivalent solvability condition. A rigorous Turing stability analysis would require us to prove the absence of roots $\lambda$ with $\Re(\lambda) \geq 0$ for the characteristic equation $F_0(\lambda) = 0$, which cannot simply be verified numerically with the argument principle from complex analysis because of the singularity at the origin. This challenging task is therefore left for further study. Similar issues were encountered when treating the in-phase mode in the PDE--ODE model.

\paragraph{Zero-crossing eigenvalue.} For this special case, a suitable ansatz for the eigenfunctions is given by
\begin{equation}\label{eq:ansatz_zero}
\Phi_n(r,\theta) = A_n \left(\frac{r}{R}\right)^n e^{in\theta}\,, \quad \Psi_n(r,\theta) = B_n \left(\frac{r}{R}\right)^n e^{in\theta}\,, \quad \phi_n(\theta) = u_n e^{in\theta}\,, \quad \psi_n(\theta) = v_n e^{in\theta}\,,
\end{equation}
where from the linearized boundary conditions \eqref{eq:linear_BC2D}, $A_n$ and $B_n$ are found to satisfy 
\begin{equation*}
A_n = -\frac{\FFc u_n + \FFg v_n}{D_c\frac{n}{R} + \FFC}\,, \quad B_n = -\frac{\GGc u_n + \GGg v_n}{D_g\frac{n}{R} + \GGG}\,.
\end{equation*}
The homogeneous linear system satisfied by the eigenvector $(u_n,v_n)^T$ is then given by
\begin{equation}\label{eq:zero}
\begin{pmatrix}
\frac{n^2 D_c^m}{R^2} - p_n(0)\FFc & - p_n(0)\FFg \\
- q_n(0) \GGc & \frac{n^2 D_g^m}{R^2} - q_n(0)\GGg
\end{pmatrix} \begin{pmatrix} u_n \\ v_n \end{pmatrix} = \begin{pmatrix} 0 \\ 0 \end{pmatrix}\,,
\end{equation}
where $p_n(0)$ and $q_n(0)$ are defined by
\begin{equation*}
p_n(0) = \frac{D_c n}{D_c n + R\FFC}\,, \quad q_n(0) = \frac{D_g n}{D_g n + R\GGG}\,.
\end{equation*}
After setting the determinant of the matrix in \eqref{eq:zero} to zero, we obtain the following necessary condition for a zero-crossing eigenvalue:
\begin{equation}\label{eq:charlambda0}
F_n(0) \equiv \left[\frac{n^2 D_c^m}{R^2} - p_n(0)\FFc \right] \left[\frac{n^2 D_c^m}{R^2} - q_n(0)\GGg \right] - p_n(0)q_n(0)\FFg\GGc = 0.
\end{equation}
When equation \eqref{eq:charlambda0} is satisfied for $n \neq 0$, the trivial steady state undergoes a pitchfork bifurcation, corresponding to the stationary Turing instability.

\subsubsection{Spatio-temporal pattern formation in the two-dimensional model}

We now explore the effect of diffusion and GEF membrane residence time on the formation of spatio-temporal patterns on a circular bulk domain when the inactive species diffuse at finite rates. As bifurcation parameters, we therefore select the bulk and membrane diffusion coefficients, as well as $\koff$, the GEF dissociation rate. The spatial discretization in our numerical method combines a finite element approach in the bulk with a finite difference approach on the boundary (see Appendix~\ref{2dnumerics} for further details). The initial conditions stimulate a specific spatial mode $n$ by adding  the corresponding eigenfunction to the uniform steady state. Finally, the stability of the uniform steady state with respect to small perturbations is predicted from a numerical eigenvalue computation, directly by solving \eqref{eq:charlambda0} and \eqref{eq:lambdaFull}.

\begin{figure}[H]
\centering
\begin{subfigure}{0.31\linewidth}
\includegraphics[width=\linewidth]{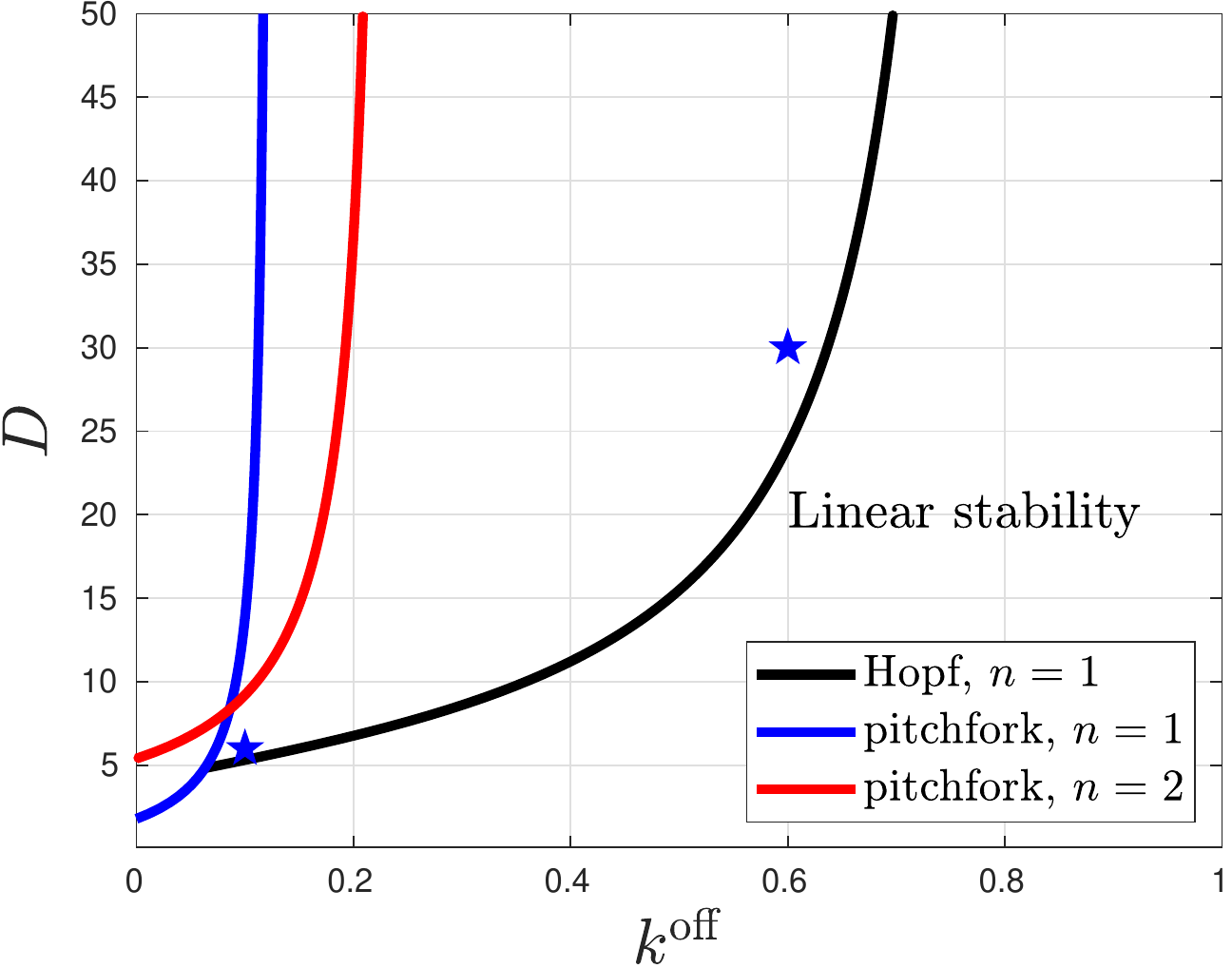} 
\caption{$(\koff,D)$ with $D_c^m = D_g^m = 0.1$.}
\end{subfigure}
\begin{subfigure}{0.31\linewidth}
\includegraphics[width=\linewidth]{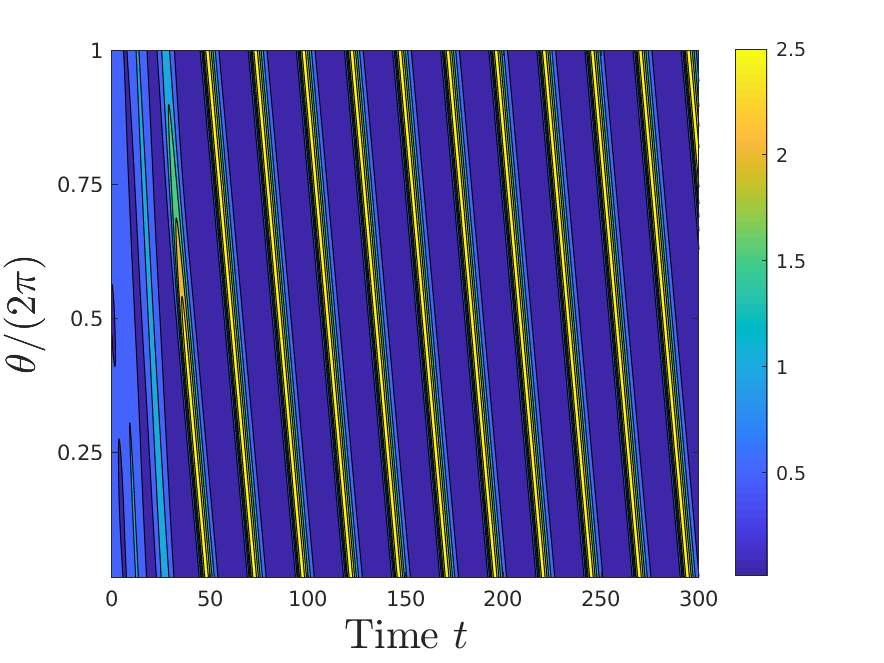} 
\caption{$\koff = 0.6,\, D=30$.}
\end{subfigure}
\begin{subfigure}{0.31\linewidth}
\includegraphics[width=\linewidth]{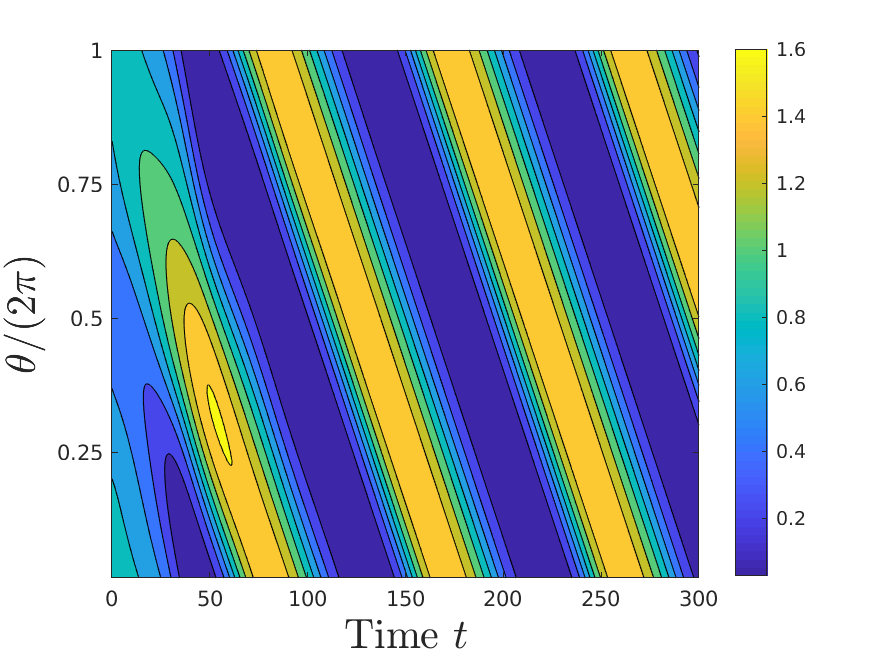} 
\caption{$\koff = 0.1,\, D=6$.}
\end{subfigure}
\caption{\label{fig:wave} Panel (a): stability boundaries in the $D$ versus $\koff$ parameter plane. Panels (b-c): numerically computed traveling wave solutions of active Cdc42 along the circular membrane. The blue stars in panel (a) indicate where each simulation is performed. As the Bogdanov--Takens bifurcation point is approached, we remark a slowing down of the phase velocity of the traveling wave. Membrane diffusion coefficients are equal to $D_c^m = D_g^m = 0.1$, while other parameter values are given in Table~\ref{table:parm}.}
\end{figure}

We present in panel (a) of Fig.~\ref{fig:wave} a stability diagram in the $(\koff,D)$ parameter plane, where we assume that both inactive Cdc42 and GEF diffuse at the same rate $D \equiv D_c = D_g$. We remark that the stability boundaries are similar to those arising in the one-dimensional PDE--ODE model as shown in panel (a) of Fig.~\ref{fig:stability_boundaries}. In both cases, the simultaneous increase of $D_c$ and $D_g$ has a destabilizing effect and results in a reduction of the region of linear stability of the trivial steady state. However, a key difference between the two diagrams is the curve of primary instabilities. In 2-D circular geometry, this curve is composed of a pitchfork (\textcolor{blue}{blue}) and a Hopf (\textcolor{black}{black}) bifurcation segments, both associated with the $n=1$ spatial mode, that connect in a Bogdanov--Takens bifurcation point in $(\koff, D) \approx (0.06, 4.83)$. Hence, decreasing $\koff$ when the bulk diffusion level is above this threshold leads to a loss of stability of the trivial steady state through a Hopf bifurcation and traveling (or rotating) waves are expected to form. Below the Bogdanov--Takens bifurcation point, our linear stability analysis predicts a loss of stability through a pitchfork bifurcation, leading to the formation of stationary patterns with accumulation of proteins in a single location. 

\begin{figure}[H]
\centering
\begin{subfigure}{0.31\linewidth}
\includegraphics[width=\linewidth]{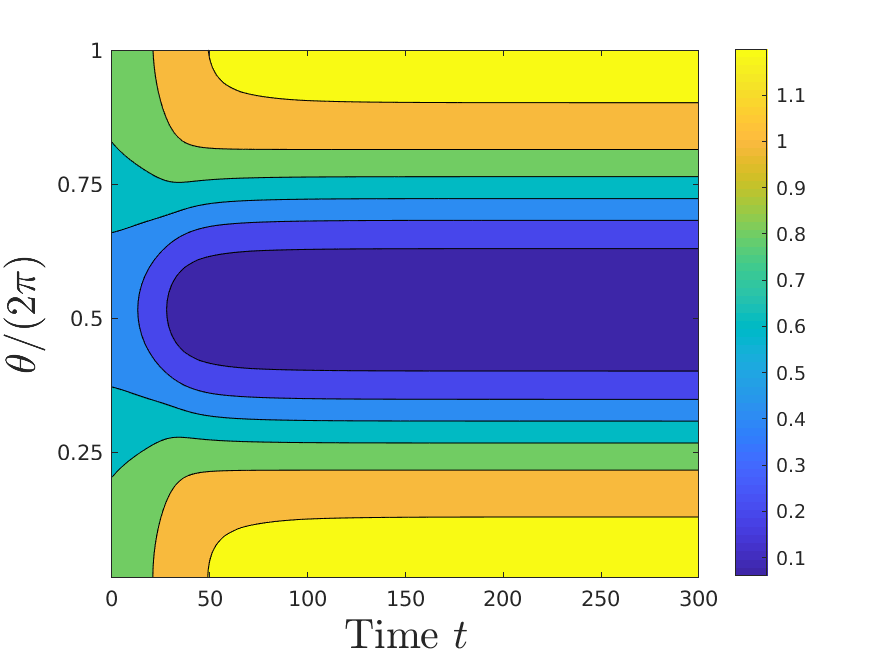} 
\caption{Kymograph of active Cdc42 level.}
\end{subfigure}
\begin{subfigure}{0.31\linewidth}
\includegraphics[width=\linewidth]{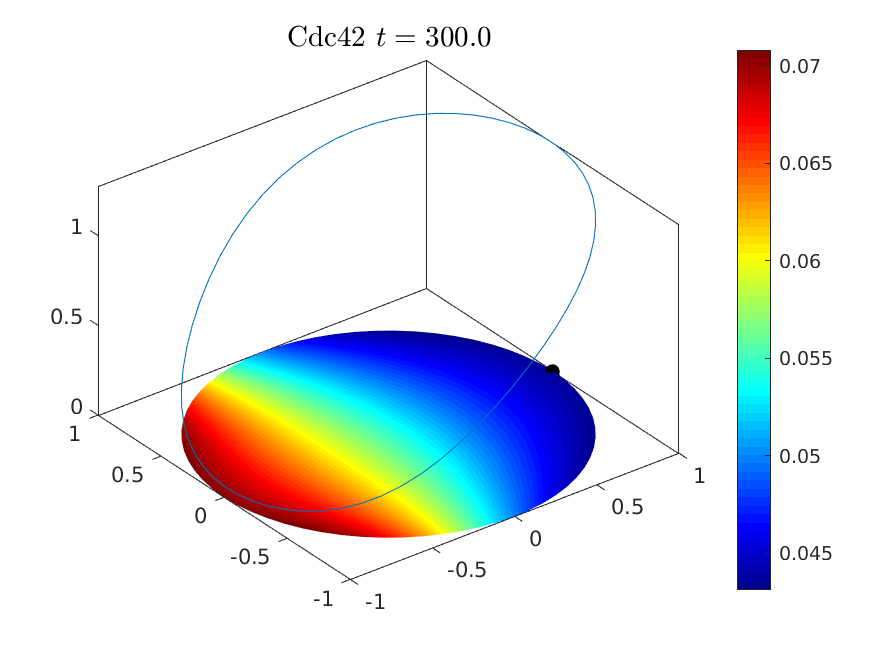}
\caption{Active and inactive Cdc42 levels.}
\end{subfigure}
\begin{subfigure}{0.31\linewidth}
\includegraphics[width=\linewidth]{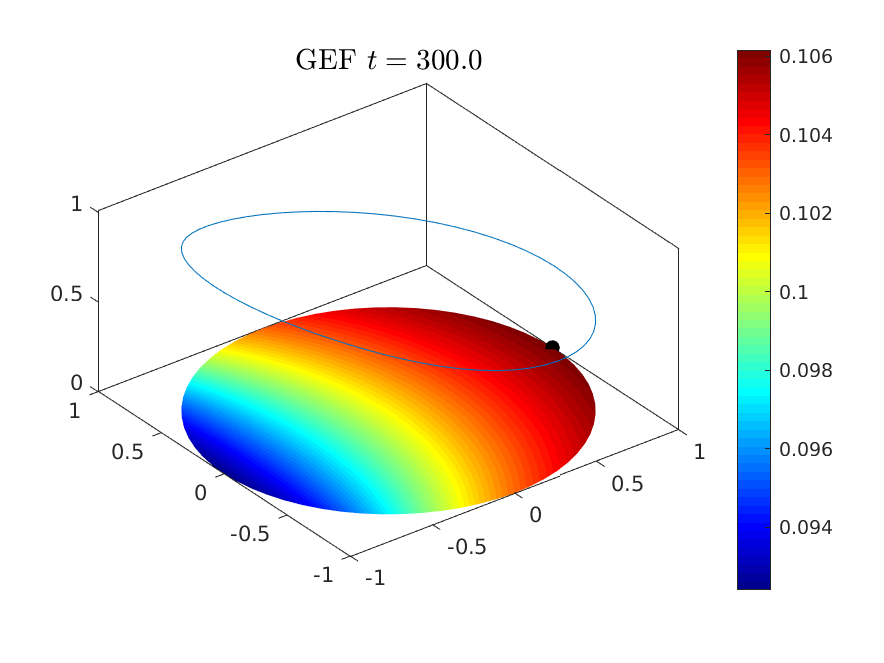}
\caption{Active and inactive GEF levels.}
\end{subfigure}
\caption{\label{fig:pattern_n1} Stable polarized state with $\koff = 0.05$ and $D = 4$ (other parameters are the same as in the caption of Fig.~\ref{fig:wave}). In panels (b-c), the level of membrane-bound (active) proteins is given by the blue curve. The black dot in the same panels indicates the position on the disk with zero azimuthal coordinate. Because of circular symmetry, any rotations and reflections of this pattern is also a solution. See Supplemental Movie S6.}
\end{figure}

As the Hopf curve approaches the pitchfork stability boundary, the pair of critical eigenvalues tends to zero and a slowing down of the phase velocity of the traveling wave is expected. As shown in panels (b-c) of Fig.~\ref{fig:wave}, this phenomena is observed in our numerical simulations. Eventually, the wave stops and becomes a stationary pattern (see Fig.~\ref{fig:pattern_n1}). Furthermore, in the vicinity of the Bogdanov--Takens bifurcation point, it is possible to derive a four-dimensional system of ODEs approximating the dynamics of the full bulk-membrane reaction-diffusion system. Although, such a dimensionality reduction process is beyond the scope of this study, we remark that the analysis of the resulting normal form could reveal the precise boundary delimiting the regions of existence of traveling waves and stationary patterns.

For the stationary pattern in Fig.~\ref{fig:pattern_n1}, we notice a small GEF dissociation rate ($\koff = 0.05$), which causes the proteins to mostly accumulate on the membrane. Also, in contrast with coupled bulk-membrane reaction-diffusion systems without mass conservation (cf.~\cite{levine2005membrane,gomez2019,paquin2018}), we note that high (low) concentrations on the membrane match with low (high) concentrations in the bulk.

\begin{figure}[H]
\centering
\begin{subfigure}{0.31\linewidth}
\includegraphics[width=\linewidth]{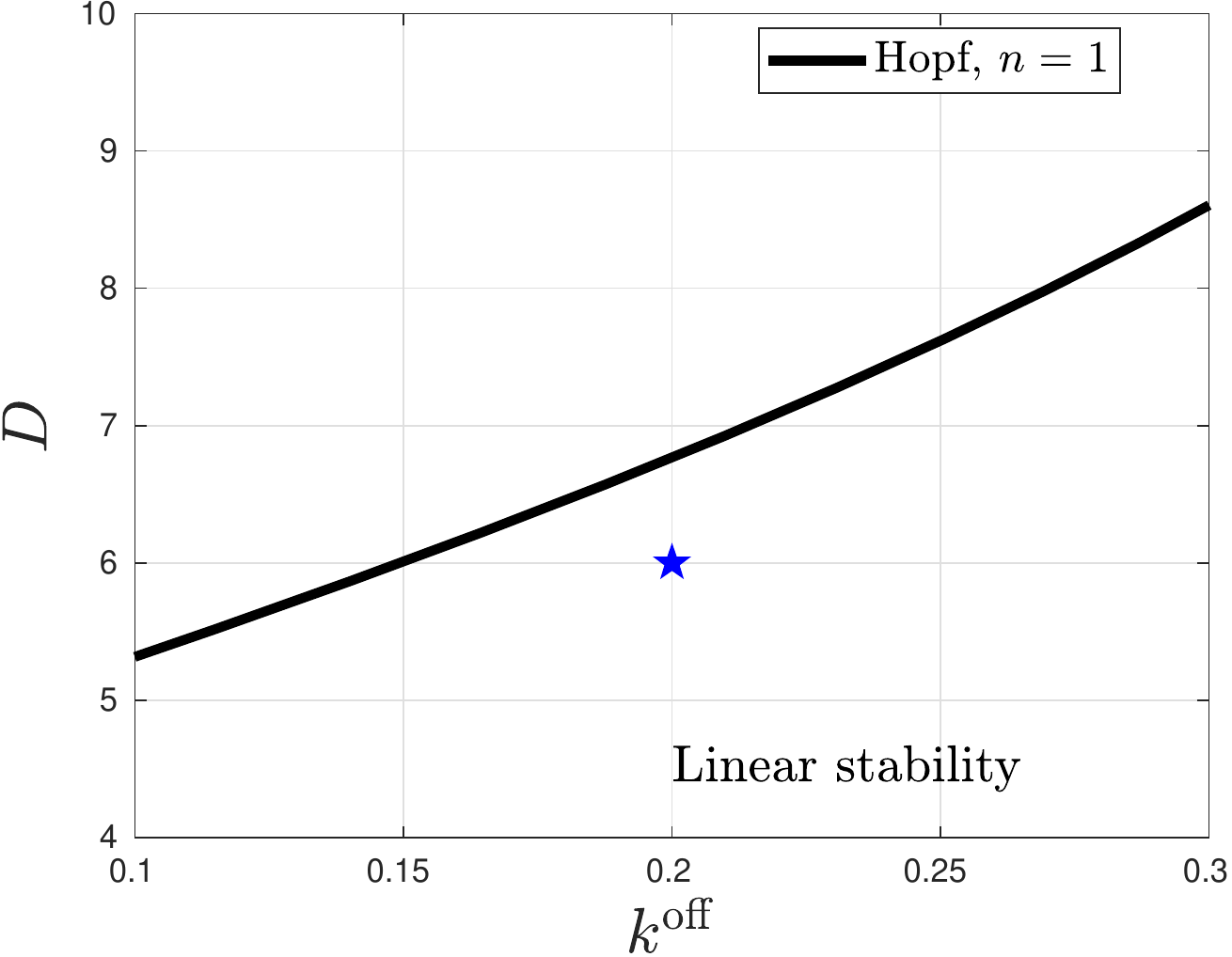}
\caption{$\koff = 0.2$, $D = 6$}
\end{subfigure}
\begin{subfigure}{0.31\linewidth}
\includegraphics[width=\linewidth]{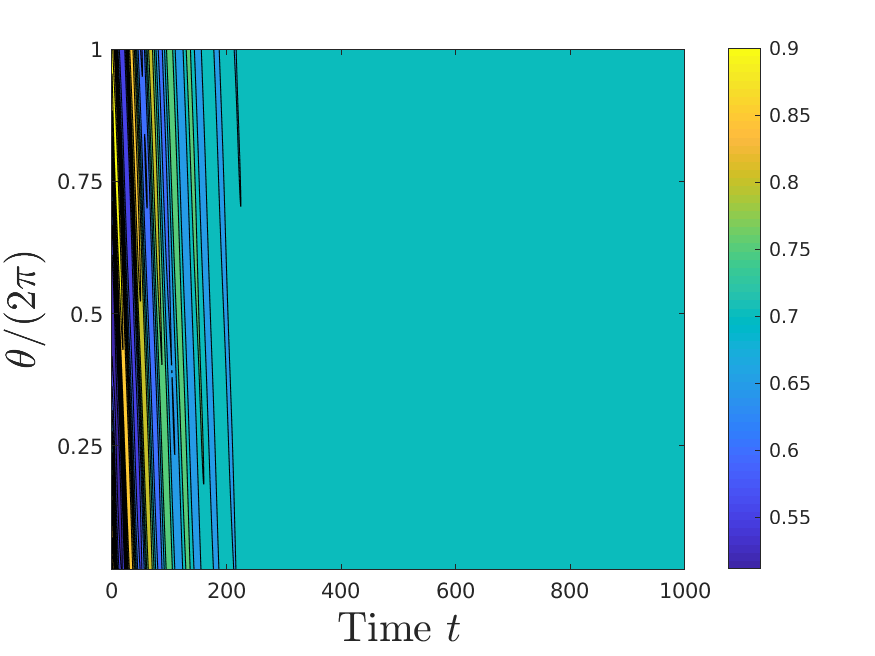}
\caption{Trivial steady state.}
\end{subfigure}
\begin{subfigure}{0.31\linewidth}
\includegraphics[width=\linewidth]{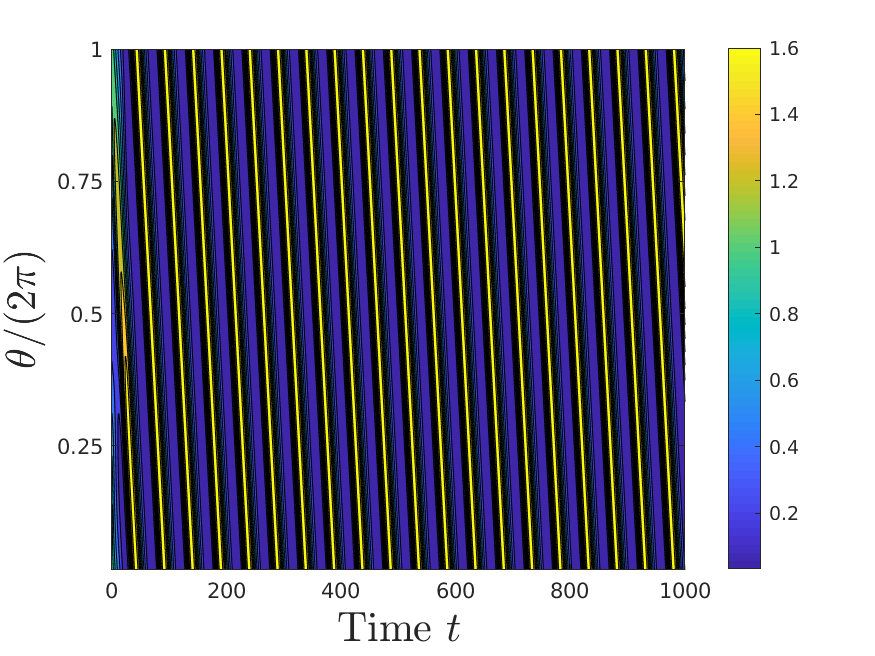}
\caption{Traveling waves.}
\end{subfigure}
\caption{\label{fig:bistability} Bistability between the trivial steady state and traveling wave solutions. The blue star just below the stability boundary in panel (a) indicates where the simulation is performed. In panel (b), we observe a small perturbation of the trivial steady state that slowly vanishes. However, as shown in panel (c), applying a bigger perturbation leads to a traveling wave. This suggests a loss of stability through a subcritical Hopf bifurcation with respect to the $n=1$ spatial mode. Other parameters are the same as in the caption of Fig.~\ref{fig:wave}.}
\end{figure}

Our numerical simulations also revealed the possible subcritical branching of the Hopf bifurcation associated with the $n=1$ spatial mode. In Fig.~\ref{fig:bistability}, we show the result of two simulations performed near the Hopf stability boundary, where the trivial steady state is expected to be linearly stable, and observe that a traveling wave can form given a large enough perturbation of the trivial steady state. We highlight that such a bistable behavior was not observed in Section \ref{sec:nonlocal}, when considering the nonlocal PDE system governing the dynamics in the well-mixed regime. This would be consistent with our findings in the one-dimensional PDE--ODE model, that the Hopf bifurcation is subcritical for low level of bulk diffusion, while it is supercritical in the well-mixed regime. The precise determination, through a normal form computation, of the diffusion level at which the criticality transition occurs, is left for further study.

\begin{figure}[H]
\centering
\begin{subfigure}{0.31\linewidth}
\includegraphics[width=\linewidth]{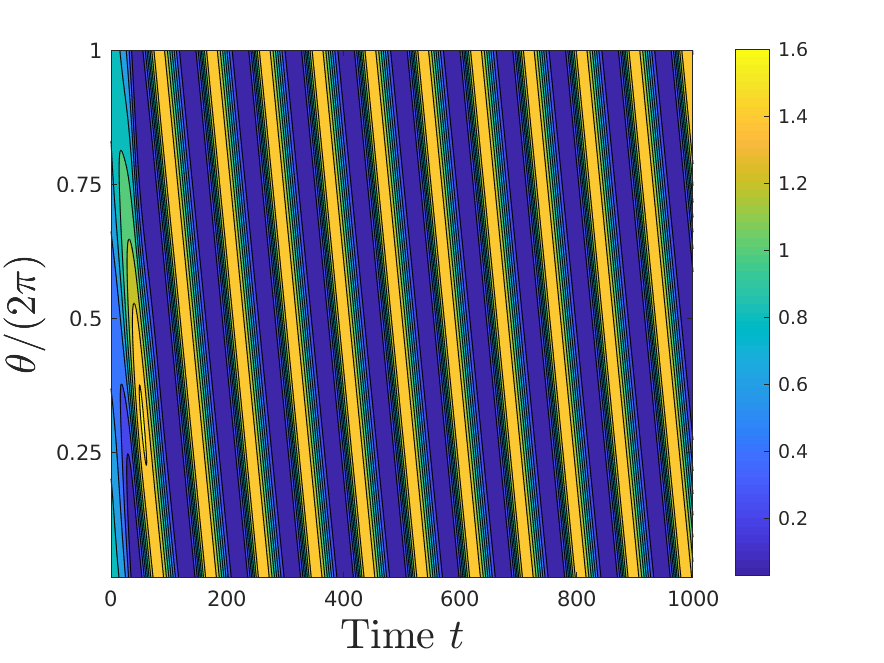}
\caption{Clockwise rotating waves.}
\end{subfigure}
\begin{subfigure}{0.31\linewidth}
\includegraphics[width=\linewidth]{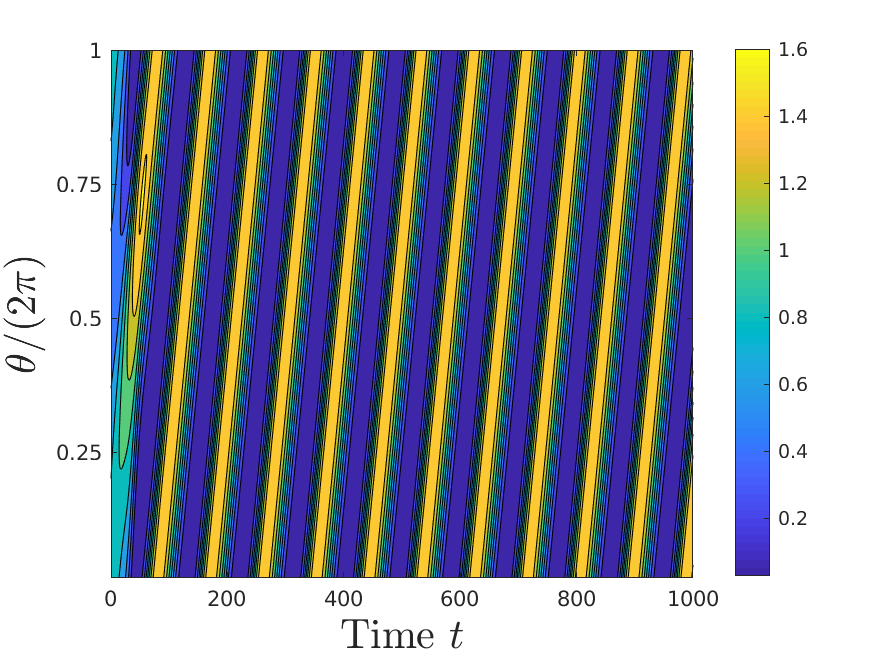} 
\caption{Anticlockwise rotating waves.}
\end{subfigure}
\begin{subfigure}{0.31\linewidth}
\includegraphics[width=\linewidth]{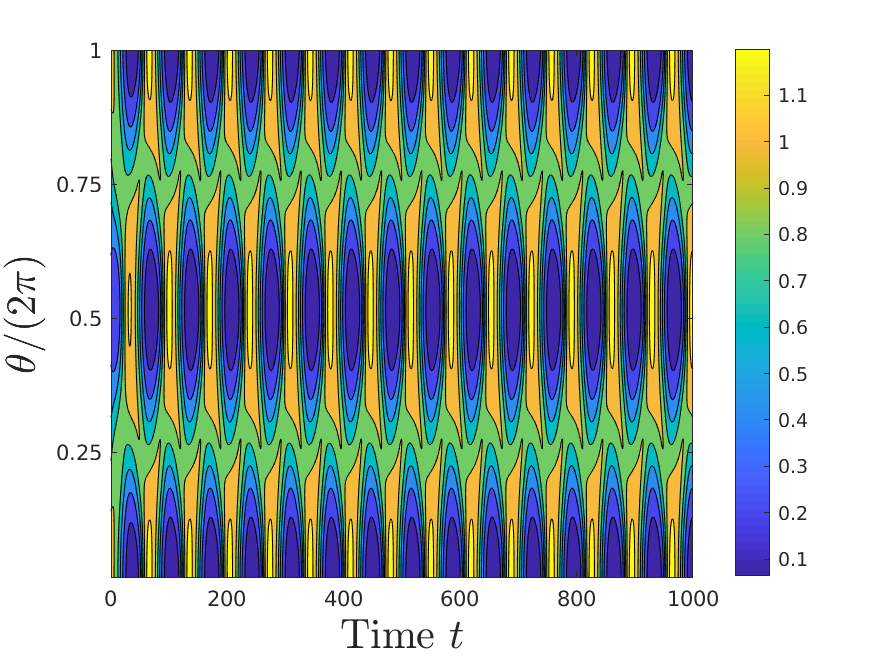} 
\caption{Standing waves.}
\end{subfigure}
\caption{\label{fig:p2p} Coexistence of rotating and standing waves when $\koff = 0.1$ and $D=6$ (other parameters are the same as in the caption of Fig.~\ref{fig:wave}). In the vicinity of a Hopf stability boundary, different initial conditions can either lead to clockwise rotating waves (panel (a)), anticlockwise rotating waves (panel (b)) or standing waves (pole-to-pole oscillations, see panel (c)). The traveling waves in each panel are shown in Supplemental Movies S7, S8 and S9.}
\end{figure}

\begin{figure}[H]
\centering
\begin{subfigure}{0.36\linewidth}
\includegraphics[width=\linewidth]{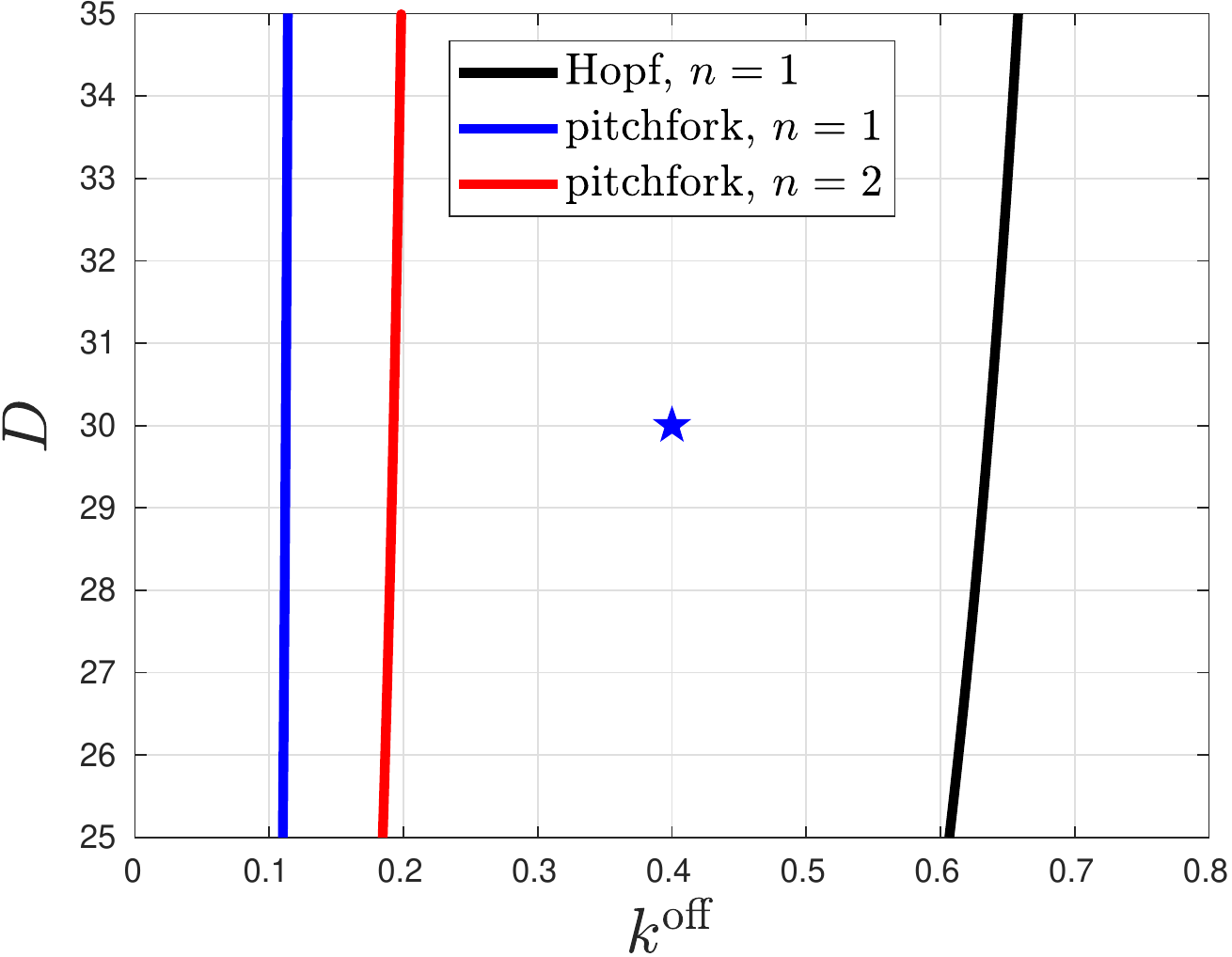}
\end{subfigure}
\begin{subfigure}{0.42\linewidth}
\includegraphics[width=\linewidth]{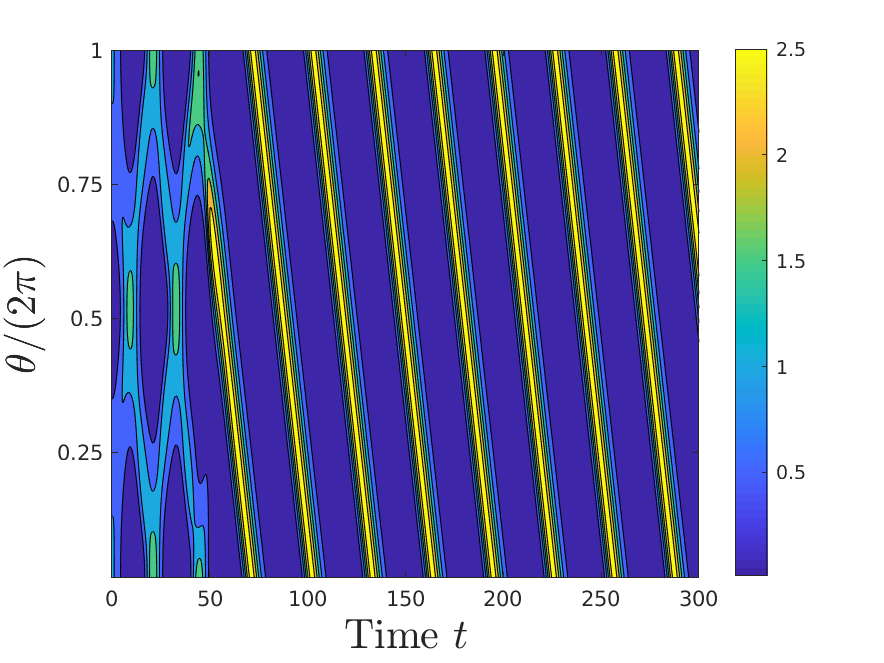}
\end{subfigure}
\caption{\label{fig:p2p_to_wave} Transition from standing to rotating waves in the highly nonlinear regime, far from any stability boundaries. The blue star in the left panel indicates where the simulation is performed (other parameters are the same as in the caption of Fig.~\ref{fig:wave}).}
\end{figure}

We now address the existence of standing waves (or pole-to-pole oscillations) in the full coupled membrane-bulk reaction-diffusion model, and recall that such oscillatory behavior is essential during the process of cell division. Numerical simulations from Fig.~\ref{fig:p2p} suggest the coexistence of pole-to-pole oscillations and traveling waves near the Hopf stability boundary. The parameter values employed here are the same as for panel (c) of Fig.~\ref{fig:wave}, although we remark that similar simulation results were obtained for different sets of parameters near the stability threshold. As shown in Fig.~\ref{fig:p2p_to_wave}, clear transitions between standing and traveling waves are observed away from the Hopf stability boundary. Overall, those findings are consistent with the dynamics in the limit of fast bulk diffusion (see Section~\ref{sec:nonlocal}).

\begin{figure}[H]
\centering
\begin{subfigure}{0.31\linewidth}
\includegraphics[width=\linewidth]{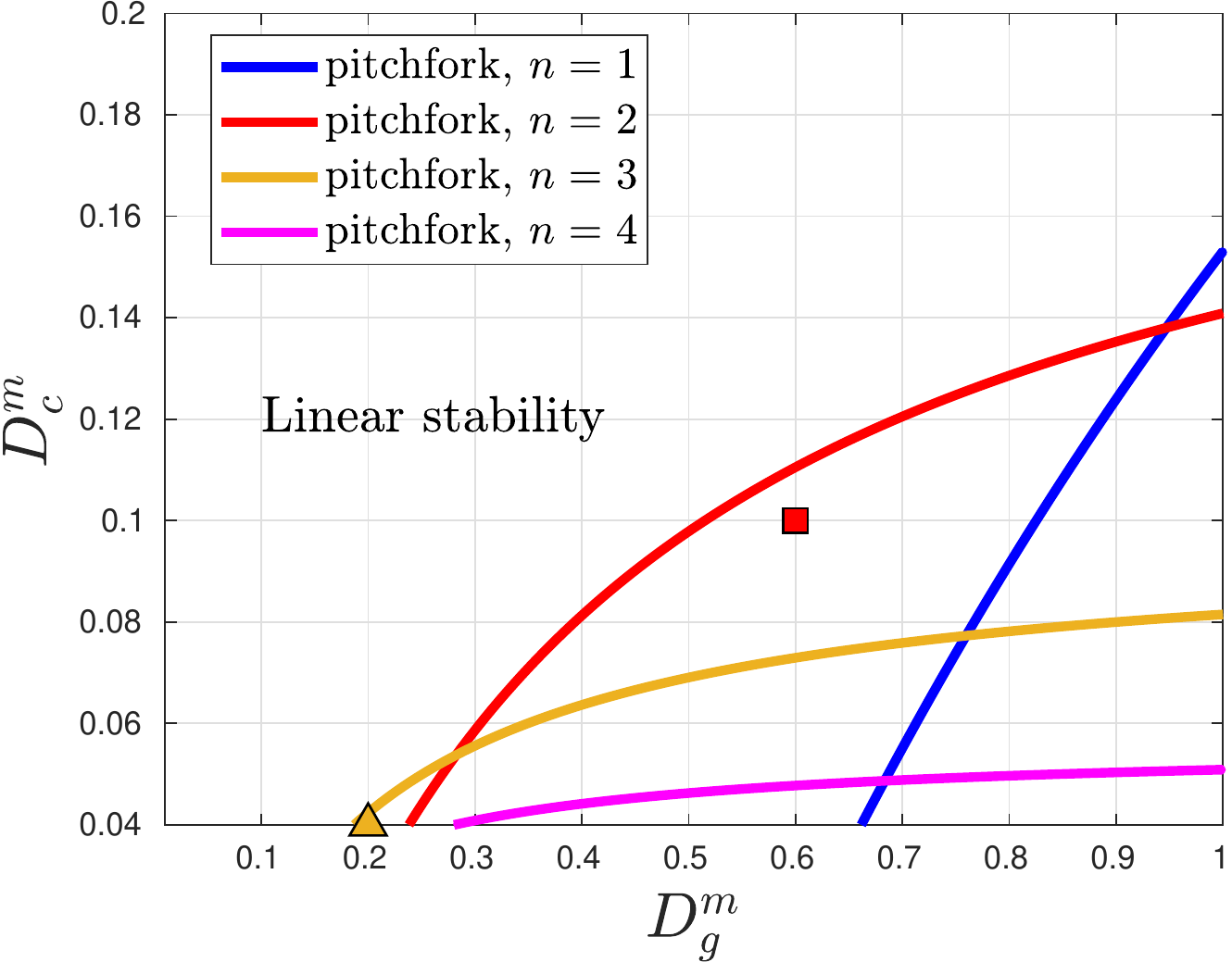}
\caption{$(D_g^m,D_c^m)$ with $\koff = 1,\,D=10$.}
\end{subfigure}
\begin{subfigure}{0.31\linewidth}
\includegraphics[width=\linewidth]{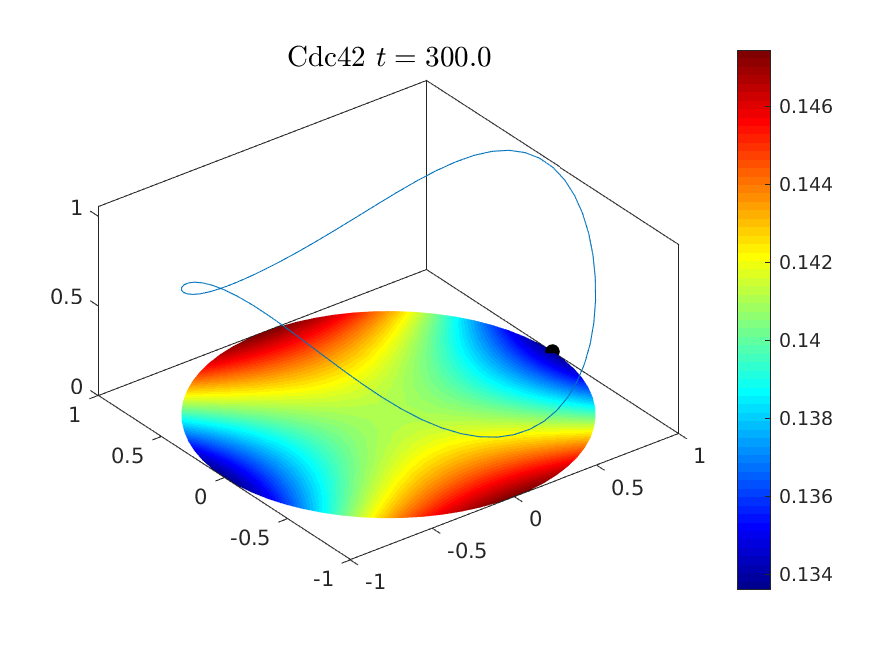}
\caption{Stationary pattern with two peaks.}
\end{subfigure}
\begin{subfigure}{0.31\linewidth}
\includegraphics[width=\linewidth]{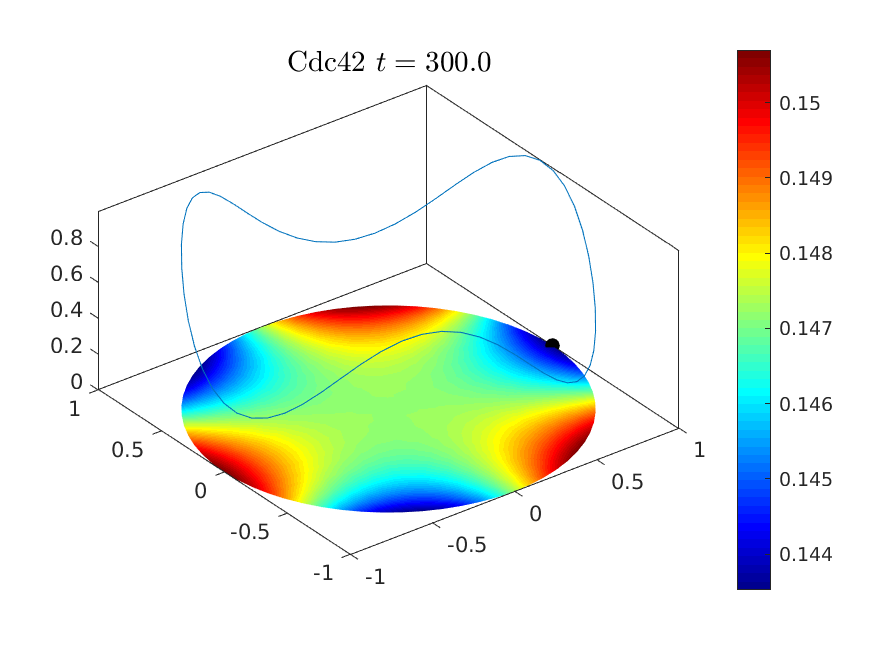}
\caption{Stationary pattern with three peaks.}
\end{subfigure}
\caption{\label{fig:multipeak} Panel (a): pitchfork stability boundaries in the $D_c^m$ versus $D_g^m$ parameter plane with $\koff = 1$, $D=10$ and other parameters taken from Table~\ref{table:parm}. Panel (b): the stationary pattern is computed near a mode $n=2$ stability boundary, as indicated by the square in panel (a) (see Supplemental Movie S10). Panel (c): the stationary pattern is computed near a mode $n=3$ stability boundary, as indicated by the triangle in panel (a) (see Supplemental Movie S11).}
\end{figure}

Next, we investigate the formation of patterns with multiple protein accumulation sites, which happens when higher spatial modes become unstable. For the purpose of avoiding any oscillatory instabilities, we set $\koff = 1,\, D=10$, and select the membrane diffusion coefficients as bifurcation parameters. Shown in panel (a) of Fig.~\ref{fig:multipeak} are four stability curves in the $D_c^m$ versus $D_g^m$ parameter plane. Each of these curves is associated with a distinct mode and computed directly from the zero-crossing eigenvalue relation \eqref{eq:charlambda0}. The resulting diagram is qualitatively similar to its counterpart in the limit of fast bulk diffusion (see panel (a) of Fig.~\ref{fig:stb_nonlocal}). Here also, we note the stabilizing effect of simultaneously decreasing the bulk diffusion coefficients. Finally, in panels (b-c) of Fig.~\ref{fig:multipeak} we show Cdc42 stationary distributions in the bulk and on the membrane when multiple accumulation sites form.

\begin{figure}[H]
\centering
\begin{subfigure}{0.31\linewidth}
\includegraphics[width=\linewidth]{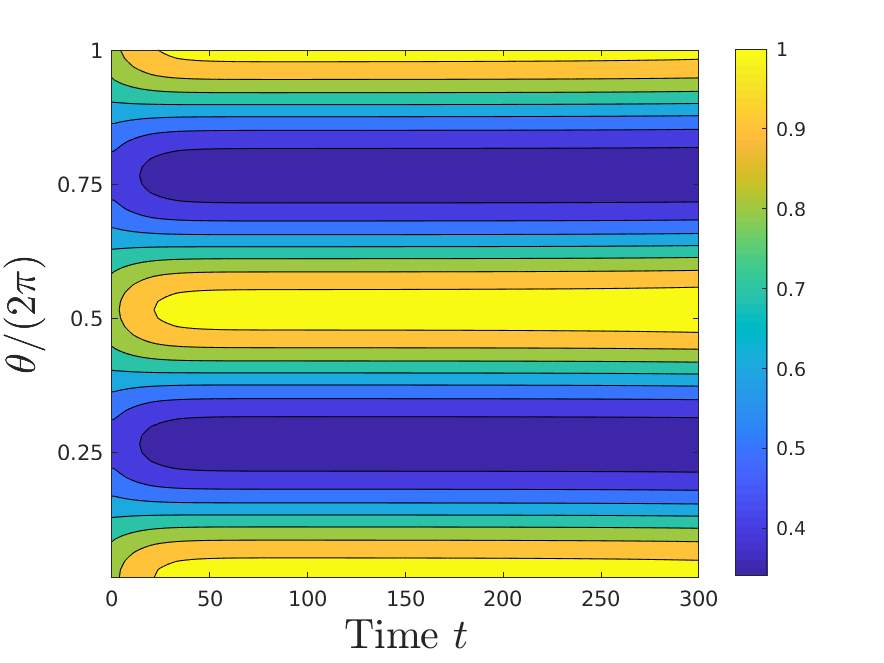} 
\caption{$D_g^m = 0.6,\,D_c^m = 0.1$}
\end{subfigure}
\begin{subfigure}{0.31\linewidth}
\includegraphics[width=\linewidth]{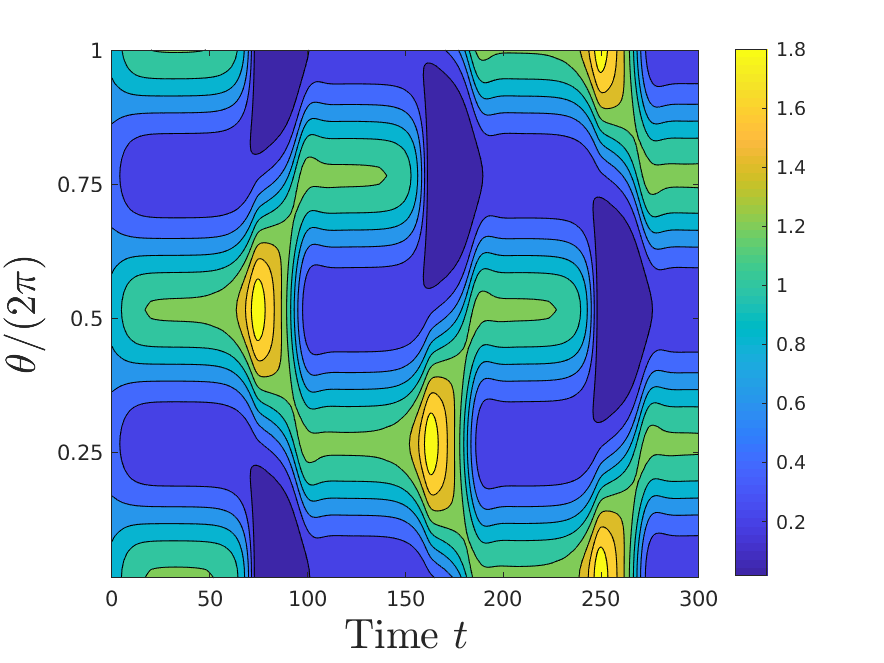} 
\caption{$D_g^m = 0.8,\,D_c^m = 0.1$}
\end{subfigure}
\begin{subfigure}{0.31\linewidth}
\includegraphics[width=\linewidth]{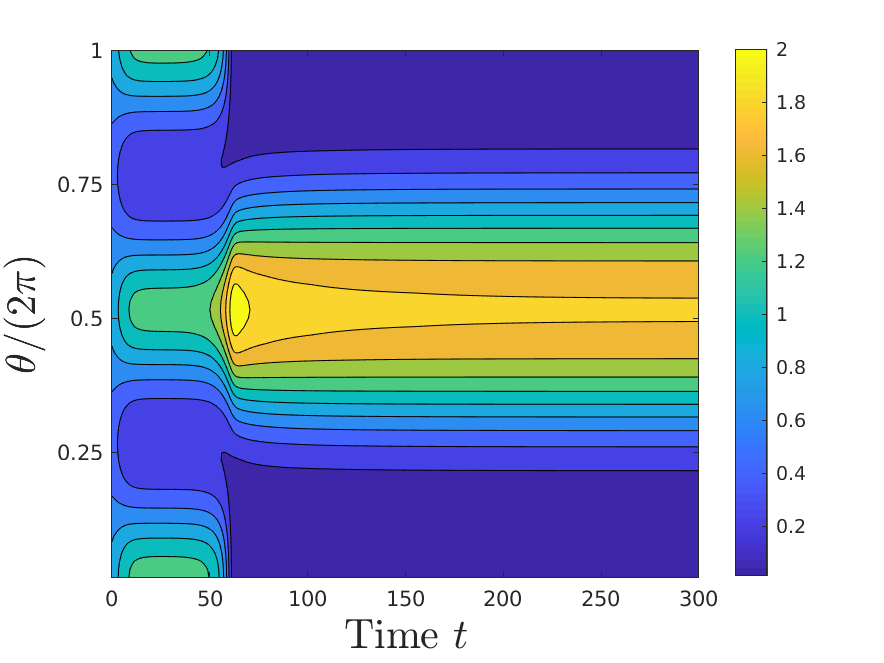}
\caption{$D_g^m = 1,\,D_c^m = 0.1$}
\end{subfigure}
\begin{subfigure}{0.31\linewidth}
\includegraphics[width=\linewidth]{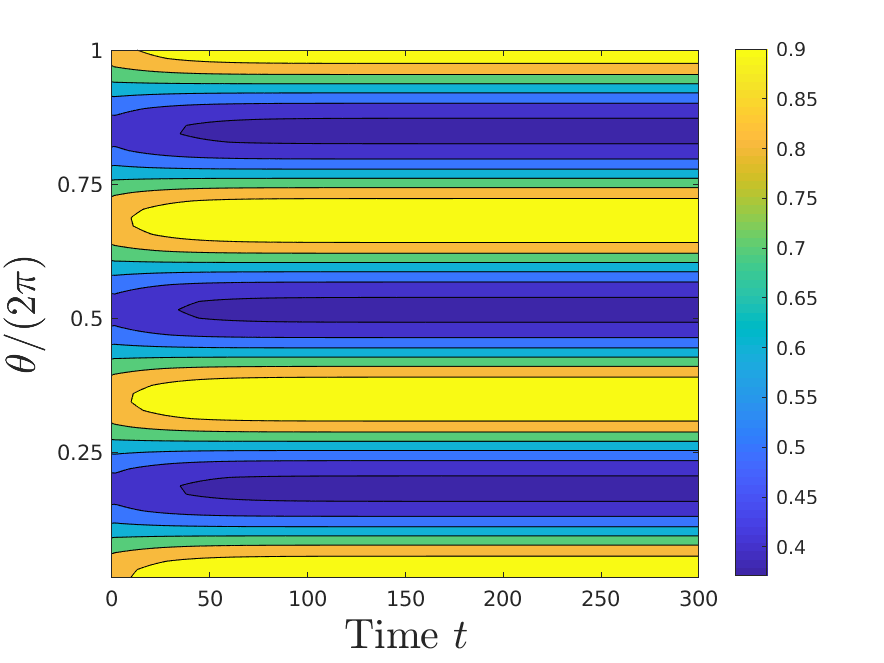} 
\caption{$D_g^m = 0.2,\,D_c^m = 0.04$}
\end{subfigure}
\begin{subfigure}{0.31\linewidth}
\includegraphics[width=\linewidth]{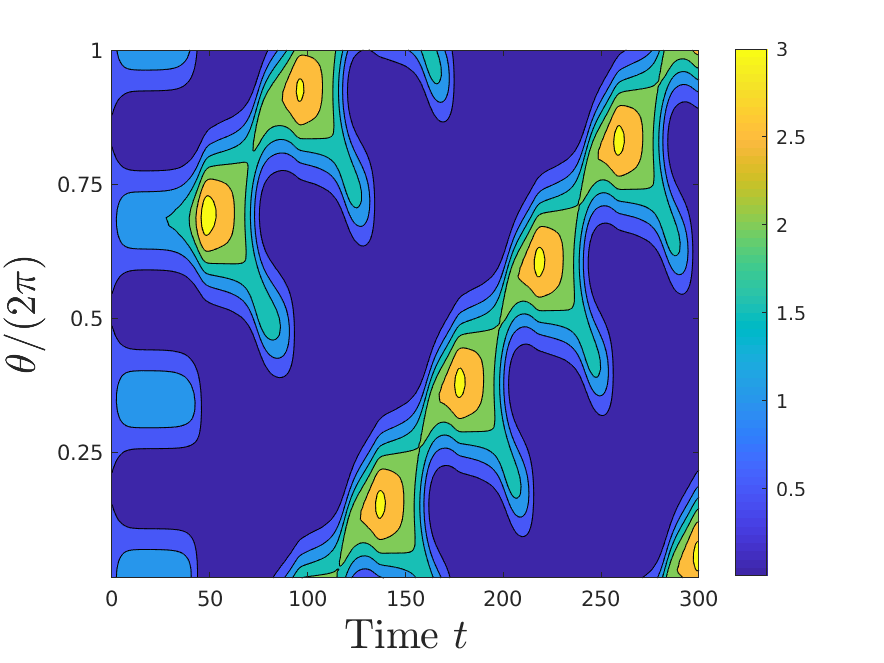}
\caption{$D_g^m = 0.4,\,D_c^m = 0.04$}
\end{subfigure}
\begin{subfigure}{0.31\linewidth}
\includegraphics[width=\linewidth]{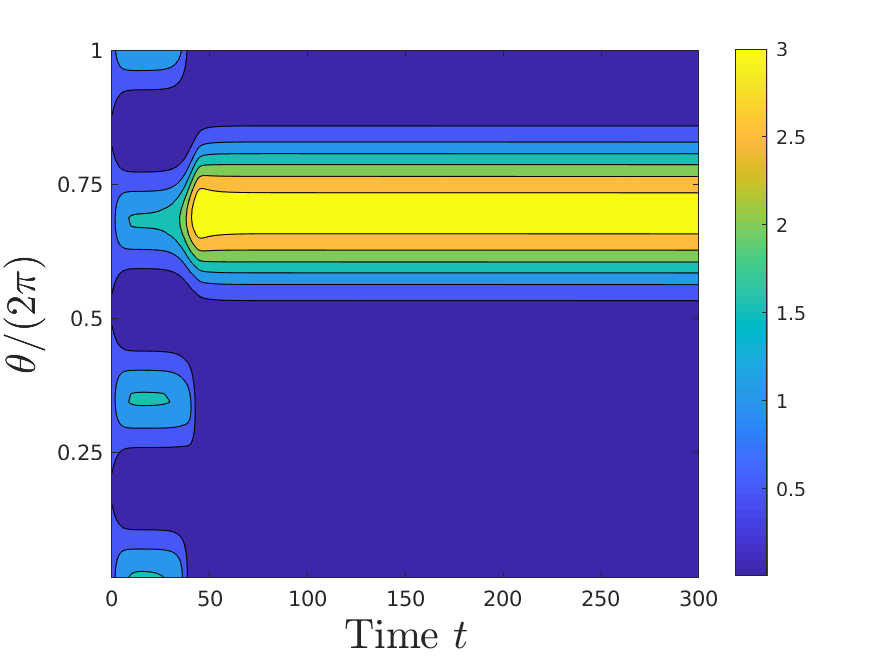}
\caption{$D_g^m = 0.6,\,D_c^m = 0.04$}
\end{subfigure}
\caption{\label{fig:pattern_n23} Effect of reducing the membrane diffusivity ratio $D_c^m/D_g^m$ on pattern formation. In panels (a-c), we observe a transition from a pattern with two peaks to a localized pattern with a single peak. The same transition is observed in panel (d-f) for an initial weakly nonlinear pattern with three peaks. Other parameters are the same as in the caption of Fig.~\ref{fig:multipeak}.}
\end{figure}

We conclude this section with simulations showing how the number of peaks of a stationary pattern can be reduced as the GEF membrane diffusion coefficient increases and as the mode $n=1$ pitchfork stability boundary is approached. More precisely, in panels (a-c) of Fig.~\ref{fig:pattern_n23}, we consider the horizontal slice $D_c^m=0.1$ and observe a transition from a two-peak to a single peak pattern. In panels (d-f) a transition between a three-peak and a single peak pattern is observed on the slice $D_c^m = 0.04$. We remark that the intermediate patterned states involve circular motions and peak switching that occur in the absence of any Hopf bifurcations. We believe this phenomena to result from nonlinear interactions between two or more spatial modes. This was also observed in the nonlocal version of this model, although here the relatively low level of bulk diffusion allows for smoother switching. Our model suggests that the number of polarized sites is dependent on membrane diffusion coefficients. As both $D_c^m$ and $D_g^m$ decrease, more polarity sites can be observed. Our findings for all three model variants are summarized in Table~\ref{table:compare}.

\section{Conclusions and discussion}

\begin{table}[h!]
\caption{Comparison of three different model geometries}
\centering
\begin{tabular}{p{3.5cm}| p{4.3cm}| p{4.3cm}| p{4.3cm}}
\hline\hline
 & \footnotesize{\textbf{1-D PDE--ODE model}} & \footnotesize{\textbf{1-D nonlocal PDE model (circle)}} & \footnotesize{\textbf{2-D membrane-bulk model (disk)}} \\
\hline
\footnotesize{\textbf{Assumptions}} & \footnotesize{finite diffusion in the bulk, well-mixed in each end compartment} & \footnotesize{well-mixed in the bulk, finite diffusion on the membrane} & \footnotesize{finite diffusion in the bulk, finite diffusion on the membrane}\\
\hline
\footnotesize{\textbf{Patterns}} & \footnotesize{symmetric or asymmetric steady states, anti-phase (pole-to-pole) oscillations} & \footnotesize{stationary patterns with one or multiple peaks, traveling or standing waves} & \footnotesize{similar as for the nonlocal model} \\
\hline
\footnotesize{\textbf{Bifurcations}} & \footnotesize{ pitchfork,  subcritical or supercritical Hopf} & \footnotesize{stationary Turing (pitchfork),  oscillatory Turing (Hopf)}  & \footnotesize{similar as for the nonlocal model} \\
\hline
\footnotesize{\textbf{Slower bulk diffusion $(D\equiv D_c = D_g)$}}& \footnotesize{(1) stability region of symmetric steady state increases, (2) supercritical Hopf bifurcation becomes subcritical} & \footnotesize{NA} & \footnotesize{(1) stability region of homogeneous steady state increases, (2) traveling waves slow down near a Bogdanov--Takens bifurcation} \\
\hline
\footnotesize{\textbf{Slower Cdc42 membrane diffusion}} & \footnotesize{NA} & \footnotesize{(1) homogeneous steady state loses stability, (2) stationary patterns or traveling waves emerge} & \footnotesize{similar as for the  nonlocal model}  \\
\hline
\footnotesize{\textbf{Slower GEF membrane diffusion}} & \footnotesize{NA} & \footnotesize{(1) stationary patterns are less localized, (2) formation of traveling waves even without Hopf bifurcations} & \footnotesize{similar as for the nonlocal model, but stationary patterns with multiple peaks can form} \\
\hline
\footnotesize{\textbf{Low  GEF dissociation rate $\koff \approx 0.1$}} & \footnotesize{anti-phase relaxation oscillations} & \footnotesize{(1) stationary patterns or traveling waves can form, (2) that interact near Bogdanov--Takens bifurcations} & \footnotesize{similar as for the nonlocal model}  \\
\hline
\footnotesize{\textbf{High  GEF dissociation rate $\koff = 1$}} & \footnotesize{no oscillations} & \footnotesize{(1) no Hopf bifurcations, (2) stationary patterns with one or several peaks, that can evolve into traveling waves} & \footnotesize{similar as for the nonlocal model, but stationary patterns are less likely to evolve into traveling waves}  \\

\end{tabular}
\label{table:compare}
\end{table}

\subsection{Conclusions}

In conclusion, we have analyzed three spatial model variants of a cell polarization model in order to explore the effect of spatial dimensionality on pattern formation. We focused on a membrane-bulk model for the Cdc42-GEF system in fission yeast, where the same reaction kinetics were considered for three, consecutively more complicated, geometries: a one-dimensional PDE--ODE model, a nonlocal one-dimensional PDE model obtained when the diffusion coefficients of the cytosolic forms $D \to\infty$ (which is often a limit considered in models for polarization \cite{goryachev2008, lo2014}), and a full two-dimensional membrane-bulk model with finite diffusion in the cytosol and on the membrane. For these models, we analyze how bulk diffusion and lateral diffusion along the membrane affect the linear stability of a homogeneous steady state. Interestingly, fast membrane diffusion of Cdc42 stabilizes the homogeneous steady state, while fast bulk diffusion of Cdc42 has a destabilizing effect. Decreasing the GEF dissociation constant $\koff$ shrinks the regions of linear stability. 

For the 1-D PDE--ODE model, we observe both symmetric and asymmetric distributions of Cdc42 at the two ends, as well as pole-to-pole oscillations. Hopf bifurcations occur when GEF dissociation is slow compared to Cdc42 dissociation, with bulk diffusion affecting the amplitude and the shape of the oscillations near the stability thresholds. When bulk diffusion is slow, relaxation oscillations emerge from a subcritical Hopf bifurcation, while weakly nonlinear oscillations emerge from a supercritical Hopf bifurcations in faster bulk diffusion regime. For the 1-D nonlocal PDE model and the 2-D membrane-bulk model, stationary or oscillatory Turing instabilities, either corresponding to pitchfork or Hopf bifurcations, can occur when membrane-bound Cdc42 diffuses slowly. All the observed patterns have a symmetry-breaking effect, either consisting of stationary patterns or traveling waves. Stationary patterns with multiple peaks are observed when the diffusion rates of both membrane-bound species decrease. These stationary patterns are sensitive to perturbations and can undergo sudden rotations or evolve into traveling waves even in the absence of Hopf bifurcations. The main results are summarized in Table ~\ref{table:compare}.

Note that there are two different limits that can result in a one-dimensional model. If the reactive compartments at the two cell ends are assumed to be well-mixed, while finite diffusion is assumed in the bulk, we obtain a one-dimensional PDE--ODE model. If we assume that the bulk domain is well-mixed and consists of a disk, with lateral diffusion allowed, we obtain a one-dimensional nonlocal PDE model on a circle.  No temporal oscillations with a uniform spatial profile are observed in the nonlocal model. Instead, the dominant patterns are traveling waves and stationary patterns. In comparison to these 1-D model reductions, the 2-D model exhibits similar spatio-temporal patterns as the 1-D nonlocal PDE model. On a 2-D circular bulk domain, there are an infinite number of spatial modes associated to the linearized system, all corresponding to the circular harmonics, and this therefore increases the likelihood of multiple interacting unstable modes. The Bogdanov--Takens bifurcation with $O(2)$ symmetry is one notable example of this richer bifurcation structure that is absent from the 1-D PDE--ODE model. One last difference consists in the absence of diffusion-driven instability in the 1-D PDE--ODE model, partly due because of the well-mixed assumption of the two end compartments.
 
\subsection{Discussion}
The role that spatial segregation between components plays in giving rise to spatio-temporal patterns in cell biology is beginning to be appreciated. By taking into account spatial segregation between nucleus and cytoplasm, oscillations can emerge in the Hes1 and p53 pathways \cite{sturrock2011}. A rigorous analysis showing that diffusion can give rise to oscillations (i.e., treating the diffusion coefficient as a bifurcation parameter) was presented in \cite{chaplain2015hopf}. Pole-to-pole oscillations of Min proteins in \textit{E. coli} have typically been studied using standard reaction-diffusion equations (reviewed in \cite{kruse2007,loose2011}). However, bulk-surface models for Min oscillations have also been proposed \cite{huang2003,halatek2012} and are more realistic in terms of protein localization. An often overlooked problem in mathematical modeling of signaling networks is the effect of cellular geometry. Initial models of spatio-temporal phenomena are often studied in an idealized 1-D setting \cite{jilkine2011comparison}. If a certain model gains acceptance, then its extension to 2-D or 3-D domains may be done, often with some corrections of parameter values or reaction terms. However, a systematic analysis of the effect of dimensionality on spatio-temporal behavior of models of cell polarization as dimensionality increases from 1-D to 2-D is not usually performed. (One notable exception is the 1-D reaction-diffusion \cite{mori2011} and 2-D membrane-bulk versions \cite{cusseddu2018} of the wave-pinning model of cell polarization.)
 
Pattern formation resulting from coupling a diffusion process in a bounded 2-D domain to a nonlinear reaction-diffusion process on the domain boundary has been studied in \cite{levine2005membrane,ratz2012turing,ratz2015turing,paquin2018}. It was first shown in \cite{levine2005membrane} that the formation of membrane-bound stationary patterns is possible even when the two species diffuse in the bulk at the same rate (membrane diffusion was neglected). R{\"a}tz and R{\"o}gers \cite{ratz2012turing,ratz2014} analyzed a coupled membrane-bulk model and its nonlocal reduction for small GTPase. Their model includes a diffusion equation for inactive Cdc42 in the bulk and a pair of reaction-diffusion equations for active and inactive Cdc42 on the membrane. From a linear stability analysis, they showed that differences between bulk and lateral diffusion rates can trigger a Turing instability. Fast bulk diffusion and slow membrane diffusion of Cdc42 can also lead to pattern formation in our model, even in the case of equal membrane diffusivities and equal bulk diffusivities. The major difference lies in the modeling assumptions about the underlying role of GEF in the activation of Cdc42 proteins. In \cite{ratz2012turing,ratz2014}, GEF is assumed to be at quasi-steady state and is expressed in terms of the concentration of active Cdc42, whereas our model includes a separate variable for GEF. Furthermore, in \cite{ratz2012turing,ratz2014} it is assumed that GEF only activates Cdc42, while our model has both  positive and negative feedback loops that involve GEF. Thus, Hopf bifurcations and oscillatory patterns can occur in our model, but are not seen in the model of R{\"a}tz and R{\"o}gers \cite{ratz2012turing,ratz2014}. Our results can also be compared with patterning in regular reaction-diffusion systems, with neither membrane-bulk coupling nor mass conservation. For a disk domain with no-flux boundary conditions, stationary patterns are expected to form over the full domain, with their building blocks in the linear regime corresponding to the eigenfunctions of the Laplacian on the disk. In our model, because diffusion was the only process assumed in the bulk, we merely observed patterning in some layer near the boundary of the bulk. Another scenario is to consider a 1-D reaction-diffusion with periodic boundary conditions, with no nonlocal coupling. There, one can verify if stationary patterns with multiple peaks exist in a wider parameter regime, and if they are more stable than in their mass-conserved membrane-bulk counterpart.

Finally, we discuss some limitations and possible extensions of our work. From a biological perspective, our model predicts that spatio-temporal patterning is more likely to emerge when the GEF dissociation rate $\koff$ is small compared to the Cdc42 dissociation rate. In the one-dimensional PDE--ODE model, Cdc42 and GEF oscillate out of phase, while the spatial model with diffusion of membrane-bound forms predicts that Cdc42 and GEF spatially segregate along the membrane. Our model also predicts that the number of polarized sites should increase as the membrane diffusion coefficients of both Cdc42 and GEF proteins decrease. While our model only involves one Cdc42 GEF, fission yeast cells have multiple GEFs. Experimental studies in fission yeast suggest that the GEF Scd1 oscillates from one pole to the other during cell growth, while no obvious oscillations of another GEF, Gef1, have been reported \cite{das2012,das2015}. Another study suggests that Gef1 is distributed in the cytosol \cite{tay2018}. Recent work suggests that positive feedback involves  Scd1, while negative feedback is  achieved through Cdc42 inhibiting Gef1 \cite{hercyk2019}. To take these findings into account, our model with a single GEF regulator would need to be replaced by a more complicated model with two GEFs, which would be harder to analyze. To limit active GEF localization to the cell tips in a full two-dimensional model, a space-dependent binding rate $\kon$ should be assumed. It may also be necessary to include GAP dependent feedback to fully model the fission yeast system \cite{goryachev2017,tay2018}. While the 2-D model geometry may not be applicable to rod-shaped fission yeast, where Cdc42 seems to preferentially bind to the poles instead of all over the membrane, it is applicable to spherical budding yeast, where positive and negative feedbacks involving Cdc42 result in stable polarization and oscillatory behaviors. Note that traveling waves of Cdc42 have been observed in budding yeast under some experimental conditions \cite{ozbudak2005}. 

For the convenience of mathematical analysis, we studied the 2-D model on a circular bulk domain, though fission yeast cells are rod-shaped. One follow-up question is to consider the effect of domain geometry and growth on the full 2-D membrane-bulk model. Another question to consider is how the patterning will change if model geometry is assumed to be three-dimensional. In a recent pattern formation study on spherical domains \cite{diegmiller2018}, Diegmiller et al.\ reduced a 3-D membrane-bulk model to a nonlocal reaction-diffusion equation on a 2-D surface by assuming that bulk diffusion is fast. Their model considers autocatalysis and predicts that a single polarity site would form. Following their framework, we can extend our 1-D nonlocal model to a 2-D surface and explore how the negative feedback may affect the number of polarity sites. Another potential extension is to take into account anomalous diffusion in the cytoplasm \cite{jeon2011} which raises challenges due to the Robin boundary conditions. Previous work \cite{henry2002} used fractional diffusion for modeling anomalous diffusion assuming an infinite domain. In a bounded domain, appropriate formulation of boundary conditions is an active area of research \cite{baeumer2018}.

From a dynamical systems perspective, a weakly nonlinear analysis could be performed to determine the criticality of the bifurcations in each of the three model variants, and to investigate how stationary patterns and traveling waves interact near Bogdanov--Takens bifurcation points. Our numerical results also revealed the formation of membrane-bound localized stationary patterns in the limit of small membrane diffusion coefficients ratio, similar to those studied in \cite{gomez2019}. There, another type of linear stability analysis can be performed, which consists of linearizing the membrane-bulk PDE system around a localized pattern constructed using matched asymptotic expansion, as opposed to linearizing around a uniform steady state. Finally, the analytical framework presented here can be used to study the spatio-temporal dynamics of signaling proteins in other cell biology contexts. 

\section*{Acknowledgments}
FPL was supported by the NSERC Doctoral Award $\#6564$. BX is supported by the Robert and Sara Lumpkins Endowment for Postdoctoral Fellows in Applied and Computational Math and Statistics at the University of Notre Dame. This work was supported by the NSF (AEL by DMS-1815216, AJ by DMS-1615800 and KLD by DGE-1313583). We would also like to thank Michael Ward and Wayne Nagata for helpful discussions.

\appendix
\section{Supplementary movies}
Supplemental movies are provided for the following model geometries:

\noindent 1) Coupled PDE--ODE model:
\begin{itemize}
 \item Supplemental movie S1 shows the spatio-temporal oscillations from Fig.~\ref{fig:pole_to_pole}.
\end{itemize}
2) Reduced nonlocal PDE model:
\begin{itemize}
 \item Supplemental movie S2 shows the formation of a stationary pattern and refers to panel (a) of Fig.~\ref{fig:WM_pattern}.
 \item Supplemental movies S3, S4 and S5 show traveling waves and refer to panels (a-b-c) of Fig.~\ref{fig:WM_wave}.
\end{itemize}
3) Full 2-D coupled membrane-bulk reaction-diffusion model:
\begin{itemize}
 \item Supplemental movie S6 shows the formation of a stationary pattern in Fig.~\ref{fig:pattern_n1}.
 \item Supplemental movies S7, S8 and S9 show the coexistence of clockwise rotating waves, anti-clockwise rotating waves and standing waves, and refer to panels (a-b-c) of Fig.~\ref{fig:p2p}.
 \item Supplemental movies S10 and S11 show the formation of stationary patterns with multiple peaks and refer to panels (b-c) of Fig.~\ref{fig:multipeak}.
\end{itemize}

\section{Parameter values}\label{sec:appendix_params}
For the reaction kinetics, we use parameter values from our previously published ODE model \cite{xu2018}. The amount of Cdc42 and the GEF, $C_{\rm tot}$ and $G_{\rm tot}$ respectively, are in arbitrary units (analogous to fluorescence), units of time are taken to be minutes and the characteristic unit of length $L=1$ is taken to be 9 $\mu m$ (comparable to the initial length of a fission yeast cell when bipolar polarity emerges). The timescale is set by $k^-=4\,\text{min}^{-1}$ \cite{das2012}. It follows that after non-dimensionalization $(x\to x/L, t\to tk^-)$ a dimensionless diffusion coefficient of $D_c = 1$ corresponds to $D_c = 81 \, \mu \text{m}^2/15\text{ sec} \approx 5.4 \, \mu m^2/\text{sec}$ in physical units. We assume the diffusion coefficient of the cytosolic form of the GEF is of the same order of magnitude. Differential mobility of the two forms of Cdc42 is required for a concentration of Cdc42 at the polarity sites, with inactive form diffusing much faster than the active form \cite{bendezu2015}. We assume that the diffusion coefficients of active membrane-bound Cdc42 and GEF are much smaller than the cytosolic diffusion coefficients. All the parameter values used in the figures are summarized in Table~\ref{table:parm}.

\begin{table}[h!]
\caption{Parameter definitions and values used for simulations.} 
\vspace{0.2cm}
\centering 
\begin{tabular}{ l c c c c c } 
\hline 
Description & Parameters & 2-D Model & 1-D Model&  \\ [0.5ex] 
\hline 
\textbf{Cell size}\\
radius (2-D) &  $R$  & $1$ & -  \\
length (1-D) &  $L$  &  - & 1 \\
\textbf{Cdc42}\\
total amount & $C_{\rm tot}$ & $1.5\pi R^2$ & $1.5L$ \\ 
diffusion coefficient & $D_c$ & $[1,\infty]$  & $[1,\infty]$ \\  
lateral diffusion & $D_c^m$ & $\approx 0.01 \times D_c$   & -   \\ \\
autocatalysis & $k_0$ & 0.1 & 0.1 \\ 
& $k_{\rm cat}$ &40  & 40\\ 
dissociation rate & $k^-$ &  1  & 1 \\ 
\textbf{GEF}\\
total amount  & $G_{\rm tot}$ & $1.5\pi R^2$ & $1.5L$ \\
diffusion coefficient & $D_g$ & $[1,\infty]$  & $[1,\infty]$  \\
lateral diffusion & $D_g^m$ & $\approx 0.01\times D_g$  &  -  \\ \\
association rate & $k^{\rm on}$ & 1 & 1  \\ 
strength of -ve feedback & $\kappa$ &8    & 8 \\ 
dissociation rate & $k^{\rm off}$ &varied & varied  \\ 
\hline  
\end{tabular}
\label{table:parm}  
\end{table}

\section{Further analysis of the nonlocal PDE model}\label{sec:appendix_nonlocal}

\begin{figure}[t!]
\centering
\begin{subfigure}{0.36\linewidth}
\includegraphics[width=\linewidth]{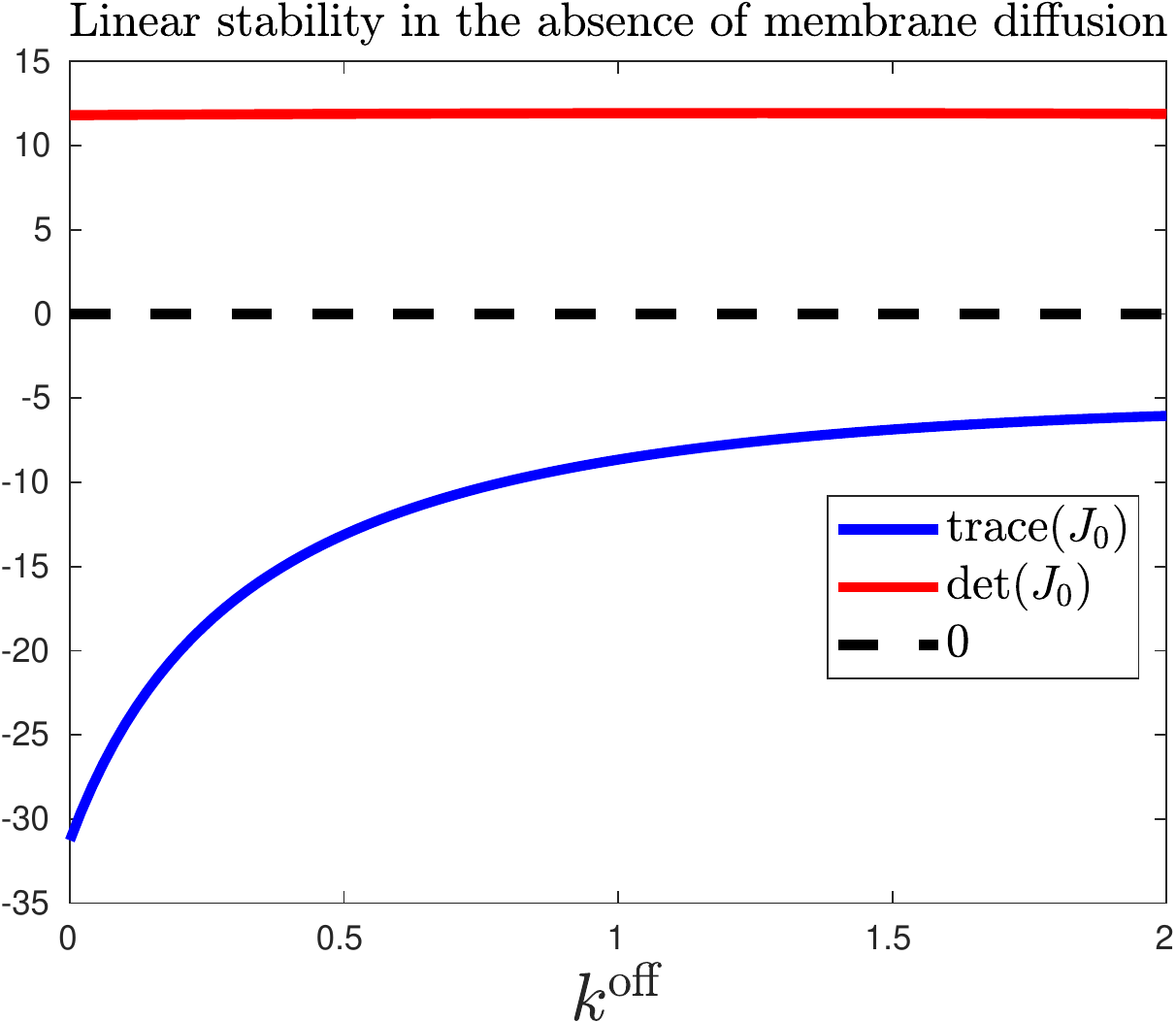}
\end{subfigure}
\caption{\label{fig:turingStability} Trace and determinant of the Jacobian matrix of the linearized system in the absence of surface diffusion (as defined in equation \eqref{eq:J_0}). Notice the negative trace and positive determinant. Parameter values are given in Table \ref{table:parm}.}
\end{figure}

We consider here the nonlocal reaction-diffusion from Section \ref{sec:nonlocal}. First, we verify numerically in Fig.~\ref{fig:turingStability} the stability condition stated in equation \eqref{eq:stability_condition} for a range of $\koff$ values employed in our simulations. This implies linear stability with respect to spatially uniform perturbations.

We then show that $\koff$ needs to be small for the oscillatory Turing instability to be possible. Let $J_n$ be the Jacobian matrix defined in \eqref{eq:J_n}, and suppose that it possesses a pair of purely imaginary eigenvalues. It therefore follows that
\begin{subequations}
\begin{align}
&\tr J_n = -(d_c(n)+d_g(n))+2\kcat c^*g^*C^*-(k^{-}+\koff)=0, \label{eq:HBa}\\
&\det J_n = [d_c(n)-2\kcat c^*g^*C^*+k^-][d_g(n)+\koff] + \eta >0, \label{eq:HBb}
\end{align}
\end{subequations}
where $d_c(n) = n^2D_c^m/R^2$ and $d_g(n) = n^2D_g^m/R^2$, with $\eta \geq 0$ as defined in \eqref{eq:eta}. Equation \eqref{eq:HBa} implies that
\begin{equation}
d_c(n)=2\kcat c^*g^*C^*-(k^{-}+\koff)-d_g(n).
\label{eq:d_c}
\end{equation}
Since $d_c(n)$ is positive, we must have
\begin{equation}\label{eq:cond1}
d_g(n) < 2\kcat c^*g^*C^*-(k^{-}+\koff),
\end{equation}
and upon substituting equation \eqref{eq:d_c} within \eqref{eq:HBb}, we obtain that 
\begin{equation}\label{eq:cond2}
-\left(d_g(n)+\koff\right)^2+\eta > 0,\, \quad \Rightarrow \quad d_g(n)<\sqrt{\eta}-\koff.
\end{equation}
Both conditions \eqref{eq:cond1} and \eqref{eq:cond2} imply that $d_g(n)$ has to be sufficiently small for a Hopf bifurcation to occur. In particular, we require
\begin{equation}\label{eq:HB}
d_g(n) < \min\left\{\sqrt{\eta}-\koff,\, 2\kcat c^*g^*C^*-k^{-}-\koff\right\}.
\end{equation}
The upper bound in \eqref{eq:HB} needs to be positive to make sure that $d_g(n) > 0$, yielding the following inequality:
\begin{equation}
\koff < \min\left\{ \sqrt{\eta},\, 2\kcat c^*g^*C^*-k^{-} \right\},
\end{equation}
which is satisfied when the GEF dissociation rate $\koff$ is sufficiently small. This is consistent with the stability diagrams as shown in Fig.~\ref{fig:stb_nonlocal}.

\section{Numerical methods}
\subsection{Numerical bifurcation analysis of the coupled PDE--ODE model}\label{sec:numerical_bifurcation_analysis}
Because of mass conservation, simple finite differences of the coupled PDE--ODE model will lead to an ill-posed system. In order to perform numerical continuation with AUTO (cf.~\cite{doedel2007}), we must spatially discretize a nonlocal formulation of the coupled PDE--ODE model. Let $U(x,t)$ and $V(x,t)$ be intermediate variables defined as
\begin{equation}\label{eq:UV}
 U(x,t) = c_1(t) + c_2(t) + \int_0^x C(s,t) ds, \quad V(x,t) = g_1(t) + g_2(t) + \int_0^x G(s,t) ds,
\end{equation}
that satisfy the following boundary conditions:
\begin{equation}
 U(0,t) = c_1(t) + c_2(t), \quad U(L,t) = \Ctot, \quad V(0,t) = g_1(t) + g_2(t), \quad V(L,t) = \Gtot.
\end{equation}
Differentiating $U(x,t)$ with respect to time yields
\begin{align*}
 \frac{\partial U}{\partial t} &= \frac{d c_1(t)}{dt} + \frac{d c_2(t)}{dt} + \int_0^x \frac{\partial C(s,t)}{\partial t} ds = \FF(C(0,t),c_1(t),g_1(t)) + \FF(C(L,t),c_2(t),g_2(t)) + \int_0^x D_c\frac{\partial^2 C(s,t)}{\partial s^2} ds, \\
 &= \FF(C(0,t),c_1(t),g_1(t)) + \FF(C(L,t),c_2(t),g_2(t)) + \left.D_c\frac{\partial C(s,t)}{\partial s}\right|_{s=x} - \left.D_c\frac{\partial C(s,t)}{\partial s}\right|_{s=0}, \\
 &= \FF(U_x(L,t),c_2(t),g_2(t)) + D_c U_{xx},
\end{align*}
and upon applying the same procedure to $V(x,t)$, we obtain the following system of nonlocal reaction-diffusion equations:
\begin{equation}
 U_t = D_c U_{xx} + \FF(U_x(L,t),c_2(t),g_2(t)), \quad V_t = D_g V_{xx} + \GG(V_x(L,t),c_2(t),g_2(t)), \quad 0 < x < L,
\end{equation}
that is coupled to the ODEs governing the dynamics of the local compartments, reformulated as
\begin{equation}
\frac{d c_i}{dt} = \FF\left(U_x(L(i-1),t),c_i(t),g_i(t)\right), \quad \frac{d g_i}{dt} = \GG\left(V_x(L(i-1),t),c_i(t),g_i(t)\right)\,,
\end{equation}
where $i=1,\,2$. Next, spatial discretization of this nonlocal PDE system yields
\begin{equation}
 \dot{\bm{W}} = \A \bm{W} + \bF(\bm{W}), \quad \bm{W} = \begin{pmatrix} \bU \\ \bV \\ c_1 \\ g_1 \\ c_2 \\ g_2 \end{pmatrix}, \quad
\bU(t) = \begin{pmatrix} U_2(t) \\ \vdots \\ U_j(t) \\ \vdots \\ U_{N-1}(t) \end{pmatrix}, \quad
\bV(t) = \begin{pmatrix} V_2(t) \\ \vdots \\ V_j(t) \\ \vdots \\ V_{N-1}(t) \end{pmatrix}
\end{equation}
where $U_j(t) \approx U((j-1)h,t)$ and $V_j(t) \approx V((j-1)h,t)$, with $h = \dfrac{L}{N-1}$ and $N$ the number of mesh points. The matrix $\A \in R^{2N \times 2N}$ and the nonlinear function $\bF$ are each defined by 
\begin{equation}
\A = \begin{pmatrix} D_c\Laplacian & 0 & 0 \\ 0 & D_g\Laplacian & 0 \\ 0 & 0 & 0 \\ \end{pmatrix}\,, \quad
\bF(\bm{W}) = 
\begin{pmatrix}
\frac{D_c((c_1+c_2)e_1 + \Ctot e_{N-2})}{h^2} + \FF\left(\frac{\Ctot - U_{N-1}}{h},c_2,g_2\right) \\
\frac{D_g((g_1+g_2)e_1 + \Gtot e_{N-2})}{h^2} + \GG\left(\frac{\Gtot - V_{N-1}}{h},c_2,g_2\right) \\
\FF\left(\frac{U_2 - c_1 - c_2}{h},c_1,g_1\right) \\
\GG\left(\frac{V_2 - g_1 - g_2}{h},c_1,g_1\right) \\
\FF\left(\frac{\Ctot - U_{N-1}}{h},c_2,g_2\right) \\
\GG\left(\frac{\Gtot - V_{N-1}}{h},c_2,g_2\right)
\end{pmatrix}\,,                 
\end{equation}
where the vectors $e_j \in \R^{N-2}$ form the standard orthonormal basis and $\Laplacian$ is the discrete Laplacian:
\begin{equation}
\Laplacian = \frac{1}{h^2}
\begin{pmatrix}
-2 & 1 & 0 & \ldots & 0 \\
1 & -2 & 1 & \ldots & 0\\
\vdots & \ddots & \ddots & \ddots & \vdots\\
0 & \ldots & 1 & -2 & 1 \\
0 & \ldots & 0 & 1 & -2 
\end{pmatrix}\,.
\end{equation}

\subsection{Numerical solution of the two-dimensional bulk-surface model}\label{2dnumerics}
In this section we describe the numerical method used to solve the coupled bulk-surface reaction-diffusion system (\ref{eq:2Dmodel}-\ref{eq:consv2dMAIN}). We have Fickian diffusion in the interior $\Omega$ and surface diffusion of membrane-bound components on $\partial\Omega$. The bulk-surface problem takes the general form
\begin{subequations}\label{eq:bulksurface}
\begin{align}
\label{eq:bulksurface_a} \frac{\partial \textbf{C}(\textbf{x},t)}{\partial t} = \textbf{D} \nabla^2 \textbf{C}(\textbf{x},t), \quad \textbf{x} \in \Omega \\ 
\label{eq:bulksurface_b} \frac{\partial \textbf{c}(\textbf{x},t)}{\partial t} = \textbf{d}\nabla_s^2 \textbf{c}(\textbf{x},t) + \textbf{f}(\textbf{C},\textbf{c}), \quad \textbf{x} \in \partial\Omega \\
 -\textbf{D}\frac{\partial \textbf{C}(\textbf{x},t)}{\partial n} = \textbf{f}(\textbf{C},\textbf{c}), \quad \textbf{x} \in \partial\Omega. \label{eq:Neumann}
\end{align}
\end{subequations}
with the following definitions:
\begin{equation}
\textbf{C} = \begin{pmatrix}
C \\ G
\end{pmatrix},
\qquad
\textbf{D} = 
\begin{pmatrix}
 D_c & 0 \\ 0 & D_g
\end{pmatrix},
 \qquad 
\textbf{c} = 
\begin{pmatrix}
c_s \\ g_s
\end{pmatrix},
\qquad
\textbf{d} = 
\begin{pmatrix}
D^m_c & 0 \\ 0 & D^m_g
\end{pmatrix}.
\qquad
\textbf{f} = 
\begin{pmatrix}
\mathcal{F} \\ \mathcal{G}
\end{pmatrix}.
\end{equation}

To numerically approximate this system, we discretize the bulk terms $\textbf{C}$ using a $\mathbb{P}^1$ finite element method \cite{2DFEM}. The surface terms $\textbf{c}$ are discretized using a finite difference approximation for the Laplace-Beltrami operator $ \nabla_s^2 \textbf{c}(\textbf{x},t)$ \cite{HUISKAMP1991} based on a modification to the more commonly used cotangent schemes \cite{Meyer2003}. The discretization of \eqref{eq:bulksurface} leads to a large system of ODEs with general form 
\begin{subequations}\label{eq:disc_system}
\begin{equation}\label{eq:matrixprob}
\M \dot{\bm{W}} = \A \bm{W} + \bF(\bm{W}),
\end{equation}
where
\begin{equation}\label{eq:twopartsystem}
\M = \begin{pmatrix}
\mathbb{K} & 0 & 0 & 0 \\ 0 & \mathbb{K} & 0 & 0 \\ 
0 & 0 & \mathbb{I} & 0 \\ 0 & 0 & 0 & \mathbb{I}
\end{pmatrix},
\qquad
\bm{W} = 
\begin{pmatrix}
C \\ G \\ c_s \\ g_s
\end{pmatrix},
\qquad
\A = 
\begin{pmatrix}
D_c\Laplacian & 0 & 0 & 0\\ 0 & D_g\Laplacian & 0 & 0 \\
0 & 0 & D_c^m \Laplacian_s & 0 \\ 0 & 0 & 0 & D_g^m\Laplacian_s 
\end{pmatrix},
\qquad
 \bF = 
\begin{pmatrix}
\textbf{G}_1 \\ 
\textbf{G}_2 \\ 
\mathcal{F} \\
 \mathcal{G}
\end{pmatrix},
\end{equation}
\end{subequations}
where $\Laplacian, \mathbb{K}$ are the stiffness and mass matrices from the finite element discretization, $\mathbb{I}$ is the identity matrix, and $\Laplacian_s$ is the discretized Laplace-Beltrami operator. The load vectors $\textbf{G}_1,\textbf{G}_2$ arise from the finite element approximations of the Neumann boundary condition \eqref{eq:Neumann} evaluated using a midpoint quadrature rule. The temporal integration of \eqref{eq:disc_system} has been performed by two independent methods with good agreement observed between them. The first method is the implicit second order Crank-Nicolson algorithm. The second is Radau IIA, a 3-stage implicit Runge-Kutta method of fifth order \cite{hairer96}. The nonlinear systems arising at each timestep are solved by Newton iterations.

\begin{figure}[H]
\centering
\begin{subfigure}{0.31\linewidth}
\includegraphics[width=\linewidth]{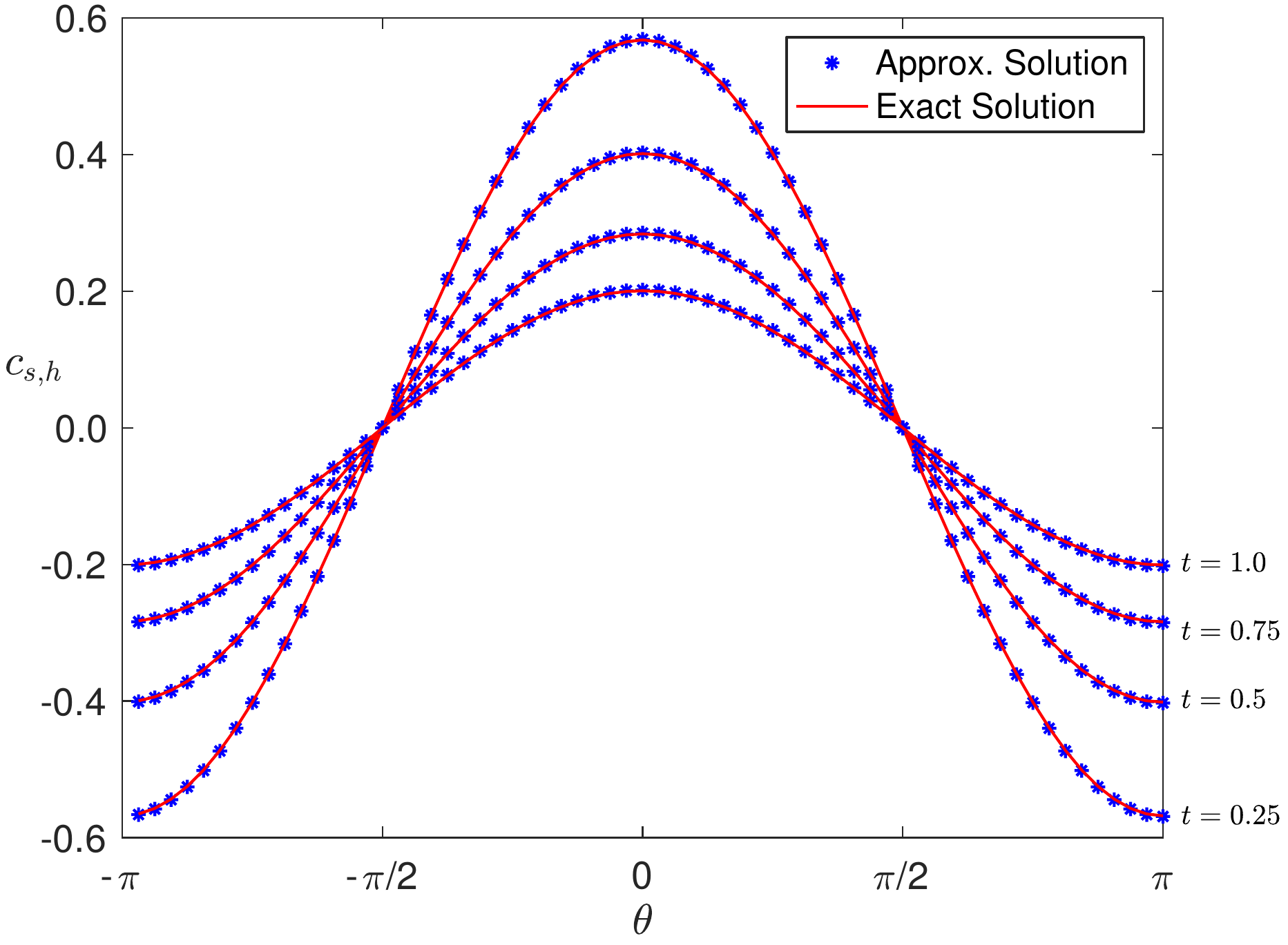}
\caption{Exact and numerical surface solutions.\label{fig:surfshots}}
\end{subfigure}\quad
\begin{subfigure}{0.31\linewidth}
\includegraphics[width=\linewidth]{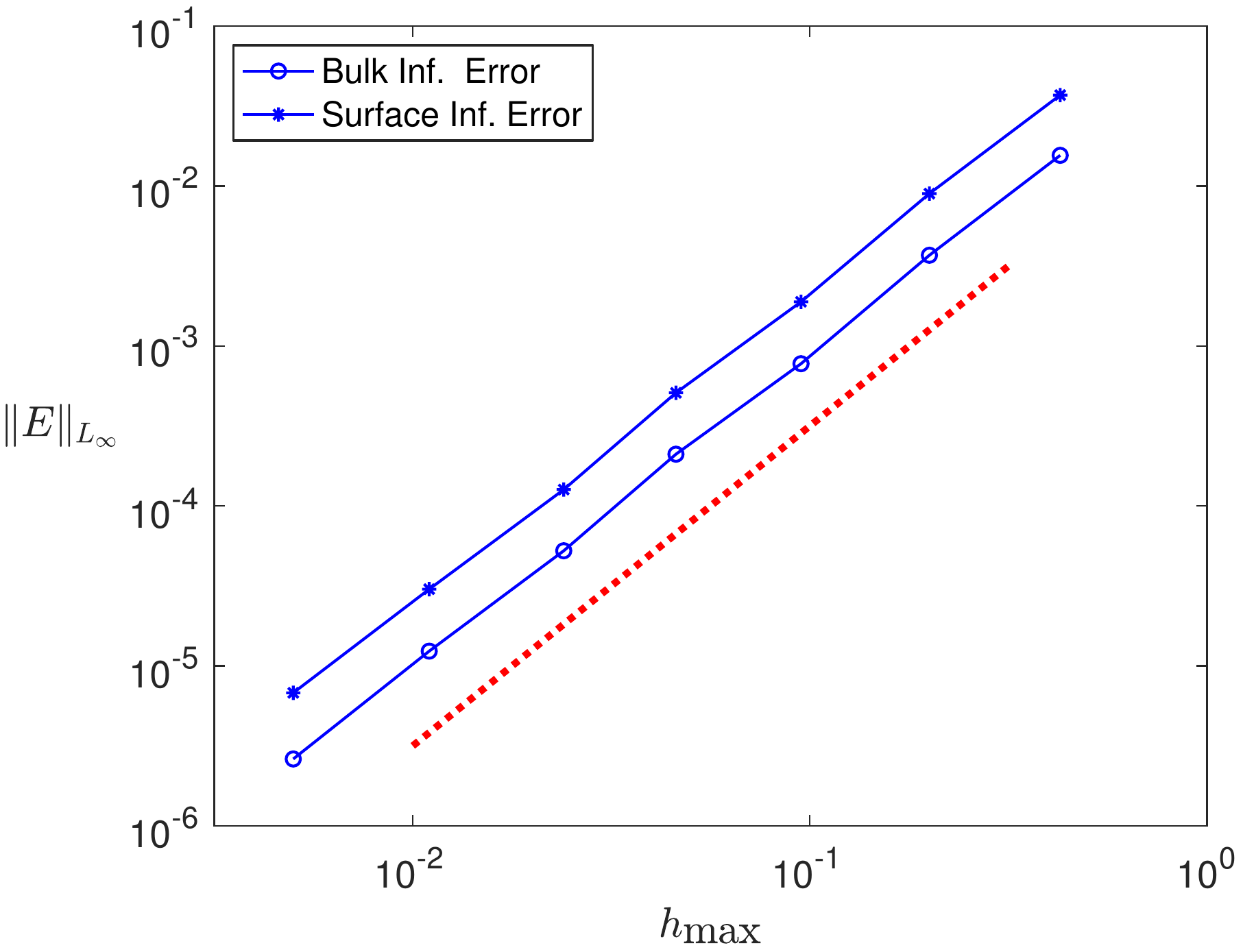}
\caption{Spatial convergence rates. \label{fig:spacecon}}
\end{subfigure}\quad
\begin{subfigure}{0.31\linewidth}
\includegraphics[width=\linewidth]{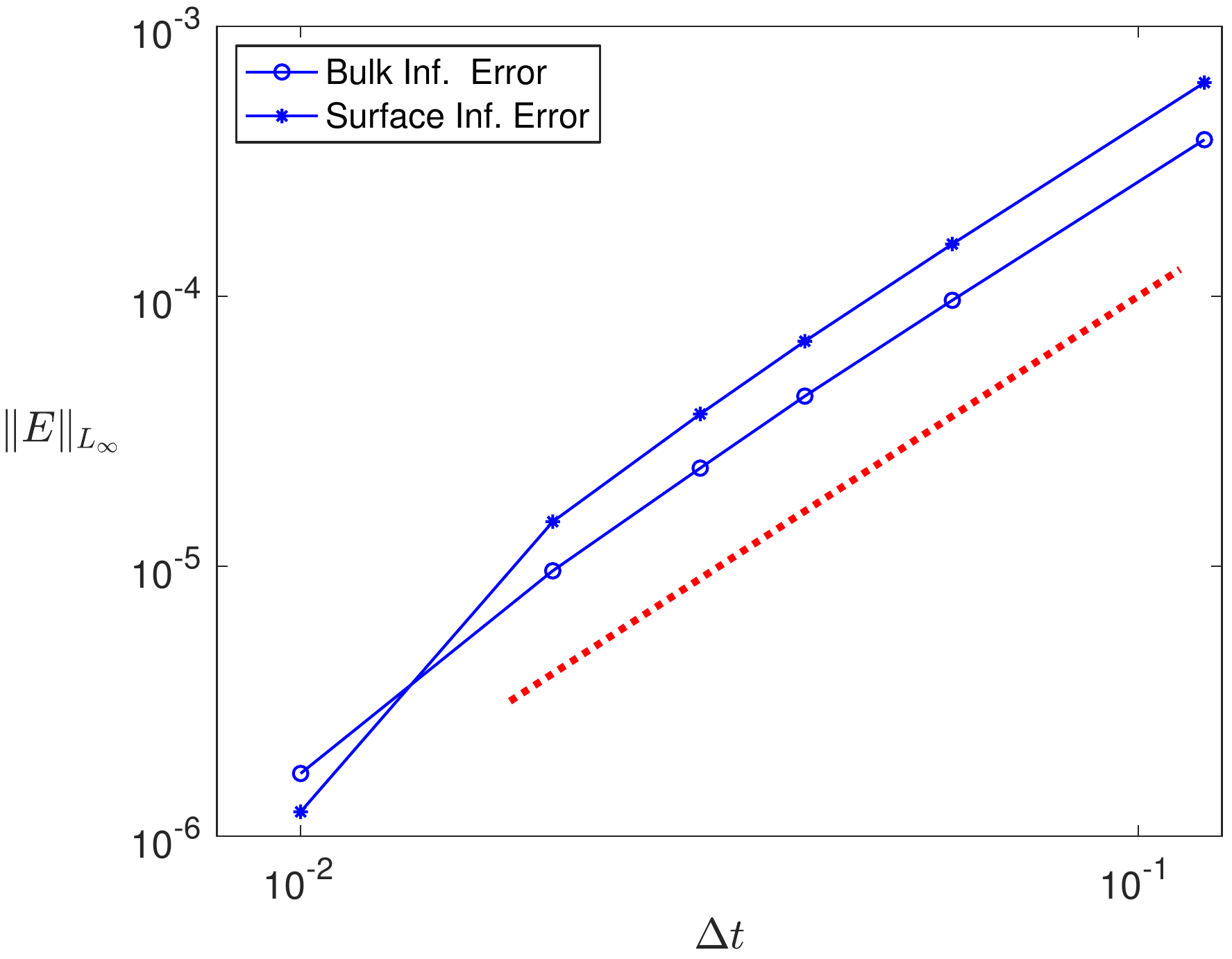}
\caption{Temporal convergence rates.\label{fig:tempcon}}
\end{subfigure}
\caption{Convergence study of the numerical method for the exactly solvable problem \eqref{eq:exampleprob}. Dotted red lines of slope $2$ added for comparison.} 
\label{fig:examples}
\end{figure}

In the case of Crank-Nicholson time integration, we demonstrate the expected second order convergence (in space and time) by using the following example (\cite{macdonald2016}) on the unit disk:
\begin{subequations}\label{eq:exampleprob}
\begin{align}
\frac{\partial C}{\partial t} = \nabla^2 C, \quad \textbf{x} \in \Omega, \\
\frac{\partial c_s}{\partial t} = \nabla_s^2 c_s + C - c_s, \quad \textbf{x} \in \partial\Omega, \\
-\frac{\partial C}{\partial n} = C - c_s, \quad \textbf{x} \in \partial\Omega. \label{eq:neumannex}
\end{align}
\end{subequations}
After discretization of \eqref{eq:exampleprob} with maximum edge length $h = h_{\text{max}}$, the numerical solution $(\textbf{C}_h , \textbf{c}_{s,h} )$ satisfies
\begin{subequations}
\begin{equation}
\M \dot{\bm{W}} = \A \bm{W} + \bF(\bm{W}),
\end{equation}
where
\begin{equation}
\mathbb{M} = \begin{pmatrix}
\mathbb{K} & 0\\ 0 & \mathbb{I}
\end{pmatrix},
\qquad
\bm{W} = 
\begin{pmatrix}
\textbf{C}_h \\ \textbf{c}_{s,h} 
\end{pmatrix},
 \qquad 
\A = 
\begin{pmatrix}
\Laplacian & 0\\ 0 & \Laplacian_s
\end{pmatrix},
\qquad
\bF = 
\begin{pmatrix}
\textbf{G}\\ \textbf{C}_h - \textbf{c}_{s,h}
\end{pmatrix},
\end{equation}
\end{subequations}
where $\textbf{G}$ is the Neumann boundary integral \eqref{eq:neumannex} evaluated using a midpoint quadrature rule. In polar coordinates $(r,\theta)$, the example problem \eqref{eq:exampleprob} has the exact solution
\begin{subequations}\label{eq:exactsol}
\begin{align}
C(r,\theta,t) = J_1(rk)e^{-k^2t} \cos \theta, \\[5pt]
c_s(\theta,t) = \frac{J_1(k)}{2-k^2}e^{-k^2t}\cos\theta, 
\end{align} 
\end{subequations}
where $J_1(z)$ is a Bessel function of the first kind and $k \approx 1.1777$. To show the accuracy of the surface approximation, we plot the numerical surface solution and the exact solution for various time points in Figure \ref{fig:examples} (for $h_{\text{max}} = 0.1, \Delta t = 10^{-3}$). To test spatial convergence, we fix $\Delta t = 10^{-4}$ and solve \eqref{eq:exampleprob} for varying levels of spatial refinement. Figure \ref{fig:spacecon} shows the expected second order convergence results for the bulk and surface error. Similarly, we test temporal convergence by fixing $h_{\text{max}} = 0.005$ and integrating \eqref{eq:exampleprob} to the final time $t=0.12$ using a sequence of decreasing timesteps. Figure \ref{fig:tempcon} shows the expected second order time convergence for the bulk and surface errors. 

\subsection{Mass conservation}\label{subsec:appendix_mass}
\begin{figure}[H]
\centering
\begin{subfigure}{0.4\linewidth}
\includegraphics[width=\linewidth]{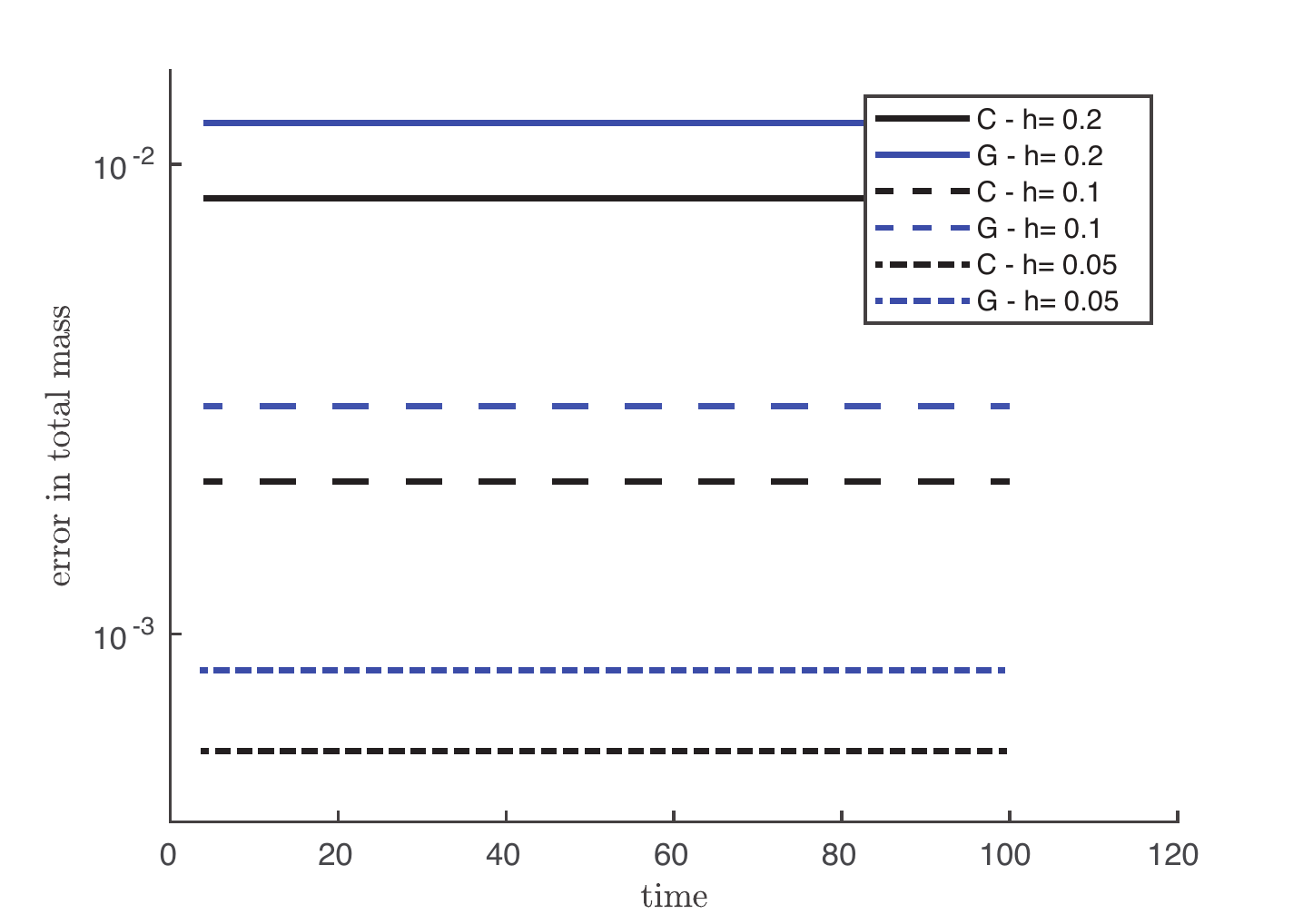}
\end{subfigure}
\caption{\label{fig:mass_conservation} Error in the total mass for the numerically computed traveling wave shown in Fig.~\ref{fig:p2p}.}
\end{figure}

In this section, we demonstrate that the 2-D coupled membrane-bulk model as described in Section \ref{subsec:2_D_membrane_bulk} conserves mass. We differentiate the conservation law for Cdc42 \eqref{eq:consv2d} with respect to time and obtain, after application of the divergence theorem on manifold with boundaries,
\begin{subequations}
\begin{align*}
\frac{d \Ctot}{dt} &= \int_{\Omega} D_c \nabla^2 Cd\x+\int_{\partial\Omega} \left(D_c^m\nabla_s^2 c+\FF[C, c, g] \right) d\x = \int_{\partial\Omega} \left(D_c (\n\cdot\nabla C) + D_c^m\nabla_s^2 c+\FF[C, c, g] \right) d\x\,. \\
\end{align*}
The first and third terms within the integral vanish because of the boundary condition, and we are left with the integral of the surface diffusion term over $\partial \Omega$,
\begin{align*}
\frac{d \Ctot}{dt} &= \int_{\partial\Omega} D_c^m\nabla_s^2 c \, d\x = 0\,,
\end{align*}
\end{subequations}
which vanishes because of the divergence theorem on manifolds without boundaries. The same procedure also applies to $\Gtot$, the total amount of GEF. Finally, we provide evidence that our numerical method preserves mass in Fig.~\ref{fig:mass_conservation}. As expected, the mismatch between the numerical and the ``true" mass decreases as the mesh size $h$ is refined. Quadratic convergence is roughly observed, with the error divided by 4 each time the maximum step size $h$ is halved.

\bibliographystyle{plainnat}
\bibliography{references}

\begin{thebibliography}{59}
\providecommand{\natexlab}[1]{#1}
\providecommand{\url}[1]{\texttt{#1}}
\expandafter\ifx\csname urlstyle\endcsname\relax
  \providecommand{\doi}[1]{doi: #1}\else
  \providecommand{\doi}{doi: \begingroup \urlstyle{rm}\Url}\fi

\bibitem[Baeumer et~al.(2018)Baeumer, Kov{\'a}cs, Meerschaert, and
  Sankaranarayanan]{baeumer2018}
Boris Baeumer, Mih{\'a}ly Kov{\'a}cs, Mark~M Meerschaert, and Harish
  Sankaranarayanan.
\newblock Boundary conditions for fractional diffusion.
\newblock \emph{Journal of Computational and Applied Mathematics},
  336:\penalty0 408--424, 2018.

\bibitem[Bendez{\'u} et~al.(2015)Bendez{\'u}, Vincenzetti, Vavylonis, Wyss,
  Vogel, and Martin]{bendezu2015}
Felipe~O Bendez{\'u}, Vincent Vincenzetti, Dimitrios Vavylonis, Romain Wyss,
  Horst Vogel, and Sophie~G Martin.
\newblock Spontaneous {Cdc42} polarization independent of {GDI}-mediated
  extraction and actin-based trafficking.
\newblock \emph{PLoS biology}, 13\penalty0 (4):\penalty0 e1002097, 2015.

\bibitem[Cerone et~al.(2012)Cerone, Nov{\'a}k, and Neufeld]{cerone2012}
Luca Cerone, B{\'e}la Nov{\'a}k, and Zolt{\'a}n Neufeld.
\newblock Mathematical model for growth regulation of fission yeast
  schizosaccharomyces pombe.
\newblock \emph{PloS one}, 7\penalty0 (11):\penalty0 e49675, 2012.

\bibitem[Chaplain et~al.(2015)Chaplain, Ptashnyk, and
  Sturrock]{chaplain2015hopf}
Mark Chaplain, Mariya Ptashnyk, and Marc Sturrock.
\newblock Hopf bifurcation in a gene regulatory network model: {M}olecular
  movement causes oscillations.
\newblock \emph{Mathematical Models and Methods in Applied Sciences},
  25\penalty0 (06):\penalty0 1179--1215, 2015.

\bibitem[Chiou et~al.(2018)Chiou, Ramirez, Elston, Witelski, Schaeffer, and
  Lew]{chiou2018}
Jian-Geng Chiou, Samuel~A Ramirez, Timothy~C Elston, Thomas~P Witelski, David~G
  Schaeffer, and Daniel~J Lew.
\newblock Principles that govern competition or co-existence in {Rho-GTPase}
  driven polarization.
\newblock \emph{PLoS computational biology}, 14\penalty0 (4):\penalty0
  e1006095, 2018.

\bibitem[Csik{\'a}sz-Nagy et~al.(2008)Csik{\'a}sz-Nagy, Gy{\H{o}}rffy, Alt,
  Tyson, and Nov{\'a}k]{csikasz2008}
Attila Csik{\'a}sz-Nagy, B{\'e}la Gy{\H{o}}rffy, Wolfgang Alt, John~J Tyson,
  and B{\'e}la Nov{\'a}k.
\newblock Spatial controls for growth zone formation during the fission yeast
  cell cycle.
\newblock \emph{Yeast}, 25\penalty0 (1):\penalty0 59--69, 2008.

\bibitem[Cusseddu et~al.(2018)Cusseddu, Edelstein-Keshet, Mackenzie, Portet,
  and Madzvamuse]{cusseddu2018}
Davide Cusseddu, Leah Edelstein-Keshet, John~A Mackenzie, St{\'e}phanie Portet,
  and Anotida Madzvamuse.
\newblock A coupled bulk-surface model for cell polarisation.
\newblock \emph{Journal of theoretical biology}, 2018.

\bibitem[Dangelmayr and Knobloch(1987)]{dangelmayr1987}
Gerhard Dangelmayr and Edgar Knobloch.
\newblock The {Takens-Bogdanov} bifurcation with {O}(2)-symmetry.
\newblock \emph{Philosophical Transactions of the Royal Society of London.
  Series A, Mathematical and Physical Sciences}, 322\penalty0 (1565):\penalty0
  243--279, 1987.

\bibitem[Das et~al.(2012)Das, Drake, Wiley, Buchwald, Vavylonis, and
  Verde]{das2012}
Maitreyi Das, Tyler Drake, David~J Wiley, Peter Buchwald, Dimitrios Vavylonis,
  and Fulvia Verde.
\newblock Oscillatory dynamics of {Cdc42 GTPase} in the control of polarized
  growth.
\newblock \emph{Science}, 337\penalty0 (6091):\penalty0 239--243, 2012.

\bibitem[Das et~al.(2015)Das, Nu{\~n}ez, Rodriguez, Wiley, Rodriguez,
  Sarkeshik, Yates~III, Buchwald, and Verde]{das2015}
Maitreyi Das, Illyce Nu{\~n}ez, Marbelys Rodriguez, David~J Wiley, Juan
  Rodriguez, Ali Sarkeshik, John~R Yates~III, Peter Buchwald, and Fulvia Verde.
\newblock Phosphorylation-dependent inhibition of {Cdc42 GEF Gef1} by 14-3-3
  protein {Rad24} spatially regulates {Cdc42 GTPase} activity and oscillatory
  dynamics during cell morphogenesis.
\newblock \emph{Molecular biology of the cell}, 26\penalty0 (19):\penalty0
  3520--3534, 2015.

\bibitem[Diegmiller et~al.(2018)Diegmiller, Montanelli, Muratov, and
  Shvartsman]{diegmiller2018}
Rocky Diegmiller, Hadrien Montanelli, Cyrill~B Muratov, and Stanislav~Y
  Shvartsman.
\newblock Spherical caps in cell polarization.
\newblock \emph{Biophysical journal}, 115\penalty0 (1):\penalty0 26--30, 2018.

\bibitem[Doedel et~al.(2007)Doedel, Champneys, Fairgrieve, Kuznetsov, Oldeman,
  Paffenroth, Sandstede, Wang, and Zhang]{doedel2007}
E.~J. Doedel, A.~R. Champneys, T.~Fairgrieve, Y.~Kuznetsov, B.~Oldeman,
  R.~Paffenroth, B.~Sandstede, X.~Wang, and C.~Zhang.
\newblock Auto07p: Continuation and bifurcation software for ordinary
  differential equations.
\newblock \emph{Technical report, Concordia University}, 2007.

\bibitem[Etienne-Manneville and Hall(2002)]{etienne2002}
Sandrine Etienne-Manneville and Alan Hall.
\newblock {Rho GTPases} in cell biology.
\newblock \emph{Nature}, 420\penalty0 (6916):\penalty0 629, 2002.

\bibitem[Freisinger et~al.(2013)Freisinger, Kl{\"u}nder, Johnson, M{\"u}ller,
  Pichler, Beck, Costanzo, Boone, Cerione, Frey, and
  Wedlich-S{\"o}ldner]{freisinger2013}
Tina Freisinger, Ben Kl{\"u}nder, Jared Johnson, Nikola M{\"u}ller, Garwin
  Pichler, Gisela Beck, Michael Costanzo, Charles Boone, Richard~A Cerione,
  Erwin Frey, and Roland Wedlich-S{\"o}ldner.
\newblock Establishment of a robust single axis of cell polarity by coupling
  multiple positive feedback loops.
\newblock \emph{Nature communications}, 4:\penalty0 1807, 2013.

\bibitem[Funken et~al.(2011)Funken, Praetorius, and Wissgott]{2DFEM}
S.~Funken, D.~Praetorius, and P.~Wissgott.
\newblock Efficient implementation of adaptive {P}1-{FEM} in {M}atlab.
\newblock \emph{Computational Methods in Applied Mathematics}, 11\penalty0
  (4):\penalty0 460--490, 2011.

\bibitem[Gomez et~al.(2019)Gomez, Ward, and Wei]{gomez2019}
Daniel Gomez, Michael~J. Ward, and Juncheng Wei.
\newblock The linear stability of symmetric spike patterns for a bulk-membrane
  coupled {G}ierer-{M}einhardt model.
\newblock \emph{SIAM J. Appl. Dyn. Syst.}, 18\penalty0 (2):\penalty0 729--768,
  2019.

\bibitem[Gomez-Marin et~al.(2007)Gomez-Marin, Garcia-Ojalvo, and
  Sancho]{gomez2007}
Alejandro Gomez-Marin, Jordi Garcia-Ojalvo, and Jos{\'e}~M Sancho.
\newblock Self-sustained spatiotemporal oscillations induced by membrane-bulk
  coupling.
\newblock \emph{Physical review letters}, 98\penalty0 (16):\penalty0 168303,
  2007.

\bibitem[Goryachev and Leda(2017)]{goryachev2017}
Andrew~B Goryachev and Marcin Leda.
\newblock Many roads to symmetry breaking: molecular mechanisms and theoretical
  models of yeast cell polarity.
\newblock \emph{Molecular biology of the cell}, 28\penalty0 (3):\penalty0
  370--380, 2017.

\bibitem[Goryachev and Pokhilko(2008)]{goryachev2008}
Andrew~B Goryachev and Alexandra~V Pokhilko.
\newblock Dynamics of {C}dc42 network embodies a {T}uring-type mechanism of
  yeast cell polarity.
\newblock \emph{FEBS letters}, 582\penalty0 (10):\penalty0 1437--1443, 2008.

\bibitem[Gou et~al.(2015)Gou, Li, Nagata, and Ward]{gou2015_SIAM}
J~Gou, YX~Li, W~Nagata, and MJ~Ward.
\newblock Synchronized oscillatory dynamics for a {1-D} model of membrane
  kinetics coupled by linear bulk diffusion.
\newblock \emph{SIAM Journal on Applied Dynamical Systems}, 14\penalty0
  (4):\penalty0 2096--2137, 2015.

\bibitem[Gou et~al.(2017)Gou, Chiang, Lai, Ward, and Li]{gou2017}
Jia Gou, Wei-Yin Chiang, Pik-Yin Lai, Michael~J Ward, and Yue-Xian Li.
\newblock A theory of synchrony by coupling through a diffusive chemical
  signal.
\newblock \emph{Physica D: Nonlinear Phenomena}, 339:\penalty0 1--17, 2017.

\bibitem[Hairer and Wanner(1996)]{hairer96}
Ernst Hairer and Gerhard Wanner.
\newblock \emph{Solving Ordinary Differential Equations II: Stiff and
  Differential Algebraic Problems}.
\newblock Springer, 1996.

\bibitem[Halatek and Frey(2012)]{halatek2012}
Jacob Halatek and Erwin Frey.
\newblock Highly canalized {MinD} transfer and {MinE} sequestration explain the
  origin of robust {MinCDE}-protein dynamics.
\newblock \emph{Cell reports}, 1\penalty0 (6):\penalty0 741--752, 2012.

\bibitem[Halatek et~al.(2018)Halatek, Brauns, and Frey]{halatek2018self}
Jacob Halatek, Fridtjof Brauns, and Erwin Frey.
\newblock Self-organization principles of intracellular pattern formation.
\newblock \emph{Philosophical Transactions of the Royal Society B: Biological
  Sciences}, 373\penalty0 (1747), 2018.

\bibitem[Henry and Wearne(2002)]{henry2002}
Bruce~Ian Henry and Susan~L Wearne.
\newblock Existence of turing instabilities in a two-species fractional
  reaction-diffusion system.
\newblock \emph{SIAM Journal on Applied Mathematics}, 62\penalty0 (3):\penalty0
  870--887, 2002.

\bibitem[Hercyk et~al.(2019)Hercyk, Rich-Robinson, Mitoubsi, Harrell, and
  Das]{hercyk2019}
Brian~S Hercyk, Julie Rich-Robinson, Ahmad~S Mitoubsi, Marcus~A Harrell, and
  Maitreyi~E Das.
\newblock A novel interplay between {GEFs} orchestrates {Cdc42} activity during
  cell polarity and cytokinesis.
\newblock \emph{Journal of cell science}, 2019.

\bibitem[Huang et~al.(2003)Huang, Meir, and Wingreen]{huang2003}
Kerwyn~Casey Huang, Yigal Meir, and Ned~S Wingreen.
\newblock Dynamic structures in {Escherichia} coli: spontaneous formation of
  mine rings and mind polar zones.
\newblock \emph{Proceedings of the National Academy of Sciences}, 100\penalty0
  (22):\penalty0 12724--12728, 2003.

\bibitem[Huiskamp(1991)]{HUISKAMP1991}
Geertjan Huiskamp.
\newblock Difference formulas for the surface laplacian on a triangulated
  surface.
\newblock \emph{Journal of Computational Physics}, 95\penalty0 (2):\penalty0
  477 -- 496, 1991.
\newblock ISSN 0021-9991.
\newblock \doi{https://doi.org/10.1016/0021-9991(91)90286-T}.
\newblock URL
  \url{http://www.sciencedirect.com/science/article/pii/002199919190286T}.

\bibitem[Jeon et~al.(2011)Jeon, Tejedor, Burov, Barkai, Selhuber-Unkel,
  Berg-S{\o}rensen, Oddershede, and Metzler]{jeon2011}
Jae-Hyung Jeon, Vincent Tejedor, Stas Burov, Eli Barkai, Christine
  Selhuber-Unkel, Kirstine Berg-S{\o}rensen, Lene Oddershede, and Ralf Metzler.
\newblock In vivo anomalous diffusion and weak ergodicity breaking of lipid
  granules.
\newblock \emph{Physical review letters}, 106\penalty0 (4):\penalty0 048103,
  2011.

\bibitem[Jilkine and Edelstein-Keshet(2011)]{jilkine2011comparison}
Alexandra Jilkine and Leah Edelstein-Keshet.
\newblock A comparison of mathematical models for polarization of single
  eukaryotic cells in response to guided cues.
\newblock \emph{PLoS computational biology}, 7\penalty0 (4):\penalty0 e1001121,
  2011.

\bibitem[Jilkine et~al.(2011)Jilkine, Angenent, Wu, and
  Altschuler]{jilkine2011}
Alexandra Jilkine, Sigurd~B Angenent, Lani~F Wu, and Steven~J Altschuler.
\newblock A density-dependent switch drives stochastic clustering and
  polarization of signaling molecules.
\newblock \emph{PLoS computational biology}, 7\penalty0 (11):\penalty0
  e1002271, 2011.

\bibitem[Johnson et~al.(2011)Johnson, Jin, and Lew]{johnson2011}
Jayme~M Johnson, Meng Jin, and Daniel~J Lew.
\newblock Symmetry breaking and the establishment of cell polarity in budding
  yeast.
\newblock \emph{Current opinion in genetics \& development}, 21\penalty0
  (6):\penalty0 740--746, 2011.

\bibitem[Kokkoris et~al.(2014)Kokkoris, Castro, and Martin]{kokkoris2014}
Kyriakos Kokkoris, Daniela~Gallo Castro, and Sophie~G Martin.
\newblock The {T}ea4--{PP}1 landmark promotes local growth by dual {C}dc42
  {GEF} recruitment and {GAP} exclusion.
\newblock \emph{J Cell Sci}, 127\penalty0 (9):\penalty0 2005--2016, 2014.

\bibitem[Kruse et~al.(2007)Kruse, Howard, and Margolin]{kruse2007}
Karsten Kruse, Martin Howard, and William Margolin.
\newblock An experimentalist's guide to computational modelling of the min
  system.
\newblock \emph{Molecular microbiology}, 63\penalty0 (5):\penalty0 1279--1284,
  2007.

\bibitem[Levine and Rappel(2005)]{levine2005membrane}
Herbert Levine and Wouter-Jan Rappel.
\newblock Membrane-bound turing patterns.
\newblock \emph{Physical Review E}, 72\penalty0 (6):\penalty0 061912, 2005.

\bibitem[Lo et~al.(2014)Lo, Park, and Chou]{lo2014}
Wing-Cheong Lo, Hay-Oak Park, and Ching-Shan Chou.
\newblock Mathematical analysis of spontaneous emergence of cell polarity.
\newblock \emph{Bulletin of mathematical biology}, 76\penalty0 (8):\penalty0
  1835--1865, 2014.

\bibitem[Loose et~al.(2011)Loose, Kruse, and Schwille]{loose2011}
Martin Loose, Karsten Kruse, and Petra Schwille.
\newblock Protein self-organization: lessons from the min system.
\newblock \emph{Annual review of biophysics}, 40:\penalty0 315--336, 2011.

\bibitem[MacDonald et~al.(2016)MacDonald, Mackenzie, Nolan, and
  Insall]{macdonald2016}
G.~MacDonald, J.A. Mackenzie, M.~Nolan, and R.H. Insall.
\newblock A computational method for the coupled solution of
  reaction--diffusion equations on evolving domains and manifolds: Application
  to a model of cell migration and chemotaxis.
\newblock \emph{Journal of Computational Physics}, 309:\penalty0 207 -- 226,
  2016.
\newblock ISSN 0021-9991.

\bibitem[Madzvamuse et~al.(2015)Madzvamuse, Chung, and
  Venkataraman]{madzvamuse2015}
Anotida Madzvamuse, Andy~HW Chung, and Chandrasekhar Venkataraman.
\newblock Stability analysis and simulations of coupled bulk-surface
  reaction--diffusion systems.
\newblock \emph{Proc. R. Soc. A}, 471\penalty0 (2175):\penalty0 20140546, 2015.

\bibitem[Martin and Arkowitz(2014)]{martin2014}
Sophie~G Martin and Robert~A Arkowitz.
\newblock Cell polarization in budding and fission yeasts.
\newblock \emph{FEMS microbiology reviews}, 38\penalty0 (2):\penalty0 228--253,
  2014.

\bibitem[Meyer et~al.(2003)Meyer, Desbrun, Schr{\"o}der, and Barr]{Meyer2003}
Mark Meyer, Mathieu Desbrun, Peter Schr{\"o}der, and Alan~H. Barr.
\newblock Discrete differential-geometry operators for triangulated
  2-manifolds.
\newblock In Hans-Christian Hege and Konrad Polthier, editors,
  \emph{Visualization and Mathematics III}, pages 35--57, Berlin, Heidelberg,
  2003. Springer Berlin Heidelberg.

\bibitem[Mori et~al.(2008)Mori, Jilkine, and Edelstein-Keshet]{mori2008}
Yoichiro Mori, Alexandra Jilkine, and Leah Edelstein-Keshet.
\newblock Wave-pinning and cell polarity from a bistable reaction-diffusion
  system.
\newblock \emph{Biophysical journal}, 94\penalty0 (9):\penalty0 3684--3697,
  2008.

\bibitem[Mori et~al.(2011)Mori, Jilkine, and Edelstein-Keshet]{mori2011}
Yoichiro Mori, Alexandra Jilkine, and Leah Edelstein-Keshet.
\newblock Asymptotic and bifurcation analysis of wave-pinning in a
  reaction-diffusion model for cell polarization.
\newblock \emph{SIAM journal on applied mathematics}, 71\penalty0 (4):\penalty0
  1401--1427, 2011.

\bibitem[Otsuji et~al.(2007)Otsuji, Ishihara, Kaibuchi, Mochizuki, Kuroda,
  et~al.]{otsuji2007}
Mikiya Otsuji, Shuji Ishihara, Kozo Kaibuchi, Atsushi Mochizuki, Shinya Kuroda,
  et~al.
\newblock A mass conserved reaction--diffusion system captures properties of
  cell polarity.
\newblock \emph{PLoS computational biology}, 3\penalty0 (6):\penalty0 e108,
  2007.

\bibitem[Ozbudak et~al.(2005)Ozbudak, Becskei, and
  Van~Oudenaarden]{ozbudak2005}
Ertugrul~M Ozbudak, Attila Becskei, and Alexander Van~Oudenaarden.
\newblock A system of counteracting feedback loops regulates {Cdc42p} activity
  during spontaneous cell polarization.
\newblock \emph{Developmental cell}, 9\penalty0 (4):\penalty0 565--571, 2005.

\bibitem[Paquin-Lefebvre et~al.(2019)Paquin-Lefebvre, Nagata, and
  Ward]{paquin2018}
Fr\'{e}d\'{e}ric Paquin-Lefebvre, Wayne Nagata, and Michael~J. Ward.
\newblock Pattern {F}ormation and {O}scillatory {D}ynamics in a
  {T}wo-{D}imensional {C}oupled {B}ulk-{S}urface {R}eaction-{D}iffusion
  {S}ystem.
\newblock \emph{SIAM J. Appl. Dyn. Syst.}, 18\penalty0 (3):\penalty0
  1334--1390, 2019.

\bibitem[Rappel and Edelstein-Keshet(2017)]{rappel2017mechanisms}
Wouter-Jan Rappel and Leah Edelstein-Keshet.
\newblock Mechanisms of cell polarization.
\newblock \emph{Current opinion in systems biology}, 3:\penalty0 43--53, 2017.

\bibitem[R{\"a}tz(2015)]{ratz2015turing}
Andreas R{\"a}tz.
\newblock Turing-type instabilities in bulk--surface reaction--diffusion
  systems.
\newblock \emph{Journal of Computational and Applied Mathematics},
  289:\penalty0 142--152, 2015.

\bibitem[R{\"a}tz and R{\"o}ger(2012)]{ratz2012turing}
Andreas R{\"a}tz and Matthias R{\"o}ger.
\newblock Turing instabilities in a mathematical model for signaling networks.
\newblock \emph{Journal of mathematical biology}, 65\penalty0 (6-7):\penalty0
  1215--1244, 2012.

\bibitem[R{\"a}tz and R{\"o}ger(2014)]{ratz2014}
Andreas R{\"a}tz and Matthias R{\"o}ger.
\newblock Symmetry breaking in a bulk--surface reaction--diffusion model for
  signalling networks.
\newblock \emph{Nonlinearity}, 27\penalty0 (8):\penalty0 1805, 2014.

\bibitem[Sturrock et~al.(2011)Sturrock, Terry, Xirodimas, Thompson, and
  Chaplain]{sturrock2011}
Marc Sturrock, Alan~J Terry, Dimitris~P Xirodimas, Alastair~M Thompson, and
  Mark~AJ Chaplain.
\newblock Spatio-temporal modelling of the {H}es1 and p53-{M}dm2 intracellular
  signalling pathways.
\newblock \emph{Journal of theoretical biology}, 273\penalty0 (1):\penalty0
  15--31, 2011.

\bibitem[Tay et~al.(2018)Tay, Leda, Goryachev, and Sawin]{tay2018}
Ye~Dee Tay, Marcin Leda, Andrew~B Goryachev, and Kenneth~E Sawin.
\newblock Local and global {Cdc42} guanine nucleotide exchange factors for
  fission yeast cell polarity are coordinated by microtubules and the
  tea1--tea4--pom1 axis.
\newblock \emph{J Cell Sci}, 131\penalty0 (14):\penalty0 jcs216580, 2018.

\bibitem[Turing(1952)]{turing1952}
Alan~M. Turing.
\newblock The chemical basis of morphogenesis.
\newblock \emph{Philosophical Transactions of the Royal Society of London.
  Series B, Biological Sciences}, 237\penalty0 (641):\penalty0 37--72, 1952.

\bibitem[van Gils and Mallet-Paret(1986)]{vangils1986}
Stephan~A. van Gils and John Mallet-Paret.
\newblock Hopf bifurcation and symmetry: travelling and standing waves on the
  circle.
\newblock \emph{Proc. Roy. Soc. Edinburgh Sect. A}, 104\penalty0
  (3-4):\penalty0 279--307, 1986.

\bibitem[Wu and Lew(2013)]{wu2013}
Chi-Fang Wu and Daniel~J Lew.
\newblock Beyond symmetry-breaking: competition and negative feedback in
  {GTPase} regulation.
\newblock \emph{Trends in cell biology}, 23\penalty0 (10):\penalty0 476--483,
  2013.

\bibitem[Xu and Bressloff(2016)]{xu2016}
Bin Xu and Paul~C Bressloff.
\newblock A {PDE}-{DDE} model for cell polarization in fission yeast.
\newblock \emph{SIAM Journal on Applied Mathematics}, 76\penalty0 (5):\penalty0
  1844--1870, 2016.

\bibitem[Xu and Bressloff(2017)]{xu2017}
Bin Xu and Paul~C Bressloff.
\newblock A theory of synchrony for active compartments with delays coupled
  through bulk diffusion.
\newblock \emph{Physica D: Nonlinear Phenomena}, 341:\penalty0 45--59, 2017.

\bibitem[Xu and Jilkine(2018)]{xu2018}
Bin Xu and Alexandra Jilkine.
\newblock Modeling the dynamics of {Cdc42} oscillation in fission yeast.
\newblock \emph{Biophysical Journal}, 114\penalty0 (3):\penalty0 711 -- 722,
  2018.

\bibitem[Xu et~al.(2019)Xu, Kang, and Jilkine]{xu2018BMB}
Bin Xu, Hye-Won Kang, and Alexandra Jilkine.
\newblock Comparison of deterministic and stochastic regime in a model for
  {C}dc42 oscillations in fission yeast.
\newblock \emph{Bulletin of Mathematical Biology}, 81\penalty0 (5):\penalty0
  1268--1302, 2019.

\end{thebibliography}

\end{document}